%% file: draft.tex
\documentclass[a4paper]{llncs}

\usepackage{amsfonts}
\usepackage{amsmath}
\usepackage{amssymb}
\usepackage{mathabx}
\usepackage{leftidx}
\usepackage{epsfig}
\usepackage{paralist}
\usepackage{graphics}
\usepackage{txfonts}
\usepackage{framed}
\usepackage{makecell}
\usepackage{url}
\usepackage[colorlinks]{hyperref}
\usepackage{proof}

\input commands

\newif\ifLongVersion\LongVersiontrue

\begin{document}
%%%%%%%%%%%%%%%%%%%%%%%%%%%%%%%%%%%%%%%%%%%%%%%%%%%%%%%%%%%%%%%%%%%%%%%%%%%%%%%

\title{Local Reasoning about Parametric and Reconfigurable Component-based Systems}
\author{Marius Bozga \and Radu Iosif \and Joseph Sifakis}
\institute{VERIMAG, CNRS, Universit\'e de Grenoble}

\maketitle

\input abstract
\input body

%%%%%%%%%%%%%%%%%%%%%%%%%%%%%%%%%%%%%%%%%%%%%%%%%%%%%%%%%%%%%%%%%%%%%%%%%%%%%%%
\bibliographystyle{splncs03}
\bibliography{refs}
%%%%%%%%%%%%%%%%%%%%%%%%%%%%%%%%%%%%%%%%%%%%%%%%%%%%%%%%%%%%%%%%%%%%%%%%%%%%%%%

%%%%%%%%%%%%%%%%%%%%%%%%%%%%%%%%%%%%%%%%%%%%%%%%%%%%%%%%%%%%%%%%%%%%%%%%%%%%%%%
\end{document}

%% file: commands.tex
\newtheorem{fact}{Fact}

\newcommand\todo[1]{
}

\newcommand{\bigO}{\mathcal{O}}

\newcommand{\pspace}{$\mathsf{PSPACE}$}

\newcommand{\expspace}{$\mathsf{EXPSPACE}$}

\newcommand{\set}[1]{\{ #1 \}}
\newcommand{\tuple}[1]{\langle #1 \rangle}
\renewcommand{\vec}[1]{\overline #1}

\newcommand{\isdef}{\stackrel{\scriptscriptstyle{\mathsf{def}}}{=}}

\newcommand{\len}[1]{{\left|{#1}\right|}}
\newcommand{\card}[1]{{||{#1}||}}

\newcommand{\arrow}[2]{\xrightarrow{{\scriptscriptstyle #1}}_{{\scriptscriptstyle #2}}}

\newcommand{\nat}{{\mathbb{N}}}

\renewcommand{\paragraph}[1]{\noindent{\bf #1}}

\newcommand{\I}{\mathcal{I}}

\newcommand{\vars}{\mathsf{Var}}

\newcommand{\pvars}{\mathsf{PVars}}
\newcommand{\ivars}{\mathsf{IVars}}

\newcommand{\preds}{\mathsf{Pred}}

\newcommand{\fv}[1]{\mathrm{fv}({#1})}

\newcommand{\behave}[2]{[{#1}]_{#2}}

\newcommand{\ports}{\mathsf{Ports}}
\newcommand{\psym}{\mathsf{PSym}}
\newcommand{\pport}{\mathsf{p}}
\newcommand{\qport}{\mathsf{q}}

\newcommand{\portsof}[1]{\mathsf{P}({#1})}
\newcommand{\bound}[1]{\mathrm{bnd}({#1})}

\newcommand{\arch}{\mathcal{A}}
\newcommand{\archset}{\mathsf{Arch}}

\newcommand{\states}{\mathsf{Q}}

\newcommand{\rules}{\rightarrow}

\newcommand{\nil}{\mathsf{nil}}
\newcommand{\emp}{\mathsf{emp}}
\newcommand{\inter}{\mathrel{\mbox{\hspace*{-0.005em}$\multimap$\hspace*{-0.005em}}}}
\newcommand{\exinter}{\mathrel{\mbox{\hspace*{-0.005em}$\overset{\scriptscriptstyle{\exists}}{\multimap}$\hspace*{-0.005em}}}}

\newcommand{\closeinter}{\mathrel{\mbox{$-\!\raisebox{.5pt}{$\scriptstyle{\square}$}$\hspace*{-0.005em}}}}
\newcommand{\closexinter}{\mathrel{\mbox{\hspace*{-0.005em}$\overset{\scriptscriptstyle{\exists}}{-}\!\raisebox{.5pt}{$\scriptstyle{\square}$}$\hspace*{-0.005em}}}}

\newcommand{\wand}{
\mathrel{\mbox{$\hspace*{-0.03em}\mathord{-}\hspace*{-0.66em}
 \mathord{-}\hspace*{-0.36em}\mathord{*}$\hspace*{-0.005em}}}
}
\newcommand{\has}[1]{\mathrm{has}({#1})}
\newcommand{\ieinter}{\mathrel{\mbox{\hspace*{-0.005em}$\stackrel{\scriptscriptstyle{\exists}}{\hookrightarrow}\!\!\circ$\hspace*{-0.005em}}}}
\newcommand{\iecloseinter}{\mathrel{\mbox{\hspace*{-0.005em}$\stackrel{\scriptscriptstyle{\exists}}{\hookrightarrow}\!\!\Box$\hspace*{-0.005em}}}}

\newcommand{\type}[2]{\mathrm{type}({#1}) \geq {#2}}
\newcommand{\test}[1]{\mathsf{TestForm}({#1})}
\newcommand{\stest}[2]{\mathsf{TestForm}({#1},{#2})}
\newcommand{\tfeq}[1]{\simeq_{#1}}

\newcommand{\X}{\mathcal{X}}

\newcommand{\intersect}[2]{\interproj{#2}{(#1)}}
\newcommand{\gtype}[3]{\interproj{#3}{\set{{#1}}} \sqcap {#2}}
\newcommand{\vtype}[3]{\mathrm{vt}_{#3,#2}({#1})}
\newcommand{\veq}[2]{\approx_{#1}^{#2}}
\newcommand{\vmap}[2]{\tau_{{#1},{#2}}}
\newcommand{\stfeq}[2]{\cong_{#1}^{#2}}

\newcommand{\interproj}[2]{{#1}^{\scriptscriptstyle{\cap{#2}}}}
\newcommand{\subsetproj}[2]{{#1}^{\scriptscriptstyle{\subseteq{#2}}}}
\newcommand{\nsubsetproj}[2]{{#1}^{\scriptscriptstyle{\nsubseteq{#2}}}}
\newcommand{\viseq}[1]{\sim_{#1}}
\newcommand{\closure}{\lhd}
\newcommand{\existmod}[1]{\langle{#1}\rangle}
\newcommand{\univmod}[1]{[{#1}]}
\newcommand{\B}{\mathtt{B}}
\newcommand{\bhas}{\mathtt{h}}
\newcommand{\bopen}{\mathtt{o}}
\newcommand{\bclose}{\mathtt{c}}
\newcommand{\boolarch}[1]{\mathbb{A}({#1})}
\newcommand{\btr}[2]{\mathtt{tr}({#1},{#2})}

\newcommand{\ids}{\mathsf{Id}}
\newcommand{\finmap}{\rightharpoonup_{\mathit{fin}}}
\newcommand{\nodes}{\mathsf{Nodes}}
\newcommand{\ivar}{\mathsf{IVar}}
\newcommand{\ipred}{\mathsf{IPred}}
\newcommand{\nvar}{\mathsf{NVar}}
\newcommand{\npred}{\mathsf{NPred}}
\newcommand{\pfunc}{\mathsf{PFun}}
\newcommand{\deploy}{\leadsto}
\newcommand{\mapset}{\mathsf{Maps}}

\newcommand{\seplog}{$\mathsf{SL}$}
\newcommand{\slam}{$\mathsf{SL}_\mathit{am}$}
\newcommand{\slmap}[1]{$\mathsf{SL}^{#1}_m$}
\newcommand{\slarch}{$\mathsf{SL}_a$}
\newcommand{\sil}{$\mathsf{SIL}$}
\newcommand{\psil}{$\mathsf{SIL}^+$}
\newcommand{\psilk}[1]{$\mathsf{SIL}^+_{#1}$}
\newcommand{\ssil}{$\mathsf{SIL}^*$}
\newcommand{\qbf}{$\mathsf{QBF}$}

\newcommand{\dom}[1]{\mathrm{dom}(#1)}
\newcommand{\intf}[1]{\mathrm{inter}(#1)}

\newcommand{\modif}[1]{\mathrm{modif}({#1})}
\newcommand{\pto}{\mapsto}

\renewcommand{\proof}[1]{\ifLongVersion \noindent\emph{Proof}: {#1} \vspace*{\baselineskip}\else\fi}

%% file: abstract.tex
We introduce a logical framework for the specification and
verification of component-based systems, in which finitely many
component instances are active, but the bound on their number is not
known. Besides specifying and verifying parametric systems, we
consider the aspect of dynamic reconfiguration, in which components
can migrate at runtime on a physical map, whose shape and size may
change. We describe such parametric and reconfigurable architectures
using resource logics, close in spirit to Separation Logic, used to
reason about dynamic pointer structures. These logics support the
principle of local reasoning, which is the key for writing modular
specifications and building scalable verification algorithms, that
deal with large industrial-size systems.

%% file: body.tex
\section{Introduction}

We consider distributed computing systems consisting of white-box
components, whose interfaces are sets of communication ports. A port
controls an internal transition of the component and interacts with
zero or more ports belonging to other components. The behavior of a
component is a finite-state machine, whose transitions are labeled
with ports, that abstracts the behavior of a real-life hardware or
software component. An architecture describes all possible
interactions in a system, however it gives no information regarding
the partial order in which they may execute. The global behavior of
the system is determined by the composition of the local behaviors of
each component, in the natural sense: an interaction represents a set
of actions that are executed simultaneously, whereas different
interactions occur interleaved.

We aim at providing a framework for the modular specification and
verification of such component-based systems. The building blocks of
this framework are: \begin{compactenum}
\item partial \emph{architectures}, defined by a \emph{domain} (set of
  ports) and a set of \emph{interactions} between ports from the
  domain and external ports,
\item a \emph{composition} operation on architectures, 
\item a \emph{modular} composition of \emph{behaviors}, that mirrors
  the composition of architectures and agrees with the global behavior
  described by interactions,
\item a \emph{separation logic} of architectures, that supports the
  principle of \emph{local reasoning} and allows to describe the local
  updates induced by reconfiguration actions.
\end{compactenum}

We describe architectures using a resource logic that views the active
components of the system as resources, which can be dynamically
created and disposed of, and whose interaction scheme can be changed
at runtime. Typically, reasoning about resources requires a notion of
locality, which is captured by the separating connectives (conjunction
and implication) of Separation Logic. In a nutshell, the advantages of
modeling systems using Separation Logic are: \begin{compactitem}
\item elegant and concise definitions of parametric architectures with
  recursive patterns. In particular, a recursive definition of an
  architecture provides support for verification, in terms of hints
  for automatic generation of network invariants, used to prove safety
  properties of the system (deadlock freedom, mutual exclusion).
\item correctness proofs of reconfiguration sequences, based on the
  principle of local reasoning: only a small region of the system
  where the update takes place, needs to be considered by the proof,
  instead of the entire system.
\end{compactitem}
In order to have practical applications, a system modeling and proof
framework requires a certain degree of automation. Altough complete
automation is, in general, impossible due to the inherent
undecidability limits, defining decidable fragments of the logic and
studying their computational complexity constitute important
ingredients for building provers that can handle dynamically
reconfigurable concurrent/distributed systems.

\begin{figure}[htbp]
  \centerline{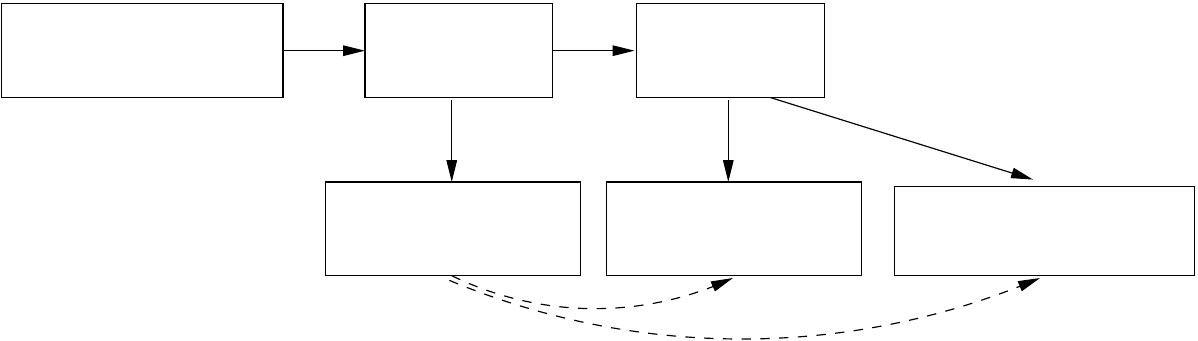}
  \caption{\label{fig:roadmap} Roadmap}
\end{figure}

\subsection{Roadmap}
The organization and reading flow of this paper are depicted in Figure
\ref{fig:roadmap}. A solid edge between two sections A and B indicates
that one needs to read A entirely before reading B. A dashed edge
between A and B indicates that some results of A are used by B but
reading of A is not necessary to understand B. Section \ref{sec:arch}
introduces the concept of architecture and defines the composition of
architectures. Section \ref{sec:sil} gives the formal syntax and
semantics of the Separation Logic of Interactions (\sil), used to
describe architectures and Section \ref{sec:decidability} deals with
the decidability of two fragments of quantifier-free \sil. In Section
\ref{sec:slarch} we extend \sil\ with component identifiers and
recursive predicates, in order to describe parametric component-based
systems, consisting of an arbitrary number of replicated components
(\slarch). From here on, the reading flow splits in two separate
directions, namely Section \ref{sec:behaviors} introduces component
behaviors and tackles the verification of safety properties (such as
deadlock freedom, mutual exclusion, etc.) using the method of network
invariants, and Section \ref{sec:reconf} introduces a framework for
specifying and verifying dynamically reconfigurable systems, using a
combination of two Separation Logics: classical \seplog\ interpreted
over graphs, for describing the physical map and \slarch\ for
describing the virtual architecture.

\section{Architectures}
\label{sec:arch}

Let $\ports$ be a countably infinite alphabet of \emph{ports}. An
\emph{interaction} is a finite set $I \in 2^\ports$ of ports.  An
\emph{architecture} is a pair $\arch = \tuple{D, \set{I_1, \ldots,
    I_k}}$, where $\dom{\arch} \isdef D \in 2^\ports$ is a finite set
of ports, called the \emph{domain} of $\arch$ and $\intf{\arch} =
\set{I_1, \ldots, I_k}$ is a set of interactions, such that $I_i \cap
D \neq \emptyset$, for all $i=1,\ldots,k$. An interaction $I \in \I$
is said to be \emph{closed} if $I \subseteq D$ and \emph{open},
otherwise. Intuitively, only closed interactions are executable in a
given architecture, because the domain provides all the required
ports. An architecture is \emph{closed} if it contains only closed
interactions, and \emph{open}, otherwise. We write $\archset$ for the
set of architectures.

\begin{example}\label{ex:arch}
  Consider the architectures $\arch_1 = \tuple{\set{p},
    \set{\set{p,q}}}$ and $\arch_2 = \tuple{\set{q},
    \set{\set{p,q}}}$. Intuitively, $\arch_1$ offers the port $p$,
  which is the only port in its domain, and requires the port $q$ in
  order to perform the interaction $\set{p,q}$. On the other hand,
  $\arch_2$ offers the port $q$ and requires $p$ to perform the same
  interaction $\set{p,q}$. In this case $\arch_1$ and $\arch_2$ have a
  match and their composition has domain $\set{p,q}$ and the only
  interaction $\set{p,q}$, which is closed and thus
  executable. \hfill$\blacksquare$
\end{example}

We move on to the formal definition of the composition of
architectures. Because ports are viewed as resources distributed among
architectures, we define composition only for architectures with
disjoint domains. Allowing non-disjoint architectures to compose would
require using multisets as architecture domains\footnote{Ports would
  be lost in composition, if domains are not disjoint and the domain
  of the composition is the union of domains.} and unnecessarily
complicate the upcoming definitions. Two architectures $\arch_1 =
\tuple{D_1, \I_1}$ and $\arch_2 = \tuple{D_2, \I_2}$ are
\emph{disjoint} if and only if $D_1 \cap D_2 = \emptyset$. For
disjoint architectures, we define the following composition:
\[\arch_1 \uplus \arch_2 \isdef \tuple{D_1 \cup D_2, (\I_1 \cap \I_2) \cup
  (\I_1 \cap 2^{\overline{D}_2}) \cup (\I_2 \cap
  2^{\overline{D}_1})}\] where $\overline{D}_i \isdef \ports \setminus
D_i$ is the complement of $D_i$, for $i=1,2$. The composition
preserves the interactions of $\arch_i$ that are disjoint from the
domain of $\arch_{3-i}$, for $i=1,2$. However, an interaction $I$ of
$\arch_i$ that has a nonempty intersection with the domain of
$\arch_{3-i}$ is kept in the composition if it matches an interaction
of $\arch_i$, i.e.\ formally $I \in \I_1 \cap \I_2$.

Recall that we require an interaction to be closed in order to be
executable. Since the domain of an architecture is enlarged by
composition, certain interactions may become closed, even if they do
not match interactions from the other arty. To understand this point,
consider the following example.

\begin{example}\label{ex:arch-comp}
Let $\arch_1 = \tuple{\set{p}, \set{\set{p,q}}}$ and $\arch_2 =
\tuple{\set{q}, \set{q,r}}$ be architectures. Since the domain of the
composition is $\dom{\arch_1 \uplus \arch_2} = \set{p,q}$, the
interaction $\set{p,q}$ of $\arch_1$ is closed in $\arch_1 \uplus
\arch_2$. However, this interaction is not executable, because it is
not matched by any interaction from $\arch_2$. This is because
$\arch_2$ provides the required port $q$, but in a different
interaction context $\set{q,r}$, that does not match $\set{p,q}$. The
natural choice is thus to remove the interaction $\set{p,q}$ from
$\intf{\arch_1 \uplus \arch_2}$. The remaining interaction $\set{q,r}$
is kept because it might become executable in a future composition
with an architecture $\arch_3$, provided that $r \in \dom{\arch_3}$
and $\set{q,r} \in \intf{\arch_3}$. \hfill$\blacksquare$
\end{example}

%% Note that $\I_1 \cap \I_2$ contains the
%% interactions that are common to both $\arch_1$ and $\arch_2$, whereas
%% $\I_1 \cap 2^{\overline{D}_2}$ is the set of interactions of $\arch_1$
%% that do not use ports of $\arch_2$ and, symmetrically, $\I_2 \cap
%% 2^{\overline{D}_1}$ is the set of interactions of $\arch_2$ that do
%% not use ports of $\arch_1$. 

We show that composition is well-defined and has natural algebraic
properties:

\begin{proposition}\label{prop:comp}
  Given disjoint architectures $\arch_1$ and $\arch_2$, their
  composition $\arch_1 \uplus \arch_2$ is again an
  architecture. Moreover, the composition is commutative, associative
  and has neutral element $\tuple{\emptyset, \emptyset}$. 
\end{proposition}
\proof{ Let $\arch_i = \tuple{D_i, \I_i}$, for all $i=1,2,3$. We have
  $\arch_1 \uplus \arch_2 = \tuple{D_1 \cup D_2, (\I_1 \cap \I_2) \cup
    (\I_1 \cap 2^{\overline{D}_2}) \cup (\I_2 \cap
    2^{\overline{D}_1})}$. Let $I \in (\I_1 \cap \I_2) \cup (\I_1 \cap
  2^{\overline{D}_2}) \cup (\I_2 \cap 2^{\overline{D}_1})$ be an
  interaction of $\arch_1 \uplus \arch_2$. To prove that $\arch_1
  \uplus \arch_2$ is an architecture, we distinguish the
  cases: \begin{compactitem}
  \item if $I \in \I_1 \cap \I_2$ then $I \cap D_1 \neq \emptyset$ and $I \cap D_2 \neq
    \emptyset$, hence $I \cap (D_1 \cup D_2) \neq \emptyset$.
  \item if $I \in \I_1 \cap 2^{\overline{D}_2}$ then $I \cap D_1 \neq
    \emptyset$, hence $I \cap (D_1 \cup D_2) = \emptyset$.
  \item if $I \in \I_2 \cap 2^{\overline{D}_1}$ then $I \cap D_2 \neq
    \emptyset$, hence $I \cap (D_1 \cup D_2) = \emptyset$.
  \end{compactitem}
  Commutativity of $\uplus$ follows from the symmetry of its
  definition. Associativity is proved by computing:
  \[\begin{array}{rcl}
  (\arch_1 \uplus \arch_2) \uplus \arch_3 & = &
  \tuple{D_1 \cup D_2, (\I_1 \cap \I_2) \cup (\I_1 \cap 2^{\overline{D}_2}) \cup (\I_2 \cap 2^{\overline{D}_1})} \uplus \tuple{D_3, \I_3} \\
  & = & \begin{array}{cl} \langle D_1 \cup D_2 \cup D_3, & (\I_1 \cap \I_2 \cap \I_3) \cup (\I_1 \cap I_3 \cap 2^{\overline{D}_2}) \cup (\I_2 \cap \I_3 \cap 2^{\overline{D}_1}) ~\cup \\
  & (\I_1 \cap \I_2 \cap 2^{\overline{D_3}}) \cup (\I_1 \cap 2^{\overline{D}_2} \cap 2^{\overline{D_3}}) \cup (\I_2 \cap 2^{\overline{D}_1} \cap 2^{\overline{D_3}}) ~\cup \\
    & (\I_3 \cap 2^{\overline{D}_1} \cap 2^{\overline{D}_2}) \rangle 
    \end{array} \\
  & = & \tuple{D_1, \I_1} \uplus \tuple{D_2 \cup D_3, (\I_2 \cap \I_3) \cup (\I_2 \cap 2^{\overline{D}_3}) \cup (\I_3 \cap 2^{\overline{D}_2})} \\
  & = & \arch_1 \uplus (\arch_2 \uplus \arch_3) 
  \end{array}\]
  Finally, $\arch_1 \uplus \tuple{\emptyset, \emptyset} = \tuple{D_1,
    (\I_1 \cap \emptyset) \cup (\I_1 \cap 2^{\ports}) \cup (\emptyset
    \cap 2^{\overline{D}_1})} = \tuple{D_1, \I_1}$.  \qed}

Sometimes it is convenient to define the \emph{closure} of an
architecture as the architecture obtained by removing all open
interactions. Formally, we define closure by means of a transitive
relation:

\begin{definition}\label{def:closure}
  Given architectures $\arch_i = \tuple{D_i, \I_i}$, for $i=1,2$, we
  have $\arch_1 \closure \arch_2$ if and only if the following
  hold: \begin{compactenum}
  \item\label{it1:closure} $D_1 = D_2$ and
  \item\label{it2:closure} $\I_1 = \subsetproj{\I_2}{D_1}$.
  \end{compactenum}
\end{definition}
Note that $\closure$ becomes the identity relation on \emph{closed}
architectures, i.e.\ architectures $\arch = \tuple{D, \I}$ with the
property that $\I = \subsetproj{\I}{D}$.

\section{Separation Logic of Interactions}
\label{sec:sil}

We introduce a first logic to describe architectures, as defined in
the previous section. Let $\pvars = \set{x,y,\ldots}$ be a countably
infinite set of variables, ranging over ports. For each port $p \in
\ports$ we consider a logical constant symbol with the same name and
let $\psym$ be the set of such constants\footnote{We use the same
  symbol for a port and its corresponding constant symbol, with the
  convention that constant symbols are only used within logical
  formulae.}. The \emph{Separation Logic of Interactions} (\sil) is
the set of formulae $\phi$ generated by the following syntax:
\[\begin{array}{rclcl}
t & := & p \in \psym \mid x \in \pvars & \hspace*{1cm} & \text{ port terms} \\
b & := & t \mid \overline{b}_1 \mid b_1 \cdot b_2 & \hspace*{1cm} & \text{ boolean terms} \\
\phi & := & t_1=t_2 \mid \emp \mid t \inter b \mid t \closeinter b \mid t \closexinter b \mid && \text{ atomic propositions} \\
&& \langle\phi_1\rangle \mid \phi_1 \wedge \phi_2 \mid \neg\phi_1 \mid \phi_1 * \phi_2 \mid \phi_1 \wand \phi_2 \mid \exists x_{\in \pvars} ~.~ \phi_1 && \text{ formulae} 
\end{array}\]
The derived connectives $\phi_1 \vee \phi_2$ and $\phi_1 \rightarrow
\phi_2$ are defined as usual and we write $\top$ ($\bot$) for $x=x$
($\neg x=x$), where the choice of $x \in \pvars$ is not important. The
set of ports that occur in a formula $\phi$ is denoted as
$\portsof{\phi}$ and is defined recursively on the structure of
$\phi$, as usual.

To describe interactions, we use boolean terms built from port terms,
connected with conjunction ($b_1 \cdot b_2$) and negation
($\overline{b}$). Boolean disjunction is defined as usual $b_1 + b_2
\isdef \overline{\overline{b}_1 \cdot \overline{b}_2}$. Intuitivelly,
$p\cdot q$ (written simply $pq$) denotes interactions in which both
$p$ and $q$ occur, $p+q$ interactions in which $p$ or $q$ occurs,
whereas $p\overline{q}$ denotes interactions in which $p$ occurs, but
not $q$, such as $\set{p,r}$. These boolean descriptors of
interactions are used within atomic propositions that describe
architectures with singleton domain, as illustrated by the following
example.

\begin{example}\label{ex:atomic-prop}
  The atomic proposition $p \inter qr$ describes those architectures
  $\arch$ with domain $\dom{\arch} = \set{p}$, whose interactions $I
  \in \intf{\arch}$ contain both $q$ and $r$. Moreover, by definition
  of architectures, $p$ belongs to every interaction, if any. For
  instance $\tuple{\set{p}, \emptyset}$, $\tuple{\set{p},
    \set{\set{p,q,r}}}$ and $\tuple{\set{p}, \set{\set{p,q,r},
      \set{p,q,r,s}}}$ are all models of $p \inter qr$. Note that some
  interactions might contain ports other than $p$, $q$ and $r$.

  On the other hand, the atomic proposition $p \closeinter r(q+s)$
  specifies those architectures whose domain is $\set{p}$ and each
  interaction is either $\set{p,q,r}$ or $\set{p,r,s}$, but not both:
  $\tuple{\set{p}, \emptyset}$, $\tuple{\set{p}, \set{p,q,r}}$ and
  $\tuple{\set{p}, \set{p,r,s}}$. Since $p$ belongs to every
  interaction, these must be \emph{minimal} boolean models of the
  propositional formula $pr(q+s)$.

  Finally, the atomic proposition $p \closexinter qr$ specifies those
  architectures whose domain is $\set{p}$ and whose interaction set
  contains at least one minimal model of $pqr$, for instance
  $\tuple{\set{p}, \set{\set{p,q}, \set{p,q,r}}}$ but not
  $\tuple{\set{p}, \set{\set{p,q}}}$. \hfill$\blacksquare$
\end{example}

Formally, a boolean term is interpreted over a valuation $\nu : \pvars
\rightarrow \ports$ and set of ports $I \subseteq \ports$, by the
relation $I \vdash_\nu b$, defined recursively on the structure of
$b$:
\[\begin{array}{rclcl}
I & \vdash_\nu & p & \iff & p \in I \\
I & \vdash_\nu & x & \iff & \nu(x) \in I \\
I & \vdash_\nu & \overline{b}_1 & \iff & I \not\vdash_\nu b_1 \\
I & \vdash_\nu & b_1b_2 & \iff & I \vdash_\nu b_1 \text{ and } I \vdash_\nu b_2 \\
\end{array}\]
We write $I \vdash^\mu_\nu b$ if and only if $I \vdash_\nu b$ and $I'
\not\vdash_\nu b$ for all $I' \subsetneq I$, i.e.\ $I$ is a minimal
model of $b$, in the propositional sense. For a port term $t \in
\pvars \cup \psym$ and a valuation $\nu : \pvars \rightarrow \ports$,
we write $\nu(t)$ for $\nu(t)$ if $t\in\pvars$ and $t$, if
$t\in\psym$.

The semantics of \sil\ formulae is defined in terms of valuations $\nu
: \pvars \rightarrow \ports$ and architectures $\arch = \tuple{D,\I}$,
by a satisfaction relation $\tuple{D,\I} \models_\nu \phi$ defined
recursively on the structure of $\phi$ as follows:

\[\begin{array}{rclcl}
\tuple{D,\I} & \models_\nu & \emp & \iff & D = \emptyset \text{ and } \I = \emptyset  \\[2mm]
\tuple{D,\I} & \models_\nu & t \inter b & \iff & D = \set{\nu(t)} \text{ and for all interactions } 
I \in \I \text{, we have }  I \vdash_\nu t\cdot b \\[2mm]
\tuple{D,\I} & \models_\nu & t \closeinter b & \iff & D = \set{\nu(t)} \text{ and for all interactions } 
I \in \I \text{, we have }  I \vdash^\mu_\nu t \cdot b \\[1mm]
\tuple{D,\I} & \models_\nu & t \closexinter b & \iff & D = \set{\nu(t)} \text{ and for some interaction }
I \in \I \text{, we have } I \vdash^\mu_\nu t \cdot b \\[2mm]
\tuple{D,\I} & \models_\nu & \existmod{\phi_1} & \iff & \text{there exists } \arch_1 \text{ such that }
\arch_1 \closure \tuple{D,\I} \text{ and } \arch_1 \models_\nu \phi_1 \\[2mm]
\tuple{D,\I} & \models_\nu & \phi_1 \wedge \phi_2 & \iff & \tuple{D,\I} \models_\nu \phi_1 \text{ and } \tuple{D,\I} \models_\nu \phi_2 \\[2mm]
\tuple{D,\I} & \models_\nu & \neg\phi_1 & \iff & \tuple{D,\I} \not\models_\nu \phi_1 \\[2mm]
\tuple{D,\I} & \models_\nu & \phi_1 * \phi_2 & \iff & \text{ there exist disjoint architectures $\arch_i$, such that} \\[2mm]
&&&& \text{$\tuple{D,\I} = \arch_1 \uplus \arch_2$ and $\arch_i \models_\nu \phi_i$, for all $i=1,2$.} \\[2mm]
\tuple{D,\I} & \models_\nu & \phi_1 \wand \phi_2 & \iff & \text{ for each architecture $\arch_1$ disjoint from $\tuple{D,\I}$ such that} \\[2mm]
&&&& \text{$\arch_1 \models_\nu \phi_1$, we have $\arch_1 \uplus \tuple{D,\I} \models_\nu \phi_2$.} \\[2mm]
\tuple{D,\I} & \models & \exists x ~.~ \phi_1 & \iff & \tuple{D,\I} \models_{\nu[x \leftarrow p]} \phi_1 \text{, for some port $p \in \ports$}
\end{array}\]

Note that it is possible to define the existential counterpart of $p
\inter b$, as the derived formula $p \exinter b \isdef p \inter p
\wedge \neg (p \inter \overline{b})$. Since $p \inter p$ defines those
architectures with domain $\set{p}$, the meaning of $p \exinter b$ is
the set of architectures with domain $\set{p}$ and interaction set
containing at least one (not necessarily minimal) boolean model of
$b$.

As a remark, the $\existmod{.}$ connective is the existential modality
with respect to the closure relation $\closure$ between
architectures. Sometimes, this connective can be used instead of $p
\closeinter b$ to describe closed interactions. Consider for instance
the formula $\existmod{p \inter q * q \exinter p}$ whose only model is
the architecture $\tuple{\set{p,q}, \set{\set{p,q}}}$, equivalently
defined by the formulae $p \closeinter q * q \closexinter p$, $p
\closeinter q * q \exinter p$ or $p \inter q * q \closexinter p$.
However, the existential modality becomes more interesting in
combination with recursive predicates (introduced next in
\S\ref{sec:slarch}), as one can use it to define closed interactions
of unbounded size (Example \ref{ex:sm}).

\paragraph{\em Remark}
Using negation, one can also define the universal modality as
$\univmod{\phi} \isdef \neg\existmod{\neg\phi}$, with the meaning
\emph{"every open extension of the current architecture must be a
  model of $\phi$"}. However, we are currently not aware of any
interesting property that may use the universal modality. \hfill$\blacksquare$

\section{Component-based Architectures}
\label{sec:slarch}

The main purpose of using Separation Logic is the modeling of
component-based systems consisting of finitely unbounded numbers of
replicated components interacting according to a recursive pattern. We
capture this aspects by the following extension of the \sil\ logic
introduced previously in \S\ref{sec:sil}: \begin{compactitem}
\item the components are identified by the elements of an infinite
  countable set $\ids$, ranged over by index variables $\ivars =
  \set{i, j, k, \ldots}$.
\item the ports are associated to components via functions of type
  $\ids \rightarrow \ports$, ranged over by the function symbols
  $\pfunc = \set{\pport,\qport,\ldots}$. Intuitively, the term
  $\pport(i)$ represents the $\pport$ ports of the $i$-th
  component. We formally require that $\pport(i)=\qport(j)$ if and
  only if $i=j$ and $\pport$ and $\qport$ are exactly the same
  function symbol.
\item recursive interaction patterns are defined by means of predicate
  symbols $P(i_1, \ldots, i_n)$, ranging over relations of type
  $\ids^n$, where $n$ is the arity of $P$, denoted $\#(P)$. We
  interpret these predicate symbols as the least solution of a system
  of inductive definitions, whose rules are written using a subset of
  the logic, defined next.
\end{compactitem}
The syntax of this extended logic, called \slarch\ in the rest of
the paper, is given below:
\[\begin{array}{rclcl}
t & := & p \in \psym \mid x \in \pvars \mid \pport(i),~ \pport \in \pfunc,~ i \in \ivars && \text{ port terms} \\
% b & := & t \mid \overline{b}_1 \mid b_1 \cdot b_2 & \hspace*{1cm} & \text{ boolean terms} \\
\phi & := & t_1=t_2 \mid \emp \mid t \inter b \mid t \closeinter b \mid t \closexinter b \mid P(i_1, \ldots, i_{\#(P)}) && \text{ atomic propositions} \\
&& \langle\phi_1\rangle \mid \phi_1 \wedge \phi_2 \mid \neg\phi_1 \mid \phi_1 * \phi_2 \mid \phi_1 \wand \phi_2 \mid \exists x_{\in \pvars} ~.~ \phi_1 \mid \exists i_{\in \ivars} ~.~ \phi_1 && \text{ formulae} 
\end{array}\]
The definition of boolean terms $b$ is the same as for \sil, thus
omitted.

A \emph{rule} is a pair written as $P(i_1, \ldots, i_{\#(P)})
\leftarrow \rho$, where $P(i_1, \ldots, i_{\#(P)})$ is a predicate
atoms and $\rho$, called the body of the rule, is a \slarch\ formula
generated by the syntax:
\[\rho := i = j \mid i \neq j \mid \emp \mid P(i_1, \ldots, i_{\#(P)}) \mid
\pport(i) \inter b \mid \pport(i) \exinter b \mid \pport(i) \closeinter b \mid
\pport(i) \closexinter b \mid \rho_1 * \rho_2 \mid \existmod{\rho_1}\]
Since this fragment of \slarch\ has no explicit negation, we consider
$\pport(i) \exinter b$ to be an atomic proposition, rather than a
derived formula.

Using \slarch, a component-based system is described by the following
methodology:
\begin{compactenum}
\item write a single predicate for each component type, which describes 
  the local interactions of that component with its neighbourhood,
\item compose one or more component predicates in a recursive
  pattern, that is usually described by a single predicate.
\end{compactenum}
This way of specifying architectures resembles the way in which
programmers design recursive data structures (lists, trees and
variations thereof), by specifying first the local links between a
memory cell and its neighbours, before encapsulating this local
specification into a recursively defined pattern. The following
example provides some intuition, before moving on with the
presentation of the formal details.

\begin{example}\label{ex:sem-task}
Consider the parametric system from Figure \ref{fig:ex:intro},
consisting of a $\mathsf{Semaphore}$ and a number of replicated
$\mathsf{Task}$s. Each task interacts with the semaphore either by
synchronizing its $\mathsf{t}\mathit{(ake)}$ port with the
$\mathsf{p}\mathit{(roberen)}$ port of the semaphore, or by
synchronizing its $\mathsf{l}\mathit{(eave)}$ port with the
$\mathsf{v}\mathit{(erhogen)}$ port of the semaphore.
  \begin{figure}[htbp]
    \centerline{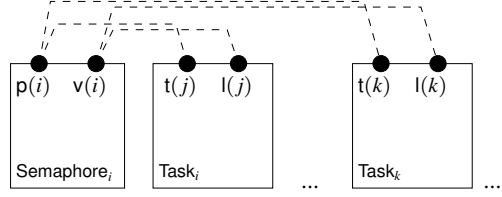}
    \caption{\label{fig:ex:intro} Semaphore and Tasks}
  \end{figure}
To describe this system in \slarch, we define predicates for each of
the two component types, describing their local interactions:
\[\begin{array}{rcl}
\mathsf{Semaphore}(i) & \leftarrow & \exists j ~.~ \mathsf{p}(i) \closeinter \mathsf{t}(j) * \mathsf{v}(i) \closeinter \mathsf{l}(j)
\\
\mathsf{Task}(i,j) & \leftarrow & \mathsf{t}(i) \exinter \mathsf{p}(j) * \mathsf{l}(i) \exinter \mathsf{v}(j) 
\\
\mathsf{Sys}(i,j,k) & \leftarrow & i=k * \mathsf{Semaphore}(j) \\
& \leftarrow & \exists \ell ~.~ \mathsf{Task}(i,j) * \mathsf{Sys}(\ell,j,k)
\end{array}\]
Intuitively, each component type ($\mathsf{Semaphore}$,
$\mathsf{Task}$) is given the self-reference $i$ as
argument. $\mathsf{Semaphore}$ choses nondeterministically a
$\mathsf{Task}$ to interact with, whereas $\mathsf{Task}$ is given a
reference $j$ to the $\mathsf{Semaphore}$ it interacts with. Note that
the composition between the atomic formulae $\mathsf{p}(i) \closeinter
\mathsf{t}(j)$ and $\mathsf{t}(j) \exinter \mathsf{p}(i)$ results in a
closed interaction involving only $\mathsf{p}(i)$ and $\mathsf{t}(j)$
(similar for $\mathsf{v}(i)$ and $\mathsf{l}(j)$). 

Finally, $\mathsf{Sys}(i,j,k)$ is a recursive pattern whose arguments
are understood as follows: $i$ and $k$ are the indices of the first
and last $\mathsf{Task}$ in the system, whereas $j$ is the reference
to the unique $\mathsf{Semaphore}$, specified by the base rule
$\mathsf{Sys}(i,j,k) \leftarrow i=k * \mathsf{Semaphore}(j)$. The
unfolding of the recursive rule $\mathsf{Sys}(i,j,k) \leftarrow
\exists \ell ~.~ \mathsf{Task}(i,j) * \mathsf{Sys}(\ell,j,k)$ creates
arbitrarily many replicas of the component type
$\mathsf{Task}$. \hfill$\blacksquare$
\end{example}

The definition of the semantics for \slarch\ requires an
\emph{interpretation} of the predicate symbols, which is a function
$\X : \preds \rightarrow \bigcup_{\alpha=1}^\infty 2^{\ids^\alpha
  \times \archset}$ associating each predicate symbol $P \in \preds$ a
set of pairs $\tuple{(k_1, \ldots, k_{\#(P)}), \arch}$, where $k_1,
\ldots, k_{\#(P)} \in \ids$ are component indices and $\arch$ is an
architecture. Moreover, because there are two types of quantified
variables in \slarch, we consider valuations $\nu : \pvars \cup \ivars
\rightarrow \ports \cup \ids$, such that $\nu(x) \in \ports$ if $x \in
\pvars$ and $\nu(x) \in \ids$ if $x \in \ivars$. The semantics of
\slarch\ is given by a satisfaction relation $\models^\X_\nu$, whose
definition is analogous to the one of $\models_\nu$ for \sil, except
for the interpretation of predicate symbols, which is the following:
\[\arch \models^\X_\nu P(i_1, \ldots, i_{\#(P)}) \iff \tuple{(\nu(i_1), \ldots, \nu(i_{\#(P)})), \arch} \in \X(P)\]
A set of rules of this form is called a \emph{system of
  definitions}. From now on, we shall assume a given system of
definitions $\Phi$, that contains one or more rules for each predicate
symbol used in a \slarch\ formula. Then a system of definitions $\Phi$
defines the following function $\mathbb{X}_\Phi$ on interpretations:
\[\mathbb{X}_\Phi(\X) \isdef \lambda P ~.~ \set{\tuple{(k_1,\ldots,k_{\#(P)}),\arch} \mid
  \arch \models^\X_{\nu[i_1 \leftarrow k_1] \ldots [i_{\#(P)}
      \leftarrow k_{\#(P)}]} \rho,~ P(i_1, \ldots, i_{\#(P)})
  \leftarrow \rho \in \Phi}\] The set of interpretations, partially
ordered by pointwise set inclusion, forms a complete
lattice. Moreover, $\mathbb{X}_\Phi$ is monotone and continuous for
each system $\Phi$, thus it has a least fixed point, denoted as
$\mu\mathbb{X}_\Phi$. In the following, we assume that the
interpretation of each predicate symbol $P$, that occurs in a
\slarch\ formula is the set $\mu\mathbb{X}_\Phi(P)$ and write
$\models^\Phi_\nu$ for $\models^\X_\nu$, whenever $\X =
\mu\mathbb{X}_\Phi$.

We conclude this section with an example of a centralized
controller-slaves architecture in which the interactions occur between
an unbounded number of participants. 

\begin{example}\label{ex:sm}
  The controller-slaves architecture in Figure \ref{fig:controller}
  consists of a single interaction between the controller component
  and each of the slave components.

  \begin{figure}[htb]
    \centerline{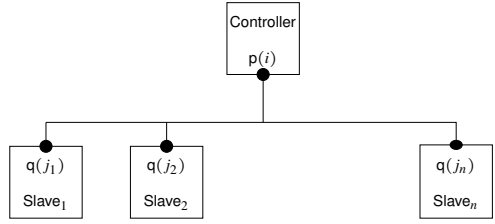}
    \caption{Controller and Slaves}
    \label{fig:controller}
  \end{figure}

  We describe such architectures using the following system of
  definitions:
  \[\begin{array}{rcl}
  \mathsf{Controller}(i,j) & \leftarrow & \pport(i) \inter \qport(j) \\
  \mathsf{Slave}(i,j) & \leftarrow & \qport(i) \inter \pport(j) \\
  \mathsf{SysRec}(i,j,k) & \leftarrow & i=k * \mathsf{Slave}(k,j) * \mathsf{Controller}(j,k) \\
  & \leftarrow & \exists \ell ~.~ \mathsf{Slave}(i,j) * \mathsf{SysRec}(\ell,j,k) \\
  \mathsf{Sys}() & \leftarrow & \exists i \exists j \exists k ~.~ \existmod{\mathsf{SysRec}(i,j,k)}
  \end{array}\]
  A $\mathsf{Controller}$ component takes the self-reference
  identifier $i$ as argument and specifies only interactions involving
  its $\pport$ port the $\qport$ of a designated $\mathsf{Slave}$
  component $j$. On the other hand, each $\mathsf{Slave}$ component
  $i$ has only interactions involving the $\mathsf{Controller}$, whose
  identifier is $j$. The $\mathsf{SysRec}$ rules create one
  $\mathsf{Controller}$ and an arbitrary number $n\geq1$ of
  $\mathsf{Slave}$ components, whereas $\mathsf{Sys}$ uses the
  existential closure modality to ensure that the (unique, if any)
  interaction between the controller and the slaves is closed.

  To understand why the models of $\mathsf{Sys}()$ are architectures
  with a single interaction, let us consider the following formula,
  describing the interactions between a controller and two slaves,
  obtained by applying the recursive rule for $\mathsf{SysRec}$ and
  the base rule once each:
  \[\begin{array}{rcl}
  \mathsf{SysRec}(i,j,k) & \Rightarrow & \exists \ell ~.~ \mathsf{Slave}(i,j) * \mathsf{SysRec}(\ell,j,k) \\
  & \Rightarrow & \exists \ell ~.~ \mathsf{Slave}(i,j) * \ell=k * \mathsf{Slave}(k,j) * \mathsf{Controller}(j,k) \\
  & \Rightarrow & \mathsf{Slave}(i,j) * \mathsf{Slave}(k,j) * \mathsf{Controller}(j,k) \\
  & \Rightarrow & \qport(i) \inter \pport(j)  * \qport(k) \inter \pport(j) * \pport(j) \inter \qport(k) 
  \end{array}\]
  Let $\arch$ be a model of the above formula.  Denoting $\pport(j) =
  p$, $\qport(i) = q$ and $\qport(k) = r$, we have $\dom{\arch} =
  \set{p,q,r}$ and $\intf{\arch}$ contains at most one interaction $I$
  such that $p,q,r \in I$. Thus any model of the formula
  $\mathsf{Sys}()$ obtained by the above unfolding of the rules
  contains at most the closed interaction
  $\set{p,q,r}$. \hfill$\blacksquare$
\end{example}

\section{Decidable Fragments of \sil}
\label{sec:decidability}

In order to automate checking the verification conditions expressed in
\sil, or its extension \slarch, we study the decidability and
computational complexity of the following decision
problems: \begin{compactitem}
\item \emph{satisfiability}: given a formula $\phi$, is there an
  architecture $\arch$ and a valuation $\nu$ such that $\arch
  \models_\nu \phi$ ?
\item \emph{entailment}: given formulae $\phi$ and $\psi$, for any
  architecture $\arch$ and valuation $\nu$, does $\arch \models_\nu
  \phi$ imply $\arch \models \psi$ ?
\end{compactitem}
Even though, in general, these problems are undecidable for \sil, in
the presence of quantifiers, we identify two nontrivial
quantifier-free fragments for which the problem is decidable. These
fragments of \sil, denoted as \psil\ and \ssil, are defined by the
syntax below, starting with the $\phi$ and $\psi$ nonterminals,
respectively:
\[\begin{array}{rcll}
\phi & := & \emp \mid p \inter b \mid p \closeinter b \mid p \exinter b \mid p \closexinter b \mid \langle\phi_1\rangle
\mid \phi_1 \wedge \phi_2 \mid \phi_1 \vee \phi_2 \mid \phi_1 * \phi_2 \mid \phi_1 \wand \phi_2 & \text{ (\psil)}\\
\psi & := & \emp \mid p \inter b \mid p \closeinter b \mid p \closexinter b \mid \psi_1 \wedge \psi_2 \mid \neg \psi_1 \mid \psi_1 * \psi_2 & \text{ (\ssil)}
\end{array}\]
Note that, because \psil\ does not have negation, we must consider the
satisfiability and entailment problems separately. On the other hand,
studying the satisfiability problem is sufficient for \ssil, because
of the negation connective allowing the encode entailment between
$\psi_1$ and $\psi_2$ and the unsatisfiability of $\psi_1 \wedge
\neg\psi_2$. The lack of negation is also the reason why we adopt the
formula $p \exinter b$ as an atomic proposition of \psil. Moreover,
since there are no port variables in \psil\ or \ssil, we omit the
valuation subscript and write $\arch \models \phi$ instead of $\arch
\models_\nu \phi$, whenever $\phi$ is a formula of \psil\ or \ssil.

The proofs of decidability for \psil\ and \ssil\ follow essentially
the same steps. First, we define an equivalence relation between
architectures that is compatible (at least) with the (de)composition
operation. Second, we define the equivalence classes of the relation
using simple \sil\ formulae belonging to a small number of patterns,
called \emph{test formulae} and show that the equivalence relation is
the same as the equivalence on a finite set of test
formulae. Consequently, each formula in the given fragment of \sil\ is
equivalent to a boolean combination of test formulae. Moreover, by
considering each test formula as a propositional variable, one can
transform the input formula into an equivalent \qbf\ formula (modulo
the interpretation of the propositional variables). The latter
transformation yields the decidability result and a characterization
of the complexity classes of the decision problems considered.

\subsection{Decidability of \psil}
\label{sec:psil}

We start by defining an equivalence relation on architectures. For any
set of ports $P \subseteq \ports$ and a set of interactions $\I
\subseteq 2^\ports$, we define the following sets of interactions:
\[\begin{array}{rclcrcl}
\interproj{\I}{P} & \isdef & \set{I \mid I \in \I,~ I \cap P \neq \emptyset}
& \hspace*{1cm} & 
\subsetproj{\I}{P} & \isdef & \set{I \mid I \in \I,~ I \subseteq P} \\
\nsubsetproj{\I}{P} & \isdef & \I \setminus \subsetproj{\I}{P} && 
\I \sqcap P & \isdef & \set{I \cap P \mid I \in \I} 
\end{array}\]

\begin{definition}\label{def:visible-equiv}
  Given architectures $\arch_i = \tuple{D_i, \I_i}$, for $i=1,2$ and a
  finite set of ports $P \in 2^\ports$, such that $D_1 \cup D_2
  \subseteq P$, we have $\arch_1 \viseq{P} \arch_2$ if and only if the
  following hold: \begin{compactenum}
  \item\label{it1:visible-equiv} $D_1 = D_2$,
  \item\label{it2:visible-equiv} $\subsetproj{\I_1}{P} = \subsetproj{\I_2}{P}$,
  \item\label{it3:visible-equiv} $\nsubsetproj{\I_1}{P} \sqcap P =
    \nsubsetproj{\I_2}{P} \sqcap P$.
  \end{compactenum}
\end{definition}
Note that the relation $\viseq{P}$ is defined only between
architectures with domain included in $P$. It is easy to check that
$\viseq{P}$ is an equivalence relation, in this case. From now on, we
shall silently assume that $\dom{\arch_1} \cup \dom{\arch_2} \subseteq
P$, whenever $\arch_1 \viseq{P} \arch_2$ holds. 

The next lemma shows that $\viseq{P}$ is compatible with the
decomposition of architectures:

\begin{lemma}\label{lemma:star}
  Let $\arch = \tuple{D, \I}$ and $\arch' = \tuple{D', \I'}$ be
  architectures and $P \in 2^\ports$ be a set of ports such that
  $\arch \viseq{P} \arch'$. Then, for any two architectures $\arch_i =
  \tuple{D_i, \I_i}$, such that $\arch = \arch_1 \uplus \arch_2$,
  there exist architectures $\arch'_i = \tuple{D'_i, \I'_i}$, for
  $i=1,2$ such that: \begin{compactenum}
  \item\label{it1:star} $\arch_i \viseq{P} \arch'_i$, for each $i=1,2$ and
  \item\label{it2:star} $\arch' = \arch'_1 \uplus \arch'_2$. 
  \end{compactenum}
\end{lemma}
\proof{
  From $\arch = \arch_1 \uplus \arch_2$ we infer that:
  \begin{equation} \tag{$\dagger$} \label{eq:split-dom}
    D_1 \cup D_2 = D \text{ and } D_1 \cap D_2 = \emptyset
  \end{equation}
  \begin{equation} \tag{$\ddagger$} \label{eq:split-int}
    \I = (\I_1 \cap \I_2) \cup (\I_1 \cap 2^{\overline{D}_2}) \cup (\I_2 \cap 2^{\overline{D}_1}) 
  \end{equation}  
  Let $D'_i \isdef D_i$ and $\I'_i \isdef \interproj{\I'}{D_i} \cup
  X_i$, where $X_i \isdef \I_i \setminus \interproj{\I}{D_i}$, for
  $i=1,2$. We prove first that $\arch_1 \viseq{P} \arch'_1$, the proof
  for $\arch_2 \viseq{P} \arch'_2$ being identical. Note that $D'_1 =
  D_1 \subseteq P$, by the definition of $D'_1$ and
  (\ref{eq:split-dom}). The two remaining points of Definition
  \ref{def:visible-equiv} are proved below:

  \noindent (\ref{it2:visible-equiv}) We compute:
  \[\begin{array}{rcl}
  \subsetproj{\I'_1}{P} & = & \subsetproj{(\interproj{\I'}{D_1} \cup X_1)}{P}
  = \subsetproj{(\interproj{\I'}{D_1})}{P} \cup \subsetproj{X_1}{P} \\
  & = & \interproj{(\subsetproj{\I'}{P})}{D_1} \cup \subsetproj{X_1}{P}  \text{, since $D_1 \subseteq P$} \\
  & = & \interproj{(\subsetproj{\I}{P})}{D_1} \cup \subsetproj{X_1}{P} \text{, since $\arch \viseq{P} \arch'$} \\
  & = & \subsetproj{(\interproj{\I}{D_1})}{P} \cup \subsetproj{X_1}{P} \text{, since $D_1 \subseteq P$} \\
  & = & \subsetproj{(\interproj{\I}{D_1} \cup (\I_1 \setminus \interproj{\I}{D_1}))}{P}
  = \subsetproj{(\interproj{\I}{D_1} \cup \I_1)}{P} 
  = \subsetproj{\I_1}{P} \text{, since $\interproj{\I}{D_1} \subseteq \I_1$.}
  \end{array}\]

   \noindent (\ref{it3:visible-equiv}) We prove that
  $\nsubsetproj{\I'_1}{P} \sqcap P =
  \nsubsetproj{(\interproj{\I'}{D_1})}{P} \sqcap P \cup
  \nsubsetproj{X_1}{P} \sqcap P = \nsubsetproj{\I_1}{P} \sqcap
  P$. ``$\subseteq$'' We distinguish the following cases: \begin{compactitem}
    \item if $I = J \cup U' \in \I'$ such that $J \subseteq P$, $J
      \cap D_1 \neq \emptyset$, $U' \neq \emptyset$ and $U' \cap P =
      \emptyset$. Clearly every interaction $I \in
      \nsubsetproj{(\interproj{\I'}{D_1})}{P}$ is of this form.  Since
      $\arch \viseq{P} \arch'$, we have
      $\nsubsetproj{(\interproj{\I}{D})}{P} \sqcap P =
      \nsubsetproj{(\interproj{\I'}{D})}{P} \sqcap P$ and thus there
      exists $U \neq \emptyset$ such that $U \cap P = \emptyset$ and
      $J \cup U \in \I$. Moreover, since $J \cap D_1 \neq \emptyset$,
      we have that $J \cup U \in \interproj{\I}{D_1}$ and $J \cup U
      \in \I_1$ follows, by (\ref{eq:split-int}). Since $U \neq
      \emptyset$ and $U \cap P = \emptyset$, we obtain $J \cup U \in
      \nsubsetproj{\I_1}{P}$ and thus $J \in \nsubsetproj{\I_1}{P}
      \sqcap P$.
      \item else $I = J \cup U' \in X_1 = \I_1 \setminus
        \interproj{\I}{D_1}$ then $J \in \nsubsetproj{\I_1}{P} \sqcap
        P$ is immediate.
  \end{compactitem}
  ``$\supseteq$'' Let $I = J \cup U \in \I_1$,
  such that $U \neq \emptyset$ and $U \cap P = \emptyset$. Then
  $I \cap D_1 \neq \emptyset$ and since $D_1 \subseteq P$, we
  have $J \cap D_1 \neq \emptyset$. We distinguish the following
  cases: \begin{compactitem}
  \item if $J \cup U \in \I$, because $\arch \viseq{P} \arch'$, we
    have $\nsubsetproj{\I}{P} \sqcap P = \nsubsetproj{\I'}{P} \sqcap
    P$, thus there exists $U' \neq \emptyset$ such that $U' \cap P =
    \emptyset$ and $J \cup U' \in \I'$. Moreover, since $J \cap D_1
    \neq \emptyset$, we have $J \cup U' \in \interproj{\I'}{D_1}
    \subseteq \I'_1$ and $J \in \nsubsetproj{\I'_1}{P} \sqcap P$.
  \item else $J \cup U \not\in \I$ then $J \cup U \in \I_1 \setminus
    \interproj{\I}{D_1} = X_1 \subseteq \I'_1$ and $J \in
    \nsubsetproj{\I'_1}{P} \sqcap P$.
  \end{compactitem}

  \vspace*{\baselineskip}\noindent
  Finally, we prove that $\arch' = \arch'_1 \uplus \arch'_2$.  We
  start by proving the following facts:
   \begin{fact}\label{fact:x-disjoint}
     $X_1 \cap X_2 = \emptyset$
   \end{fact}
   \proof{
     \[\begin{array}{rcl}
     X_1 \cap X_2 & = & (\I_1 \setminus \interproj{\I}{D_1}) \cap (\I_2 \setminus \interproj{\I}{D_2}) \\
     & = & (\I_1 \cap \I_2) \setminus (\interproj{\I}{D_1} \cup \interproj{\I}{D_2}) \\
     & = & (\I_1 \cap \I_2) \setminus \interproj{\I}{(D_1 \cup D_2)} \\
     & = & (\I_1 \cap \I_2) \setminus \I \\
     & = & \emptyset \text{, the last step follows from (\ref{eq:split-int}) \qed}
     \end{array}\]
   }

   \begin{fact}\label{fact:x-inter}
     For all $I \in X_i$, we have $I \cap D_1 \neq \emptyset$ and $I
     \cap D_2 \neq \emptyset$, for $i=1,2$.
   \end{fact}
   \proof{ We prove the case $i=1$, the proof of the other case being
     identical. Let $I \in X_1$. Then $I \in \I_1$, thus $I \cap D_1
     \neq \emptyset$, by the assumption that $\arch_1$ is an
     architecture. Suppose $I \cap D_2 = \emptyset$. By
     (\ref{eq:split-int}), we have $I \in \I$, thus $I \cap (D_1 \cup
     D_2) \neq \emptyset$. Since $I \cap D_2 = \emptyset$, we have $I
     \cap D_1 \neq \emptyset$, thus $I \in \interproj{\I}{D_1}$, which
     contradicts with $I \in X_1 = \I_1 \setminus
     \interproj{\I}{D_1}$. \qed}

   \noindent
   We have: \[\I' = (\interproj{\I'}{D_1} \cap \interproj{\I'}{D_2})
   \cup (\I' \cap 2^{\overline{D}_2}) \cup (\I' \cap
   2^{\overline{D}_1})\] and compute, successively:
     \[\begin{array}{rcl}
     \I'_1 \cap 2^{\overline{D}_2} & = & (\interproj{\I'}{D_1} \cup X_1) \cap 2^{\overline{D}_2} \\
     & = & \interproj{\I'}{D_1} \cap 2^{\overline{D}_2} \text{, by Fact \ref{fact:x-inter}} \\
     & = & \I' \cap 2^{\overline{D}_2} \text{, since $\forall I \in \I' ~.~ I \cap (D_1 \cup D_2) \neq \emptyset$} \\
     \\
     \I'_2 \cap 2^{\overline{D}_1} & = & \I' \cap 2^{\overline{D}_1} \text{, by a symmetric argument}
     \\
     \I'_1 \cap \I'_2 & = & (\interproj{\I'}{D_1} \cup X_1) \cap (\interproj{\I'}{D_2} \cup X_2) \\
     & = & (\interproj{\I'}{D_1} \cap \interproj{\I'}{D_2}) \cup
     (\interproj{\I'}{D_1} \cap X_2) \cup (\interproj{\I'}{D_2} \cap X_1) \text{, by Fact \ref{fact:x-disjoint}} \\
     & = & \interproj{\I'}{D_1} \cap \interproj{\I'}{D_2}
     \end{array}\]
     The last step follows from $\interproj{\I'}{D_1} \cap X_2
     \subseteq \interproj{\I'}{D_2}$ and $\interproj{\I'}{D_2} \cap
     X_1 \subseteq \interproj{\I'}{D_1}$, which is proved below. Let
     $I \in \interproj{\I'}{D_1} \cap X_2$ (the other case is
     symmetric). If $I \in X_2$, we have $I \cap D_2 \neq \emptyset$,
     by Fact \ref{fact:x-inter}. Then $I \in
     \interproj{\I'}{D_2}$. This concludes the proof the Lemma. \qed}

Conversely, the next lemma shows that $\viseq{P}$ is compatible with
the composition of architectures:

\begin{lemma}\label{lemma:wand}
  Let $\arch = \tuple{D, \I}$ and $\arch' = \tuple{D', \I'}$ be
  architectures and $P \in 2^\ports$ be a set of ports such that
  $\arch \viseq{P} \arch'$. Then, for any architecture $\arch_1 =
  \tuple{D_1, \I_1}$ such that $D_1 \cap D = \emptyset$ and $D_1
  \subseteq P$ there exists an architecture $\arch'_1 = \tuple{D'_1,
    \I'_1}$ such that: \begin{compactenum}
  \item\label{it1:wand} $\arch_1 \viseq{P} \arch'_1$ and
  \item\label{it2:wand} $\arch_1 \uplus \arch \viseq{P} \arch'_1 \uplus
    \arch'$.
  \end{compactenum}
\end{lemma}
\proof{
  Let $D'_1 \isdef D_1$ and $\I'_1 \isdef \subsetproj{\I_1}{P} \cup X_1 \cup Y_1$, where:
  \[\begin{array}{rcl}
  X_1 & \isdef & \set{J \cup U' \mid U' \neq \emptyset, U' \cap P = \emptyset,
    J \in \nsubsetproj{(\I_1 \cap \I)}{P} \sqcap P, J \cup U' \in \I'} \\
  Y_1 & \isdef & \set{J \cup \set{\alpha} \mid J \in \nsubsetproj{(\I_1 \setminus \I)}{P} \sqcap P,
    \forall U' ~.~ U' \neq \emptyset \wedge U' \cap P = \emptyset \Rightarrow J \cup U' \not\in X_1}
  \end{array}\]
  and $\alpha \in \ports$ is a fresh port, not occurring in either $\arch, \arch_1$ or $\arch'$.

  \vspace*{\baselineskip}\noindent We prove that $\arch_1 \viseq{P}
  \arch'_1$. Note that $D'_1 = D_1$ by definition. The two remaining
  points of Definition \ref{def:visible-equiv} are proved below:

    \noindent
  (\ref{it2:visible-equiv}) We have $\subsetproj{\I'_1}{P} =
  \subsetproj{\I_1}{P} \cup \subsetproj{X_1}{P} \cup
  \subsetproj{Y_1}{P} = \subsetproj{\I_1}{P}$, because
  $\subsetproj{X_1}{P} = \subsetproj{Y_1}{P} = \emptyset$, by
  definition.

  \noindent
  (\ref{it3:visible-equiv}) By definition of $\I'_1$, we have
  $\nsubsetproj{\I'_1}{P} = X_1 \cup Y_1$. We have to prove that $X_1
  \sqcap P \cup Y_1 \sqcap P = \nsubsetproj{\I_1}{P} \sqcap
  P$. ``$\subseteq$'' Let $I = J \cup U'$, where $J \subseteq P$ and
  $U' \neq \emptyset$, $U' \cap P = \emptyset$, be an interaction. We
  distinguish the following cases: \begin{compactitem}
  \item if $J \cup U' \in \nsubsetproj{X_1}{P}$ then $J \in
    \nsubsetproj{(\I_1 \cap \I)}{P} \sqcap P$ and $J \in
    \nsubsetproj{\I_1}{P} \sqcap P$ follows. 
  \item else $J \cup U' \in \nsubsetproj{Y_1}{P}$ then $J \in
    \nsubsetproj{(\I_1 \setminus \I)}{P} \sqcap P$ and $J \in
    \nsubsetproj{\I_1}{P} \sqcap P$ follows.
  \end{compactitem}
  ``$\supseteq$'' Let $I = J \cup U \in \nsubsetproj{\I_1}{P}$ be an
  interaction, such that $J \subseteq P$, $U \neq \emptyset$ and $U
  \cap P = \emptyset$. We distinguish the following
  cases: \begin{compactitem}
  \item if $J \cup U \in \I$ then $J \cup U \in \I_1 \cap
    \I$. Moreover, because $\arch \viseq{P} \arch'$, we have
    $\subsetproj{\I}{P} = \subsetproj{\I'}{P}$, thus there exists $U'
    \neq \emptyset$ such that $U' \cap P = \emptyset$ and $J \cup U'
    \in \I'$ and thus $J \cup U' \in X_1$, by the definition of
    $X_1$. Consequently, we have $J \in X_1 \sqcap P$ in this case.
  \item else $J \cup U \not\in \I$, then $J \cup U \in
    \nsubsetproj{(\I_1 \setminus \I)}{P}$ and $J \in
    \nsubsetproj{(\I_1 \setminus \I)}{P} \sqcap P$. We distinguish two
    cases: \begin{compactitem}
    \item if there exists $U' \neq \emptyset$ such that $U' \cap P =
      \emptyset$ and $J \cup U' \in X_1$, then $J \in X_1 \sqcap P$. 
    \item else $J \cup U' \not\in X_1$, for all $U' \neq \emptyset$
      such that $U' \cap P = \emptyset$, then $J \cup \set{\alpha} \in
      Y_1$ and $J \in Y_1 \sqcap P$.
    \end{compactitem}
  \end{compactitem}
Finally, we prove that $\arch_1 \uplus \arch \viseq{P} \arch'_1 \uplus
\arch'$. Note that $D_1 \cup D \subseteq P$ and $D_1 \cup D = D'_1
\cup D'$, by the definition of $D'_1$ and the fact that $\arch
\viseq{P} \arch'$. We are left with proving the following two points
of Definition \ref{def:visible-equiv}:

\noindent(\ref{it2:visible-equiv}) We have $\subsetproj{\I_1}{P} =
\subsetproj{\I'_1}{P}$, since $\arch_1 \viseq{P} \arch'_1$ and
$\subsetproj{\I}{P} = \subsetproj{\I'}{P}$, since $\arch \viseq{P}
\arch'$. Thus we obtain the following equalities:
\[\begin{array}{rcl}
\subsetproj{(\I_1 \cap \I)}{P} & = & \subsetproj{(\I'_1 \cap \I')}{P} \\
\subsetproj{(\I_1 \cap 2^{\overline{D}})}{P} & = & \subsetproj{(\I'_1 \cap 2^{\overline{D}})}{P} \\
\subsetproj{(\I \cap 2^{\overline{D}_1})}{P} & = & \subsetproj{(\I' \cap 2^{\overline{D}_1})}{P}
\end{array}\]

\noindent(\ref{it3:visible-equiv}) We prove the following
points: \begin{compactitem}
\item $\nsubsetproj{(\I_1 \cap \I)}{P} \sqcap P = \nsubsetproj{(\I'_1
  \cap \I')}{P} \sqcap P$: ``$\subseteq$'' Let $I = J \cup U \in \I_1
  \cap \I$, where $U \neq \emptyset$ and $U \cap P = \emptyset$ be an
  interaction. Since $\arch \viseq{P} \arch'$, we have
  $\nsubsetproj{\I}{P} \sqcap P = \nsubsetproj{\I'}{P} \sqcap P$ and
  thus there exists $U' \neq \emptyset$ such that $U' \cap P =
  \emptyset$ and $J \cup U' \in \I'$. Consequently, $J \cup U' \in X_1
  \subseteq \I'_1$ and we obtain $J \cup U' \in \I'_1 \cap \I'$, thus
  $J \in \nsubsetproj{(\I'_1 \cap \I')}{P} \sqcap P$
  follows. ``$\supseteq$'' Let $I = J \cup U' \in \I'_1 \cap \I'$,
  where $U' \neq \emptyset$ and $U' \cap P = \emptyset$ be an
  interaction.  Since $J \cup U' \in \I'_1$ then $J \cup U' \in X_1
  \cup Y_1$ and since $J \cup U' \in \I'$ it must be that $J \cup U'
  \in X_1$. By the definition of $X_1$, we obtain $J \in
  \nsubsetproj{(\I_1 \cap \I)}{P} \sqcap P$.
\item $\nsubsetproj{(\I_1 \cap 2^{\overline{D}})}{P} \sqcap P =
  \nsubsetproj{(\I'_1 \cap 2^{\overline{D}})}{P} \sqcap P$: Because $\arch_1
  \viseq{P} \arch'_1$, we have $\nsubsetproj{\I_1}{P} \sqcap P =
  \nsubsetproj{\I'_1}{P} \sqcap P$ and the result follows.
\item $\nsubsetproj{(\I \cap 2^{\overline{D}_1})}{P} \sqcap P =
  \nsubsetproj{(\I' \cap 2^{\overline{D}_1})}{P} \sqcap P$: Because
  $\arch \viseq{P} \arch'$, we have $\nsubsetproj{\I}{P} \sqcap P =
  \nsubsetproj{\I'}{P} \sqcap P$ and the result follows.  \qed
\end{compactitem}}

Finally, we show that $\viseq{P}$ is also compatible with the closure
relation on architectures:

\begin{lemma}\label{lemma:closure}
  Let $\arch = \tuple{D, \I}$ and $\arch' = \tuple{D', \I'}$ be
  architectures and $P \in 2^\ports$ be a set of ports such that
  $\arch \viseq{P} \arch'$ and let $\arch_1 = \tuple{D_1, \I_1}$ be an
  architecture such that $\arch_1 \closure \arch$. Then $\arch_1
  \closure \arch'$ as well.
\end{lemma}
\proof{ Because $\arch_1 \closure \arch$ we have $D_1 = D$ and $\I_1 =
  \subsetproj{\I}{D}$. Moreover, since $\arch \viseq{P} \arch'$, we
  have $D = D'$ and $\subsetproj{\I}{P} = \subsetproj{\I'}{P}$. Since
  $D \subseteq P$, we obtain $\subsetproj{\I}{D} = \subsetproj{\I'}{D}
  = \I_1$, hence $\arch_1 \closure \arch'$. \qed}

The following theorem shows that $\viseq{P}$ coincides with the
equivalence of architectures with respect to \psil\ formulae. The
proof of the theorem requires that every model of a \psil\ formula has
only visible ports in its domain, which is proved below:

\begin{lemma}\label{lemma:visible-domain}
  For each formula $\phi$ of \psil and each architecture $\arch =
  \tuple{D, \I}$ such that $\arch \models \phi$, we have $D
  \subseteq \portsof{\phi}$. 
\end{lemma}
\proof{
  By induction on the structure of $\phi$: \begin{compactitem}
  \item $\emp$: in this case $D = \emptyset$.
  \item $p \inter b$, $p \closeinter b$, $p \exinter b$ and $p
    \closexinter b$: in this case $D = \set{p}$.
  \item $\existmod{\phi_1}$: in this case there exists an architecture
    $\arch_1 = \tuple{D_1, \I_1}$ such that $\arch \closure \arch_1$
    and $\arch_1 \models \phi_1$. Then $D = D_1$ and $D_1 \subseteq
    \portsof{\phi_1}$, by the induction hypothesis. We conclude
    noticing that $\portsof{\existmod{\phi_1}} = \portsof{\phi_1}$.
  \item $\phi_1 \wedge \phi_2$: since $\arch \models \phi_1$, by the
    induction hypothesis we have $D \subseteq \portsof{\phi_1}
    \subseteq \portsof{\phi_1 \wedge \phi_2}$.
  \item $\phi_1 \vee \phi_2$: if $\arch \models \phi_1$, by the
    induction hypothesis we have $D \subseteq \portsof{\phi_1}
    \subseteq \portsof{\phi_1 \vee \phi_2}$. The case $\arch \models
    \phi_2$ is symmetric.
  \item $\phi_1 * \phi_2$: in this case there exists $\arch_i =
    \tuple{D_i, \I_i}$ such that $\arch = \arch_1 \uplus \arch_2$ and
    $\arch_i \models \phi_i$, for both $i=1,2$. By the induction
    hypothesis, $D_i \subseteq \portsof{\phi_i} \subseteq
    \portsof{\phi_1 * \phi_2}$, for both $i=1,2$, thus $D = D_1 \cup
    D_2 \subseteq \portsof{\phi_1 * \phi_2}$.
  \item $\phi_1 \wand \phi_2$: let $\arch_1 = \tuple{D_1, \I_1}$ be
    any architecture such that $D_1 \cap D = \emptyset$ and $\arch_1
    \models \phi_1$. Since $\arch \models \phi_1 \wand \phi_2$, we
    obtain $\arch_1 \uplus \arch \models \phi_2$. Again, by the
    induction hypothesis, $D_1 \cup D \subseteq \portsof{\phi_2}$,
    hence $D \subseteq \portsof{\phi_1 \wand \phi_2}$ follows
    immediately. \qed
\end{compactitem}}

\begin{theorem}\label{thm:psil-equiv}
  Let $\arch = \tuple{D, \I}$ and $\arch' = \tuple{D', \I'}$ be
  architectures and $P \in 2^\ports$ be a set of ports such that
  $\arch \viseq{P} \arch'$. Then, for any formula $\phi$ of \psil,
  such that $\portsof{\phi} \subseteq P$, we have $\arch \models \phi$ if and
  only if $\arch' \models \phi$.
\end{theorem}
\proof{ By induction on the structure of $\phi$: \begin{compactitem}
    \vspace*{\baselineskip}
    \item $\emp$: if $\arch \models \emp$ then $D = \emptyset$ and $\I
      = \emptyset$. Since $\arch \viseq{P} \arch'$, we obtain $D' = D
      = \emptyset$, thus $\I'=\emptyset$ must be the case, otherwise
      every interaction $I \in \I'$ would have a non-empty
      intersection with $D'$. Consequently, $\arch' \models \emp$.
      \vspace*{\baselineskip}
    \item $p \inter b$: if $\arch \models p \inter b$, then $D =
      \set{p}$ and, since $\arch \viseq{P} \arch'$, we obtain $D' = D
      = \set{p}$. Let $I \in \I'$ be an interaction. If $I \subseteq
      P$ then $I \in \subsetproj{\I'}{P} = \subsetproj{\I}{P}$,
      because $\arch \viseq{P} \arch'$. Then $I \vdash p \cdot b$,
      because $\arch \models p \inter b$. Else, $I \not\subseteq P$
      and $I \in \nsubsetproj{\I'}{P}$. Because $\arch \viseq{P}
      \arch'$, we obtain $I \cap P = J \cap P$, for some interaction
      $J \in \I$. Moreover, $J \vdash p \cdot b$ because $\arch
      \models p \inter b$ and, since $\portsof{p \inter b} \subseteq
      P$, it must be the case that $I \vdash p \cdot b$ as well.
      \vspace*{\baselineskip}
    \item $p \closeinter b$: if $\arch \models p \inter b$, then $D =
      \set{p}$ and, since $\arch \viseq{P} \arch'$, we obtain $D' = D
      = \set{p}$. Let $I \in \I'$ be an interaction. The proof in the
      case $I \subseteq P$ is given at the point above, so we consider
      that $I \not\subseteq P$. Because $\arch \viseq{P} \arch'$,
      there exists an interaction $J \in \nsubsetproj{\I}{P}$ such
      that $I \cap P = J \cap P$. Moreover, since $\arch \models p
      \closeinter b$, we have that $J \vdash_\mu p \cdot b$. Then $J
      \subseteq \portsof{p \closeinter b} \subseteq P$, which
      contradicts the fact that $J \in
      \nsubsetproj{\I}{P}$. Consequently, the only case possible is $I
      \subseteq P$, in which case $I \vdash_\mu p \cdot b$, thus
      $\arch' \models p \closeinter b$.
      \vspace*{\baselineskip}
    \item $p \exinter b$: if $\arch \models p \exinter b$, then $D =
      \set{p}$ and, since $\arch \viseq{P} \arch'$, we obtain $D' = D
      = \set{p}$. Moreover, there exists an interaction $I \in \I$
      such that $I \vdash p \cdot b$. If $I \subseteq P$ then $I \in
      \subsetproj{\I}{P} = \subsetproj{\I'}{P}$, thus $\arch' \models
      p \exinter b$. Else, $I \not\subseteq P$, hence $I \in
      \nsubsetproj{\I}{P}$. Since $\arch \viseq{P} \arch'$, there
      exists an interaction $J \in \nsubsetproj{\I'}{P}$ such that $J
      \cap P = I \cap P$. But then $J \vdash p\cdot b$, thus $\arch'
      \models p \exinter b$.
      \vspace*{\baselineskip}
    \item $p \closexinter b$: if $\arch \models p \closexinter b$,
      then $D = \set{p}$ and, since $\arch \viseq{P} \arch'$, we
      obtain $D' = D = \set{p}$. Moreover, there exists an interaction
      $I \in \I$ such that $I \vdash^\mu p \cdot b$. Then $I \subseteq
      P$ and $I \in \subsetproj{\I}{P}$ follows. Since $\arch
     \viseq{P} \arch'$, we have $I \in \subsetproj{\I'}{P}$ and thus
      $\arch' \models p \closexinter b$.
      \vspace*{\baselineskip}
    \item $\existmod{\phi_1}$: because $\arch \models
      \existmod{\phi_1}$, there exists $\arch_1$ such that $\arch
      \closure \arch_1$ and $\arch_1 \models \phi_1$. By Lemma
      \ref{lemma:closure}, we have $\arch' \closure \arch_1$, thus
      $\arch' \models \existmod{\phi_1}$. 
      \vspace*{\baselineskip}
    \item $\phi_1 \wedge \phi_2$: since $\arch \viseq{P} \arch'$ and
      $\portsof{\phi_i} \subseteq \portsof{\phi_1 \wedge \phi_2}
      \subseteq P$, by the induction hypothesis, we obtain $\arch'
      \models \phi_i$, for both $i=1,2$, thus $\arch' \models \phi_1
      \wedge \phi_2$.
      \vspace*{\baselineskip}
    \item $\phi_1 \vee \phi_2$: assume that $\arch \models \phi_1$,
      the case $\arch \models \phi_2$ being symmetric. By the
      induction hypothesis, since $\portsof{\phi_1} \subseteq
      \portsof{\phi_1 \vee \phi_2} \subseteq P$, we obtain $\arch'
      \models \phi_1$, thus $\arch' \models \phi_1 \vee \phi_2$.
      \vspace*{\baselineskip}
    \item $\phi_1 * \phi_2$: because $\arch \models \phi_1 * \phi_2$,
      there exists disjoint architectures $\arch_1$ and $\arch_2$ such
      that $\arch = \arch_1 \uplus \arch_2$ and $\arch_i \models
      \phi_i$, for both $i=1,2$. Since $\arch \viseq{P} \arch'$, by
      Lemma \ref{lemma:star}, there exist architectures $\arch'_1$ and
      $\arch'_2$, such that $\arch' = \arch'_1 \uplus \arch'_2$ and
      $\arch_i \viseq{P} \arch'_i$, for both $i=1,2$. Since
      $\portsof{\phi_i} \subseteq \portsof{\phi_1 * \phi_2} \subseteq
      P$, by the induction hypothesis, we obtain $\arch'_i \models
      \phi_i$, for both $i=1,2$, and consequently $\arch' \models
      \phi_1 * \phi_2$.
      \vspace*{\baselineskip}
    \item $\phi_1 \wand \phi_2$: Let $\arch'_1 = \tuple{D'_1, \I'_1}$
      be any architecture such that $\arch' \models \phi_1$ and $D'_1
      \cap D' = \emptyset$. Because $\phi$ is in \psil, by Lemma
      \ref{lemma:visible-domain}, we obtain $D'_1 \subseteq
      \portsof{\phi_1} \subseteq \portsof{\phi_1 \wand \phi_2}
      \subseteq P$. By Lemma \ref{lemma:wand}, there exists an
      architecture $\arch_1$ such that $\arch'_1 \viseq{P} \arch_1$
      and $\arch_1 \uplus \arch \viseq{P} \arch'_1 \uplus \arch'$. By
      the induction hypothesis, we have $\arch_1 \models \phi_1$ and,
      since $\arch \models \phi_1 \wand \phi_2$, we obtain $\arch_1
      \uplus \arch \models \phi_2$. Again, by the induction
      hypothesis, we obtain that $\arch'_1 \uplus \arch' \models
      \phi_2$, thus $\arch' \models \phi_1 \wand \phi_2$. \qed
    \end{compactitem}}

The rest of this section is concerned with the translation of any
\psil~ formula into an equivalent boolean combination of \sil~
formulae that are instances of a restricted set of patterns, called
\emph{test formulae}. As a remark, the test formulae are not \psil~
formulae, as they contain quantification, negation and equality
atoms. However, these constructs occur in a strictly controlled
context and will not be used outside test formulae.

\begin{definition}\label{def:test-formulae}
  Given a set $P \subseteq \psym$ of port symbols, $p \in P$ and $b$ a
  boolean term over the vocabulary $P$, the following are called
  \emph{test formulae}:
  \[\begin{array}{rcl}
  \has{p} & \isdef & p \inter p \wand \bot \\
  p \ieinter b & \isdef & \exists x ~.~ \bigwedge_{q \in P} x \neq q \wedge p \exinter x \cdot b ~* \top \\
  p \iecloseinter b & \isdef & p \closexinter b ~* \top 
  \end{array}\]
  Let $\test{P}$ be the set of test formulae $\phi$ such that
  $\portsof{\phi} \subseteq P$.  Given architectures $\arch_1$ and
  $\arch_2$, we write $\arch_1 \tfeq{P} \arch_2$ for $\arch_1 \models
  \phi \iff \arch_2 \models \phi$, for any $\phi \in \test{P}$.
\end{definition}

Intuitively, the test formulae $\has{p}$ are true in an architecture
$\arch$ whenever $p \in \dom{\arch}$. The test formulae $p \ieinter b$
(resp. $p \iecloseinter b$) are true in $\arch$ whenever
$\intf{\arch}$ contains an interaction $I$ such that $I \vdash b$ and
$I$ is a non-minimal (resp. minimal) model of $b$. The following lemma
states these properties formally:

\begin{lemma}\label{lemma:tf}
  Given an architecture $\arch = \tuple{D,\I}$, a set of port symbols
  $P \subseteq \psym$ and port symbol $p \in \psym$ and a boolean term
  $b$ over the vocabulary $\psym$, the following hold: \begin{compactenum}
  \item\label{it1:tf} $\arch \models \has{p} \iff p \in D$,
  \item\label{it2:tf} $\arch \models p \ieinter b \iff \text{ there
    exists } I \in \I \text{ such that } I \vdash b \text{ and } I
    \not\vdash_\mu b$,
  \item\label{it3:tf} $\arch \models p \iecloseinter b \iff \text{
    there exists } I \in \I \text{ such that } I \vdash_\mu b$.
  \end{compactenum}
\end{lemma}
\proof{ (\ref{it1:tf}) ``$\Rightarrow$'' If $\arch \models \has{p}$
  then for no disjoint architecture $\arch_1$ such that $\arch_1
  \models p \inter p$, the composition $\arch_1 \uplus \arch$ is
  defined. Since $\dom{\arch_1} = \set{p}$, the only reason for
  $\arch_1$ and $\arch$ not being disjoint is $p \in
  \dom{\arch}$. ``$\Leftarrow$'' $p \in \dom{\arch}$ means that any
  architecture $\arch_1$ such that $\dom{\arch_1} = \set{p}$ cannot
  compose with $\arch$, thus $\arch \models \has{p}$.

  \noindent (\ref{it2:tf}) For all valuations $\nu : \vars \rightarrow
  \ports$, we have $\arch \models _\nu p \ieinter b \iff \arch \models
  p \exinter x \cdot b ~* \top$, for some port $\nu(x)$ that is
  distinct from all $q \in P$. Since $I \vdash_\nu x \cdot b$, by
  induction on the structure of $b$, one shows that $I \vdash
  b$. Moreover, $I \not\vdash_\mu b$, because $I \setminus
  \set{\nu(x)} \subsetneq I$ and $I \setminus \set{\nu(x)} \vdash
  b$.

  \noindent (\ref{it3:tf}) Immediate, by the semantics of $p
  \iecloseinter b$. \qed}

Clearly, $\tfeq{P}$ is an equivalence relation between architectures.
Below we show that $\tfeq{P}$ is at least as fine as $\viseq{P}$:

\begin{lemma}\label{lemma:test-formulae-visible}
  Given a set of ports $P \in 2^{\ports}$ and two architectures
  $\arch_i = \tuple{D_i, \I_i}$, such that $D_i \subseteq P$, for each
  $i=1,2$, we have $\arch_1 \viseq{P} \arch_2$ if $\arch_1 \tfeq{P}
  \arch_2$.
\end{lemma}
\proof{ Assume that $\arch_1 \models \phi \iff \arch_2 \models \phi$
  for each $\phi \in \test{P}$ and prove the three points of
  Definition \ref{def:visible-equiv}:

  \vspace*{\baselineskip}\noindent (\ref{it1:visible-equiv}) Suppose,
  for a contradiction, that $D_1 \not\subseteq D_2$ and let $p \in D_1
  \setminus D_2$ be a port. Then, by Lemma \ref{lemma:tf}
  (\ref{it1:tf}), we have $\arch_1 \models \has{p}$ and $\arch_2
  \not\models \has{p}$, which contradicts that $\arch_1 \tfeq{P}
  \arch_2$, since $p \in D_1 \subseteq P$ and, consquently, $\has{p}
  \in \test{P}$. Then $D_1 \subseteq D_2$ and the proof for the other
  direction is symmetric.

  \vspace*{\baselineskip}\noindent (\ref{it2:visible-equiv}) Suppose,
  for a contradiction, that $\subsetproj{\I_1}{P} \not\subseteq
  \subsetproj{\I_2}{P}$ and let $I \in \subsetproj{\I_1}{P} \setminus
  \subsetproj{\I_2}{P}$ be an interaction. Because $I \in
  \subsetproj{\I_1}{P}$, we have $I \cap D_1 \neq \emptyset$ and $I
  \subseteq P$. Let $p \in I \cap D_1$ be a port and let $\set{q_1,
    \ldots, q_k} \isdef I \setminus \set{p}$. Then $p,q_1, \ldots, q_k
  \in P$, consequently $\portsof{p \iecloseinter q_1 \ldots q_k}
  \subseteq P$ and thus $p \iecloseinter q_1 \ldots q_k \in
  \test{P}$. Then, by Lemma \ref{lemma:tf} (\ref{it3:tf}), we have
  $\arch_1 \models p \iecloseinter q_1 \ldots q_k$ and $\arch_2
  \not\models q_1 \ldots q_k \iecloseinter b$, which contradicts with
  $\arch_1 \tfeq{P} \arch_2$. Then $\subsetproj{\I_1}{P} \subseteq
  \subsetproj{\I_2}{P}$. The proof for the other direction is
  symmetric.

  \vspace*{\baselineskip}\noindent (\ref{it3:visible-equiv}) Suppose,
  for a contradiction, that $\nsubsetproj{\I_1}{P} \sqcap P
  \not\subseteq \nsubsetproj{\I_2}{P} \sqcap P$ and let $I \in
  (\nsubsetproj{\I_1}{P} \sqcap P) \setminus (\nsubsetproj{\I_2}{P}
  \sqcap P)$ be an interaction. Then there exists an interaction $J
  \in \nsubsetproj{\I_1}{P}$ such that $J \setminus P \neq \emptyset$,
  $I = J \cap P$. Since $\arch_1$ is an architecture, $J \cap D_1 \neq
  \emptyset$ and let $p \in D_1 \cap J$ be a port. Because $D_1
  \subseteq P$, we have $p \in P$ and thus $p \in D_1 \cap I$. Let
  $\set{q_1, \ldots, q_k} \isdef I \setminus \set{p}$ and $\set{r_1,
    \ldots, r_m} \isdef P \setminus \set{p, q_1, \ldots, q_k}$.  By
  Lemma \ref{lemma:tf} (\ref{it2:tf}), we have $\arch_1 \models p
  \ieinter q_1 \ldots q_k \overline{r_1} \ldots
  \overline{r_m}$. Suppose, for a contradiction, that $\arch_2 \models
  p \ieinter q_1 \ldots q_k \overline{r_1} \ldots
  \overline{r_m}$. Then, by Lemma \ref{lemma:tf} (\ref{it2:tf}), there
  exists an interaction $J \in \nsubsetproj{\I_2}{P}$ such that $J
  \cap P = I \cap P$, which contradicts with $I \not\in
  \nsubsetproj{\I_2}{P} \sqcap P$. \qed}

The expressive completeness result of this section is stated below:

\begin{corollary}\label{cor:boolean-test-formulae}
  Each formula $\phi$ of \psil~ is equivalent to a finite boolean combination
  of test formulae from $\test{\portsof{\phi}}$. 
\end{corollary}
\proof{ Let $\arch$ be a models of $\phi$ and define the formula:
  \[\Phi(\arch) \isdef \bigwedge_{\begin{array}{c} \scriptstyle{\phi \in \test{\portsof{\phi}}} 
      \\[-1mm] \scriptstyle{\arch \models \phi} \end{array}} \phi
  \wedge \bigwedge_{\begin{array}{c} \scriptstyle{\phi \in
        \test{\portsof{\phi}}} \\[-1mm] \scriptstyle{\arch \not\models
        \phi} \end{array}} \neg\phi\] Since $\test{\portsof{\phi}}$ is
  finite, there are finitely many such formulae. In the following, we
  prove the equivalence $\varphi \equiv \bigvee_{\arch \models
    \varphi} \Phi(\arch)$. ``$\Rightarrow$'' Let $\arch \models
  \varphi$ be an architecture. Then clearly $\arch \models
  \Phi(\arch)$ by the definition of $\Phi(\arch)$, as a conjunction of
  formulae $\psi$, such that $\arch \models \psi$. ``$\Leftarrow$''
  Let $\arch \models \Phi(\arch')$, for some $\arch' \models
  \varphi$. Then $\arch \tfeq{\portsof{\varphi}} \arch'$, by the
  definition of $\Phi(\arch)$. By Lemma
  \ref{lemma:test-formulae-visible}, we obtain $\arch
  \viseq{\portsof{\varphi}} \arch'$ and, since $\arch' \models
  \varphi$, by Theorem \ref{thm:psil-equiv}, we have $\arch \models
  \varphi$. \qed}

\subsubsection{Translation of \psil into \qbf}

Our decision procedure for \psil\ is based on an
equivalence-preserving translation in \qbf, which enables the use of
off-the-shelf QSAT solvers to decide the satisfiability and entailment
problem for \psil. Moreover, since any \qbf\ formula is a succint
encoding of a propositional formula, we obtain a finite representation
of the set of models of a \psil\ formula, that will become useful in
designing a verification method for the safety properties of a system
described by recursive predicates (\S\ref{sec:behaviors}).

From now until the end of this section, let $P = \set{p_1, \ldots,
  p_k}$ be a set of visible ports and denote by $\B$ the following set
of boolean variables, parameterized by $P$: \begin{compactitem}
\item $\bhas(i)$ stands for the test formulae $\has{p_i}$, for all $1
  \leq i \leq k$,
\item $\bopen(i_1,\ldots,i_\ell)$, for all $1 \leq i_1 < \ldots <
  i_\ell \leq k$, stands for the following boolean combination of test
  formulae: \(\bigvee_{1 \leq h \leq \ell} \has{p_{i_h}} \wedge
  p_{i_h} \ieinter p_{i_1} \ldots p_{i_\ell}\).
\item $\bclose(i_1,\ldots,i_\ell)$, for all $1 \leq i_1 < \ldots <
  i_\ell \leq k$, stands for the following boolean combination of test
  formulae: \(\bigvee_{1 \leq h \leq \ell} \has{p_{i_h}} \wedge
  p_{i_h} \iecloseinter p_{i_1} \cdots p_{i_\ell}\).
\end{compactitem}
Clearly there are $2^{\bigO(k)}$ boolean variables in $\B$.  In the
following, we use the shorthands $\B' \isdef \set{b' \mid b \in \B}$,
$\B'' \isdef \set{b'' \mid b \in \B}$ and $\exists \B ~.~ \phi$
(resp. $\exists B'$ and $\exists B''$) for the formula obtained from
$\phi$ by existentially quantifying every boolean variable from $\B$
(resp. $\B'$ and $\B''$). We write $\vec{\imath}$
(resp. $\vec{\jmath}$) for the strictly increasing sequence $i_1 <
\ldots < i_\ell$ (resp. $j_1 < \ldots < j_\ell$). Since there are at
most $2^k$ such sequences, we obtain that $\card{\B} = \card{\B'} =
\card{\B''} = 2^{\bigO(k)}$.

Before giving the translation of an arbitrary formula of \psil\ into
\qbf, we need to introduce a number of shorthands. First, the boolean
formula below characterizes those boolean valuations of $\B$ that
define valid architectures:
\[\arch(\B) \isdef \bigwedge_{1 \leq \vec{\imath} \leq k} (\bopen(\vec{\imath}) \vee \bclose(\vec{\imath})) \rightarrow
\bigvee_{\scriptstyle{1 \leq h \leq \ell}} \bhas(i_h)\] More
precisely, if $\beta : \B \rightarrow \set{\bot,\top}$ is a boolean
valuation, such that $\beta \models \arch(\B)$, the architectures
corresponding to $\beta$ are the members of the set
$\boolarch{\beta}$, defined below:

\begin{definition}\label{def:boolarch}
For any architecture $\arch = \tuple{D,\I}$ and any boolean valuation
$\beta : \B \rightarrow \set{\bot,\top}$, we have $\arch \in
\boolarch{\beta}$ if and only if the following
hold: \begin{compactitem}
\item $D = \set{p_i \mid \beta(\bhas(i)) = \top}$,
\item $\subsetproj{\I}{P} = \set{\set{p_{i_1}, \ldots, p_{i_\ell}}
  \mid \beta(\bclose(i_1, \ldots, i_\ell)) = \top}$,
\item $\nsubsetproj{\I}{P} \sqcap P = \set{\set{p_{i_1}, \ldots,
    p_{i_\ell}} \mid \beta(\bopen(i_1, \ldots, i_\ell)) = \top}$.
\end{compactitem}
\end{definition}
It is not hard to prove that the set of boolean valuations $\set{\beta
  : \B \rightarrow \set{\bot,\top} \mid \arch \in \boolarch{\beta}}$
is closed under intersection and has a minimal element, denoted by
$\beta_\arch$ in the following.

The formula $\#(\B,\B')$ below states that the architectures defined
by the boolean sets $\B$ and $\B'$ have disjoint domains:
\[\#(\B,\B') \isdef \bigwedge_{1 \leq i \leq k} \neg(\bhas'(i) \wedge \bhas''(i))\]

\begin{lemma}\label{lemma:disjoint}
  For any valuations $\beta : \B \rightarrow \set{\bot,\top}$ and
  $\beta' : \B' \rightarrow \set{\bot,\top}$, such that $\beta \cup
  \beta' \models \#(\B,\B')$ and any architectures $\tuple{D,\I} \in
  \boolarch{\beta}$ and $\tuple{D',\I'} \in \boolarch{\beta'}$, we
  have $D \cap D' = \emptyset$.
\end{lemma}
\proof{ Since $\arch \in \boolarch{\beta}$, we have $D = \set{p \mid
    \beta(\bhas(j)) = \top}$ and $D' = \set{p \mid \beta(\bhas'(j)) =
    \top}$, by a similar argument for $\boolarch{\beta'}$. Suppose,
  for a contradiction, that there exists a port $p_j \in D \cap
  D'$. Then $\beta(\bhas(j)) = \beta'(\bhas'(j)) = \top$, 
  contradicting $\beta \cup \beta' \models \#(\B,\B')$. \qed}

The formula $\biguplus(\B,\B',\B'')$ below states that, whenever $\B'$
and $\B''$ define disjoint architectures $\arch'$ and $\arch''$, $\B$
defines their composition $\arch' \uplus \arch''$:

\[\begin{array}{rcl}
\biguplus(\B,\B',\B'') & \isdef & \bigwedge_{1 \leq i \leq k} (\bhas(i) \leftrightarrow (\bhas'(i) \vee \bhas''(i))) \\[2mm]
& \wedge & \bigwedge_{1 \leq \vec{\imath} \leq k} (\bclose'(\vec{\imath}) \wedge \bclose''(\vec{\imath}) \rightarrow \bclose(\vec{\imath})) \\[2mm]
& \wedge & \bigwedge_{1 \leq \vec{\imath} \leq k} (\bclose'(\vec{\imath}) \wedge \bigwedge_{1 \leq h \leq \ell} \neg\bhas''(i_h) \rightarrow \bclose(\vec{\imath})) \\[2mm]
& \wedge & \bigwedge_{1 \leq \vec{\imath} \leq k} (\bclose''(\vec{\imath}) \wedge \bigwedge_{1 \leq h \leq \ell} \neg\bhas'(i_h) \rightarrow \bclose(\vec{\imath})) \\[2mm]
& \wedge & \bigwedge_{1 \leq \vec{\imath} \leq k} (\bopen'(\vec{\imath}) \wedge \bigwedge_{1 \leq h \leq \ell} \neg\bhas''(i_h) \rightarrow \bopen(\vec{\imath})) \\[2mm]
& \wedge & \bigwedge_{1 \leq \vec{\imath} \leq k} (\bopen''(\vec{\imath}) \wedge \bigwedge_{1 \leq h \leq \ell} \neg\bhas'(i_h) \rightarrow \bopen(\vec{\imath})) \\[2mm]
& \wedge & \bigwedge_{1 \leq \vec{\imath} \leq k} (\neg\bclose'(\vec{\imath}) \wedge \neg\bclose''(\vec{\imath}) \rightarrow \neg\bclose(\vec{\imath})) \\[2mm]
& \wedge & \bigwedge_{1 \leq \vec{\imath} \leq k} (\bclose'(\vec{\imath}) \wedge \neg\bclose''(\vec{\imath}) \wedge \bigvee_{1 \leq h \leq \ell} \bhas''(i_h) \rightarrow \neg\bclose(\vec{\imath})) \\[2mm]
& \wedge & \bigwedge_{1 \leq \vec{\imath} \leq k} (\bclose''(\vec{\imath}) \wedge \neg\bclose'(\vec{\imath}) \wedge \bigvee_{1 \leq h \leq \ell} \bhas'(i_h) \rightarrow \neg\bclose(\vec{\imath})) \\[2mm]
& \wedge & \bigwedge_{1 \leq \vec{\imath} \leq k} (\neg\bopen'(\vec{\imath}) \wedge \neg\bopen''(\vec{\imath}) \rightarrow \neg\bopen(\vec{\imath})) \\[2mm]
& \wedge & \bigwedge_{1 \leq \vec{\imath} \leq k} (\bopen'(\vec{\imath}) \wedge \neg\bopen''(\vec{\imath}) \wedge \bigvee_{1 \leq h \leq \ell} \bhas''(i_h) \rightarrow \neg\bopen(\vec{\imath})) \\[2mm]
& \wedge & \bigwedge_{1 \leq \vec{\imath} \leq k} (\bopen''(\vec{\imath}) \wedge \neg\bopen'(\vec{\imath}) \wedge \bigvee_{1 \leq h \leq \ell} \bhas'(i_h) \rightarrow \neg\bopen(\vec{\imath})) 
\end{array}\]
Note that nothing can be stated about $\bopen(\vec{\imath})$ when
$\bopen'(\vec{\imath})$ and $\bopen''(\vec{\imath})$ both hold, because these
boolean variables denote interactions that coincide on their visible
part, whereas $\bopen(\vec{\imath})$ holds only when those interactions
coincide also on their invisible parts.

\begin{lemma}\label{lemma:composition}
  For any valuations $\beta' : \B' \rightarrow \set{\bot,\top}$ and
  $\beta'' : \B'' \rightarrow \set{\bot,\top}$, such that $\beta' \cup
  \beta'' \models \#(\B',\B'')$, the following
  hold: \begin{compactenum}
  \item\label{it1:composition} for any architectures $\arch' \in
    \boolarch{\beta'}$ and $\arch'' \in \boolarch{\beta''}$, there
    exists a valuation $\beta : \B \rightarrow \set{\bot,\top}$ such
    that $\beta \cup \beta' \cup \beta'' \models
    \biguplus(\B,\B',\B'')$ and $\arch' \uplus \arch'' \in
    \boolarch{\beta}$,
  \item\label{it2:composition} for any valuation $\beta : \B
    \rightarrow \set{\bot,\top}$ such that $\beta \cup \beta' \cup
    \beta'' \models \biguplus(\B,\B',\B'')$ and any $\arch \in
    \boolarch{\beta}$ there exist $\arch' \in \boolarch{\beta'}$ and
    $\arch'' \in \boolarch{\beta''}$, such that $\arch = \arch' \uplus
    \arch''$.
    \end{compactenum}
\end{lemma}
\proof{(\ref{it1:composition}) Let $\arch' = \tuple{D',\I'}$, $\arch''
  = \tuple{D'', \I''}$ and $\arch' \uplus \arch'' \isdef
  \tuple{D,\I}$. Let $\beta = \beta_{\arch' \uplus \arch''}$. Clearly,
  we have $\arch' \uplus \arch'' \in \boolarch{\beta}$. It remains to
  show that $\beta \cup \beta' \cup \beta'' \models
  \biguplus(\B,\B',\B'')$, by proving that each of the implications
  from the definition of $\biguplus(\B,\B',\B'')$ is valid. We prove
  the most interesting cases below and leave the rest to the
  reader: \begin{compactitem}
  \item $\bigwedge_{1 \leq i \leq k} (\bhas(i) \leftrightarrow
    (\bhas'(i) \vee \bhas''(i)))$: Because $\arch' \in
    \boolarch{\beta'}$, we have $D' = \set{p_i \mid \beta'(\bhas'(i))
      = \top}$ and, because $\arch'' \in \boolarch{\beta''}$, we have
    $D'' = \set{p_i \mid \beta''(\bhas''(i)) = \top}$. Because $D'
    \cup D'' = \set{p_i \mid \beta(\bhas(i)) = \top}$, by the
    definition of $\beta$, we have $\beta \cup \beta' \cup \beta''
    \models \bhas(i) \leftrightarrow (\bhas'(i) \vee \bhas''(i))$, for
    each $1 \leq i \leq k$.
    \vspace*{\baselineskip}
  \item $\bigwedge_{1 \leq \vec{\imath} \leq k} (\bclose'(\vec{\imath}) \wedge
    \bclose''(\vec{\imath}) \rightarrow \bclose(\vec{\imath}))$: Because $\arch'
    \in \boolarch{\beta'}$, we have $\subsetproj{\I'}{P} = \set{p(i_1,
      \ldots, i_\ell) \mid \beta'(\bclose'(i_1, \ldots, i_\ell)) =
      \top}$ and, because $\arch'' \in \boolarch{\beta''}$, we have
    $\subsetproj{\I''}{P} = \set{p(i_1, \ldots, i_\ell) \mid
      \beta''(\bclose''(i_1, \ldots, i_\ell)) = \top}$. Assume that,
    for some $1 \leq i_1, \ldots, i_\ell \leq k$, we have $\beta'
    \models \bclose'(i_1, \ldots, i_\ell)$ and $\beta'' \models
    \bclose''(i_1, \ldots, i_\ell)$. Then $\set{p_{i_1}, \ldots,
      p_{i_\ell}} \in \subsetproj{\left(\I' \cap \I''\right)}{P}$ and
    $\beta \models \bclose(i_1, \ldots, i_\ell)$ follows, by the
    definition of $\beta$.
    \vspace*{\baselineskip}
  \item $\bigwedge_{1 \leq \vec{\imath} \leq k} (\bclose'(\vec{\imath}) \wedge
    \bigwedge_{1 \leq h \leq \ell} \neg\bhas''(i_h) \rightarrow
    \bclose(\vec{\imath}))$: Because $\arch' \in \boolarch{\beta'}$, we
    have $\subsetproj{\I'}{P} = \set{p(i_1, \ldots, i_\ell) \mid
      \beta'(\bclose'(i_1, \ldots, i_\ell)) = \top}$ and, because
    $\arch'' \in \boolarch{\beta''}$, we have $D'' = \set{p_i \mid
      \beta''(\bhas''(i)) = \top}$. Assume that, for some $1 \leq i_1,
    \ldots, i_\ell \leq k$, we have $\beta' \models \bclose'(i_1,
    \ldots, i_\ell)$ and $\beta'' \not\models \bhas''(i_1, \ldots,
    i_\ell)$. Then, $\set{p_{i_1}, \ldots, p_{i_\ell}} \in
    \subsetproj{\I'}{P}$ and $\set{p_{i_1}, \ldots, p_{i_\ell}} \cap
    D'' = \emptyset$, i.e.\ $\set{p_{i_1}, \ldots, p_{i_\ell}} \in
    \subsetproj{\left(\I' \cap 2^{\overline{D''}}\right)}{P}$ By the
    definition of $\beta$, we have $\beta \models \bclose(i_1, \ldots,
    i_\ell)$. 
    \vspace*{\baselineskip}
    \item $\bigwedge_{1 \leq \vec{\imath} \leq k}
      (\bopen'(\vec{\imath}) \wedge \neg\bopen''(\vec{\imath}) \wedge
      \bigvee_{1 \leq h \leq \ell} \bhas''(i_h) \rightarrow
      \neg\bopen(\vec{\imath}))$: Because $\arch' \in
      \boolarch{\beta'}$, we have $\nsubsetproj{\I'}{P} \sqcap P =
      \set{p(i_1, \ldots, i_\ell) \mid \beta'(\bopen'(i_1, \ldots,
        i_\ell)) = \top}$ and, because $\arch'' \in
      \boolarch{\beta''}$, we have $D'' = \set{p_i \mid
        \beta''(\bhas''(i)) = \top}$ and $\nsubsetproj{\I''}{P} \sqcap
      P = \set{p(i_1, \ldots, i_\ell) \mid \beta''(\bopen''(i_1,
        \ldots, i_\ell)) = \top}$. Assume that, for some $1 \leq i_1,
      \ldots, i_\ell \leq k$, we have $\beta' \cup \beta'' \models
      \bopen'(i_1, \ldots, i_\ell) \wedge \neg\bopen''(i_1, \ldots,
      i_\ell) \wedge \bigvee_{1 \leq h \leq \ell} \bhas''(i_h)$. Then,
      by the definition of $\beta$, we have $\beta \not\models
      \bopen(i_1, \ldots, i_\ell)$, because $\set{p_{i_1}, \ldots,
        p_{i_\ell}} \not\in \nsubsetproj{\left(\I' \cap
        \I''\right)}{P} \sqcap P \cup \nsubsetproj{\left(\I' \cap
        2^{\overline{D''}}\right)}{P} \sqcap P \cup
      \nsubsetproj{\left(\I'' \cap 2^{\overline{D'}}\right)}{P} \sqcap
      P$, which is an easy check.
  \end{compactitem}

  \vspace*{\baselineskip}\noindent(\ref{it2:composition}) Let $\arch =
  \tuple{D, \I}$ be a given architecture and $\arch' = \tuple{D',
    \I'}$, $\arch'' = \tuple{D'', \I''}$, be architectures defined as
  follows:
  \[\begin{array}{rcl}
  D' & \isdef & \set{p_i \mid \beta'(\bhas'(i)) = \top} \\
  D'' & \isdef & \set{p_i \mid \beta''(\bhas''(i)) = \top} \\
  \I' & \isdef & \set{\set{i_1, \ldots, i_\ell} \mid \beta'(\bclose'(i_1, \ldots, i_\ell))=\top} \cup X' \\
  \I'' & \isdef & \set{\set{i_1, \ldots, i_\ell} \mid \beta''(\bclose''(i_1, \ldots, i_\ell))=\top} \cup X''
  \end{array}\]
  where $X'$ and $X''$ are defined below, for two distinct ports
  $\alpha', \alpha'' \in \ports \setminus
  \bigcup\I$: \begin{compactitem}
  \item for all $1 \leq i_1, \ldots, i_\ell \leq k$, such that
    $\beta'(\bopen'(i_1, \ldots, i_\ell)) = \top$, if $\set{i_1,
    \ldots, i_\ell} \cup Y \in \I$, for some $Y \subseteq \ports
    \setminus P$, we have $\set{i_1, \ldots, i_\ell} \cup Y \in \I'$,
    else $\set{i_1, \ldots, i_\ell} \cup \set{\alpha'} \in \I'$ and,
    moreover, nothing else is in $\I'$.
  \item for all $1 \leq i_1, \ldots, i_\ell \leq k$, such that
    $\beta''(\bopen''(i_1, \ldots, i_\ell)) = \top$, if $\set{i_1,
    \ldots, i_\ell} \cup Y \in \I$, for some $Y \subseteq \ports
    \setminus P$, we have $\set{i_1, \ldots, i_\ell} \cup Y \in \I''$,
    else $\set{i_1, \ldots, i_\ell} \cup \set{\alpha''} \in \I''$ and,
    moreover, nothing else is in $\I''$.
  \end{compactitem}
  It is easy to check that $\nsubsetproj{\I'}{P} \sqcap P =
  \set{\set{i_1, \ldots, i_\ell} \mid \beta'(\bopen'(i_1, \ldots,
    i_\ell)) = \top}$ and $\nsubsetproj{\I''}{P} \sqcap P =
  \set{\set{i_1, \ldots, i_\ell} \mid \beta''(\bopen''(i_1, \ldots,
    i_\ell)) = \top}$. Consequently, $\arch' \in \boolarch{\beta'}$
  and $\arch'' \in \boolarch{\beta''}$ is again an easy check. We are
  left with proving that $\arch = \arch' \uplus \arch''$. First, we
  compute:
  \[\begin{array}{rcll}
  D & = & \set{p_i \mid \beta(\bhas(i)) = \top} & \text{, because $\arch \in \boolarch{\beta}$} \\
  & = & \set{p_i \mid \beta'(\bhas'(i)) = \top \vee \beta''(\bhas''(i)) = \top} & \text{, because $\beta \cup \beta' \cup \beta'' \models \bhas(i) \leftrightarrow (\bhas'(i) \vee \bhas''(i))$} \\
  & = & \set{p_i \mid \beta'(\bhas'(i)) = \top} \cup \set{p_i \mid \beta''(\bhas''(i)) = \top} \\
  & = & D' \cup D'' & \text{, by the definitions of $D'$ and $D''$}
  \end{array}\]
  To show that $\I = (\I' \cap \I'') \cup (\I' \cap
  2^{\overline{D''}}) \cup (\I'' \cap 2^{\overline{D'}})$, we prove
  the following points: \begin{compactitem}
  \item $\subsetproj{\I}{P} = \subsetproj{\left(\I' \cap
    \I''\right)}{P} \cup \subsetproj{\left(\I' \cap
    2^{\overline{D''}}\right)}{P} \cup \subsetproj{\left(\I'' \cap
    2^{\overline{D'}}\right)}{P}$: Because $\beta \cup \beta' \cup
    \beta'' \models \biguplus(\B,\B',\B'')$, we obtain, by a simple
    rewriting of the $\biguplus(\B,\B',\B'')$
    formula: \[\beta\cup\beta'\cup\beta'' \models \bigwedge_{1 \leq
      \vec{\imath} \leq k} \left(\left((\bclose'(\vec{\imath}) \wedge
    \bclose''(\vec{\imath})) \vee \left(\bclose'(\vec{\imath}) \wedge \bigwedge_{1 \leq
      h \leq \ell} \neg\bhas''(i_h)\right) \vee \left(\bclose''(\vec{\imath}) \wedge
    \bigwedge_{1 \leq h \leq \ell} \neg\bhas'(i_h)\right)\right)
    \leftrightarrow \bclose(\vec{\imath})\right)\] Because $\arch \in
    \boolarch{\beta}$, we have $\subsetproj{\I}{P} = \set{\set{i_1,
        \ldots, i_\ell} \mid \beta(\bclose(i_1, \ldots, i_\ell)) =
      \top}$. Moreover, because $\arch' \in \boolarch{\beta'}$, we
    have $D' = \set{p_i \mid \beta'(\bhas'(i)) = \top}$ and
    $\subsetproj{\I'}{P} = \set{\set{i_1, \ldots, i_\ell} \mid
      \beta'(\bclose'(i_1, \ldots, i_\ell)) = \top}$ and, because
    $\arch'' \in \boolarch{\beta''}$, we have $D'' = \set{p_i \mid
      \beta''(\bhas''(i)) = \top}$ and $\subsetproj{\I''}{P} =
    \set{\set{i_1, \ldots, i_\ell} \mid \beta''(\bclose''(i_1, \ldots,
      i_\ell)) = \top}$, which implies the required equality.
  \item $\nsubsetproj{\I}{P} = \nsubsetproj{\left(\I' \cap
    \I''\right)}{P} \cup \nsubsetproj{\left(\I' \cap
    2^{\overline{D''}}\right)}{P} \cup \nsubsetproj{\left(\I'' \cap
    2^{\overline{D'}}\right)}{P}$: ``$\subseteq$'' Let $I \in
    \nsubsetproj{\I}{P}$ be an interaction. Because $\arch \in
    \boolarch{\beta}$, there exists $Y \subseteq \ports \setminus P$
    such that $I = \set{i_1, \ldots, i_\ell} \cup Y$, where
    $\beta(\bopen(i_1, \ldots, i_\ell)) = \top$. Since $\arch$ is an
    architecture, it must be that $\set{i_1, \ldots, i_\ell} \cap D
    \neq \emptyset$. Since $D = D' \cup D''$, by the previous point,
    we distinguish the cases below: \begin{compactitem}
    \item if $\set{i_1, \ldots, i_\ell} \cap D' \neq \emptyset$ and
      $\set{i_1, \ldots, i_\ell} \cap D'' \neq \emptyset$ then,
      because $\beta\cup\beta'\cup\beta'' \models
      \biguplus(\B,\B',\B'')$ we obtain:
      \[\begin{array}{rcl} 
      \beta'\cup\beta'' & \models & (\bopen'(i_1, \ldots, i_\ell) \vee \bopen''(i_1, \ldots, i_\ell)) \wedge \\ 
      && (\neg\bopen'(i_1,\ldots, i_\ell) \vee \bopen''(i_1, \ldots, i_\ell)) \wedge \\ 
      && (\neg\bopen''(i_1, \ldots,i_\ell) \vee \bopen'(i_1, \ldots, i_\ell))
      \end{array}\] thus $\beta'(\bopen'(i_1, \ldots, i_\ell)) =
      \beta''(\bopen''(i_1, \ldots, i_\ell)) = \top$. By the
      definition of $\I'$ and $\I''$, we obtain $I \in \I' \cap \I''$. 
    \item else, if $\set{i_1, \ldots, i_\ell} \cap D' \neq \emptyset$ and
      $\set{i_1, \ldots, i_\ell} \cap D'' = \emptyset$ then,
      because $\beta\cup\beta'\cup\beta'' \models
      \biguplus(\B,\B',\B'')$ we obtain:
      \[\begin{array}{rcl}
      \beta'\cup\beta'' & \models & (\bopen'(i_1, \ldots, i_\ell) \vee
      \bopen''(i_1, \ldots, i_\ell)) \wedge (\neg\bopen''(i_1,\ldots,
      i_\ell) \vee \bopen'(i_1, \ldots, i_\ell))
      \end{array}\]
      thus $\beta'(\bopen'(i_1, \ldots, i_\ell)) = \top$ and, by the
      definition of $\I'$, we have $I \in \I' \cap 2^{\overline{D''}}$.
    \item otherwise, if $\set{i_1, \ldots, i_\ell} \cap D' =
      \emptyset$ and $\set{i_1, \ldots, i_\ell} \cap D'' \neq
      \emptyset$ then, by a symmetric argument, we obtain $I \in \I''
      \cap 2^{\overline{D'}}$.
    \end{compactitem}
    ``$\supseteq$'' Let $I$ be an interaction such that $I \cap P =
    \set{i_1, \ldots, i_\ell}$ and $I \setminus P = Y$. We consider
    the following cases: \begin{compactitem}
    \item if $I \in \nsubsetproj{\left(\I' \cap \I''\right)}{P}$ then
      $Y \neq \set{\alpha'}$ and $Y \neq \set{\alpha''}$, thus $I =
      \set{i_1, \ldots, i_\ell} \cup Y \in \I$, by the definition of
      $\I'$ and $\I''$.
    \item else, if $\I \in \nsubsetproj{\left(\I' \cap
      2^{\overline{D''}}\right)}{P}$ then, because $\beta \cup \beta'
      \cup \beta'' \models \biguplus(\B,\B',\B'')$, we obtain
      $\beta(\bopen(i_1,\ldots,i_\ell))=\top$, thus $\set{i_1, \ldots,
        i_\ell} \in \subsetproj{\I}{P} \sqcap P$. Moreover, by the
      definition of $\I'$, we have that $Y \neq \set{\alpha'}$, thus
      $I = \set{i_1, \ldots, i_\ell} \cup Y \in \I$.
    \item the last case $\I \in \nsubsetproj{\left(\I'' \cap
      2^{\overline{D'}}\right)}{P}$ is symmetric to the previous. \qed
    \end{compactitem}
  \end{compactitem}}

Before giving the effective translation of \psil\ formulae to \qbf, we
define the formula $\closure(\B,\B')$, stating that $\B$ defines an
architecture which is the closure of an architecture defined by $\B'$:

\[\closure(\B,\B') \isdef \bigwedge_{1 \leq j \leq k} (\bhas(j) \leftrightarrow \bhas'(j))
\wedge \bigwedge_{1 \leq \vec{\imath} \leq k} (\neg\bopen(\vec{\imath})
\wedge (\bclose(\vec{\imath}) \leftrightarrow \bclose'(\vec{\imath})))\]

\begin{lemma}\label{lemma:close}
  For any valuations $\beta : \B \rightarrow \set{\bot,\top}$, such
  that $\beta \cup \beta' \models \closure(\B,\B')$ and $\beta' : \B'
  \rightarrow \set{\bot,\top}$ and any architectures such that $\arch
  \in \boolarch{\beta}$ and $\arch' \in \boolarch{\beta'}$, we have
  $\arch \closure \arch'$. 
\end{lemma}
\proof{ Let $\arch = \tuple{D,\I}$ and $\arch' =
  \tuple{D',\I'}$. Because $\beta \cup \beta' \models
  \closure(\B,\B')$, $\arch \in \boolarch{\beta}$ and $\arch' \in
  \boolarch{\beta'}$, we have $D = \set{p_i \mid \beta(\bhas(i)) =
    \top} = \set{p_i \mid \beta'(\bhas'(j)) = \top} = D'$. Moreover,
  $\subsetproj{\I}{P} = \set{\set{i_1, \ldots, i_\ell} \mid
    \beta(\bclose(i_1,\ldots,i_\ell)) = \top} = \set{\set{i_1, \ldots,
      i_\ell} \mid \beta'(\bclose'(i_1,\ldots,i_\ell)) = \top} =
  \subsetproj{\I'}{P}$ and $\nsubsetproj{\I}{P} = \emptyset$, thus $\I
  = \subsetproj{\I'}{P}$. \qed}

Let us fix the set of visible port symbols $P = \set{p_1, \ldots, p_k}
\subseteq \psym$ for the rest of this section. We view the port
symbols in $P$ as propositional variables and write $\exists P$
[$\forall P$] for $\exists p_1 \ldots \exists p_k$ [$\forall p_1
  \ldots \forall p_k$].  Given a nonempty strictly increasing sequence
$\vec{\imath} = i_1 < \ldots < i_\ell \in \set{1, \ldots, k}$ and a
boolean term $b$ over $P$, we define the propositional formulae below:
\[\begin{array}{rcl}
\pi(\vec{\imath}) & \isdef & \bigwedge_{j \in \set{i_1, \ldots, i_\ell}} p_j \wedge \bigwedge_{j \not\in \set{i_1, \ldots, i_\ell}} \neg p_j \\\\
\theta(p_i) & \isdef & p_i \hspace*{1cm} \theta(b_1 \cdot b_2) \isdef \theta(b_1) \wedge \theta(b_2) \hspace*{1cm} \theta(\overline{b}_1) \isdef \neg \theta(b_1)
\end{array}\]

The translation of a \psil\ formula in \qbf\ is defined recursively on
its structure:
\[\begin{array}{rcl}
\btr{\emp}{\B} & \isdef & \bigwedge_{1 \leq i \leq k} \neg\bhas(i)
\\[2mm]
\btr{p_i \inter b}{\B} & \isdef & \bhas(i) \wedge \bigwedge_{1 \leq j \neq i \leq k} \neg\bhas(j) \wedge 
\bigwedge_{1 \leq \vec{\imath} \leq k} \forall P ~.~ [\pi(\vec{\imath}) \wedge (\bopen(\vec{\imath}) \vee \bclose(\vec{\imath}))] \rightarrow \theta(b)
\\[2mm]
\btr{p_i \closeinter b}{\B} & \isdef & \bhas(i) \wedge \bigwedge_{1 \leq j \neq i \leq k} \neg\bhas(j) \wedge \bigwedge_{1 \leq \vec{\imath} \leq k} \neg\bopen(\vec{\imath}) ~\wedge \\[2mm]
&& \forall P ~.~ [(\pi(\vec{\imath}) \wedge \bclose(\vec{\imath})) \rightarrow \theta(b)] \wedge 
\bigwedge_{\vec{\jmath} \subsetneq \vec{\imath}} \pi(\vec{\jmath}) \rightarrow \neg\theta(b)
\end{array}\]
\[\begin{array}{rcl}
\btr{p_i \exinter b}{\B} & \isdef & \bhas(i) \wedge \bigwedge_{1 \leq j \neq i \leq k} \neg\bhas(j) \wedge 
\bigvee_{1 \leq \vec{\imath} \leq k} \exists P ~.~ \pi(\vec{\imath}) \wedge \theta(b) \wedge (\bopen(\vec{\imath}) \vee \bclose(\vec{\imath}))
\\[2mm]
\btr{p_i \closexinter b}{\B} & \isdef & \bhas(i) \wedge \bigwedge_{1 \leq j \neq i \leq k} \neg\bhas(j) ~\wedge \\[2mm]
&& \bigvee_{1 \leq \vec{\imath} \leq k} \exists P ~.~ \pi(\vec{\imath}) \wedge \theta(b) \wedge \bclose(\vec{\imath}) \wedge
\forall P ~.~ \bigwedge_{\vec{\jmath} \subsetneq \vec{\imath}} \pi(\vec{\jmath}) \rightarrow \neg\theta(b)
\end{array}\]
\[\begin{array}{rcl}
\btr{\langle\phi_1\rangle}{\B} & \isdef & \exists \B' ~.~ \arch(\B') \wedge \closure(\B,\B') \wedge \btr{\phi_1}{\B'} \\[2mm]
\btr{\phi_1 \wedge \phi_2}{\B} & \isdef & \btr{\phi_1}{\B} \wedge \btr{\phi_2}{\B} \\[2mm]
\btr{\phi_1 \vee \phi_2}{\B} & \isdef & \btr{\phi_1}{\B} \vee \btr{\phi_2}{\B} \\[2mm]
\btr{\phi_1 * \phi_2}{\B} & \isdef & \exists \B' \exists \B'' ~.~ \arch(\B') \wedge \arch(\B'') \wedge \#(\B',\B'') \wedge
\biguplus(\B,\B',\B'') \wedge \btr{\phi_1}{\B'} \wedge \btr{\phi_2}{\B''} \\[2mm]
\btr{\phi_1 \wand \phi_2}{\B} & \isdef & \forall \B' ~.~ \arch(\B') \wedge \#(\B,\B') \wedge \btr{\phi_1}{\B'} \rightarrow
\exists \B'' ~.~ \arch(\B'') \wedge \biguplus(\B'',\B,\B') \wedge \btr{\phi_2}{\B''}
\end{array}\]
Note that, for any \psil\ formula $\phi$, $\btr{\phi}{\B}$ is a QBF
formula with free variables in $\B$. The following result proves the equivalence between
\psil\ formulae and their QBF translations. 

\begin{theorem}\label{thm:psil-qbf}
  Given a \psil\ formula $\phi$, such that $\portsof{\phi} \subseteq
  P$, for any architecture $\arch = \tuple{D,\I}$, we have $\arch
  \models \phi$ if and only if $\beta_\arch \models \btr{\phi}{\B}$.
\end{theorem}
\proof{ 
  We prove first the following fact:   
  \begin{fact}\label{fact:eq-bval}
    Given an architecture $\arch$ and boolean valuations $\beta,
    \beta' : \B \rightarrow \set{\bot,\top}$, if $\arch \in
    \boolarch{\beta} \cap \boolarch{\beta'}$ then $\beta = \beta'$.
  \end{fact}
  \proof{ Necessarily $\beta$ and $\beta'$ agree on any propositional
    variable from $\B$. \qed}
  The proof goes by induction on the structure of $\phi$. We
  consider the cases below:
  \begin{compactitem}
    \vspace*{\baselineskip}
  \item $\emp$: $\arch \models \emp \iff D = \emptyset \iff
    \beta_\arch \models \bigwedge_{1 \leq i \leq k} \neg\bhas(i)$.
    \vspace{\baselineskip}
  \item $p_i \inter b$: ``$\Rightarrow$'' If $\arch \models p_i
    \inter b$ then $D = \set{p_i}$ and $I \vdash b$, for all $I \in
    \I$ and $\beta_\arch \models \btr{p_i \inter b}{\B}$ is an easy
    check. ``$\Leftarrow$'' Since $\beta_\arch \models \bhas(i)
    \wedge \bigwedge_{1 \leq j \neq i \leq k} \neg\bhas(j)$, we have
    $D = \set{p_i}$. Let $I \in \I$ be an arbitrary interaction of
    $\arch$. We distinguish two cases: \begin{compactitem}
    \item if $I \in \subsetproj{\I}{P}$ then let $\set{p_{i_1},
      \ldots, p_{i_\ell}} \isdef I$. We obtain that
      $\beta_\arch(\bclose(i_1, \ldots, i_\ell))=\top$, thus
      $\beta_\arch \models \forall P ~.~ \pi(i_1, \ldots, i_\ell)
      \rightarrow \theta(b)$, leading to $I \vdash b$.
    \item else $I \in \nsubsetproj{\I}{P}$ and let $\set{p_{i_1},
      \ldots, p_{i_\ell}} \isdef I \cap P$. We obtain
      $\beta(\bopen(i_1, \ldots, i_\ell)) = \top$, thus $\beta
      \models \forall P ~.~ \pi(i_1, \ldots, i_\ell) \rightarrow
      \theta(b)$, leading to $I \vdash b$.
    \end{compactitem}
    Consequently, we have $\arch \models p_i \inter b$. 
    \vspace*{\baselineskip}
  \item $p_i \closeinter b$: ``$\Rightarrow$'' This direction is
    similar to the above point. ``$\Leftarrow$'' Similar to the
    above point, we obtain $D = \set{p_i}$. Let $I \in \I$ be an
    arbitrary interaction of $\arch$ and $\set{p_{i_1}, \ldots,
      p_{i_\ell}} \isdef I \cap P$. Since $\beta_\arch \models
    \bigwedge_{1 \leq \vec{\imath} \leq k} \neg\bopen(i_1, \ldots,
    i_\ell)$ then $I \not\in \nsubsetproj{\I}{P}$, so the only
    possibility is $I \in \subsetproj{\I}{P}$ and thus $I =
    \set{p_{i_1}, \ldots, p_{i_\ell}}$. Then
    $\beta_\arch(\bclose(i_1, \ldots, i_\ell))=\top$, thus $\beta
    \models \forall P ~.~ \pi(i_1, \ldots, i_\ell) \rightarrow
    \theta(b)$, leading to $I \vdash b$.  Moreover, for any
    interaction $J = \set{p_{j_1}, \ldots, p_{j_m}} \subsetneq I$,
    we have $\beta \models \forall P ~.~ \pi(j_1, \ldots, j_m)
    \rightarrow \neg\theta(b)$, from which we conclude that $J
    \not\vdash b$ and, consequently $I \vdash^\mu b$. Then, we have
    $\arch \models p_i \closeinter b$.
    \vspace*{\baselineskip}
  \item $p_i \exinter b$: ``$\Rightarrow$'' $\beta_\arch \models
    \btr{p_i \exinter b}{\B}$ is an easy check. ``$\Leftarrow$''
    Similar to the above point, we obtain $D = \set{p_i}$. Let $I =
    \set{p_{i_1}, \ldots, p_{i_\ell}}$ be the interaction for which
    $\beta_\arch \models \bopen(i_1, \ldots, i_\ell) \vee
    \bclose(i_1, \ldots, i_\ell)$ holds and conclude, since $I
    \vdash b$ follows from $\beta_\arch \models \exists P ~.~
    \pi(i_1, \ldots, i_\ell) \wedge \theta(b)$.
    \vspace*{\baselineskip}
  \item $p_i \closexinter b$: ``$\Rightarrow$'' Similar to the above
    point. ``$\Leftarrow$'' Similar to the above point, we obtain $D
    = \set{p_i}$ and an interaction $I = \set{p_{i_1}, \ldots,
      p_{i_\ell}}$ such that $I \vdash b$. Moreover, for any
    interaction $J = \set{p_{j_1}, \ldots, p_{j_m}} \subsetneq I$ we
    have $\beta_\arch \models \pi(j_1, \ldots, j_m) \rightarrow
    \neg\theta(b)$, leading to $J \not\vdash b$, thus we obtain $I
    \vdash^\mu b$.
    \vspace*{\baselineskip}
  \item $\existmod{\phi_1}$: ``$\Rightarrow$'' $\arch \models
    \existmod{\phi_1}$ only if $\arch_1 \models \phi_1$, for some
    architecture $\arch_1$ such that $\arch \closure \arch_1$. By
    Lemma \ref{lemma:close}, we obtain $\beta_\arch \cup
    \beta_{\arch_1} \models \closure(\B,\B_1)$, thus $\beta_\arch
    \models \btr{\existmod{\phi_1}}{\B}$. ``$\Leftarrow$'' If
    $\beta_\arch \models \btr{\existmod{\phi_1}}{\B}$ then there
    exists a valuation $\beta' : \B' \rightarrow \set{\bot,\top}$
    such that $\beta' \models \arch(\B')$, $\beta_\arch \cup \beta'
    \models \closure(\B,\B')$ and $\beta' \models
    \btr{\phi_1}{\B'}$. Since $\beta' \models \arch(\B')$ there
    exists an architecture $\arch' \in \boolarch{\beta'}$ and, by
    Lemma \ref{lemma:close}, we obtain $\arch \closure
    \arch'$. Moreover, by the induction hypothesis, we have $\arch'
    \models \phi_1$, thus $\arch \models \existmod{\phi_1}$.
    \vspace*{\baselineskip}
  \item $\phi_1 \wedge \phi_2$: ``$\Rightarrow$'' If $\arch \models
    \phi_1 \wedge \phi_2$, by the induction hypothesis, we have
    $\beta_{\arch} \models \btr{\phi_i}{\B}$, for $i=1,2$, thus
    $\beta_\arch \models \btr{\phi_1 \wedge
      \phi_2}{\B}$. ``$\Leftarrow$'' If $\beta_\arch \models
    \btr{\phi_1 \wedge \phi_2}{\B}$, by the induction hypothesis, we
    obtain $\arch \models \phi_i$, for $i=1,2$, hence $\arch \models
    \phi_1 \wedge \phi_2$.
    \vspace*{\baselineskip}
  \item $\phi_1 \vee \phi_2$: similar to the above point, by direct
    application of the induction hypothesis.
    \vspace*{\baselineskip}
  \item $\phi_1 * \phi_2$: ``$\Rightarrow$'' If $\arch \models \phi_1
    * \phi_2$ then there exist $\arch_i = \tuple{D_i, \I_i}$ such that
    $\arch_1 \uplus \arch_2 = \arch$ and $\arch_i \models \phi_i$, for
    $i=1,2$. By the induction hypothesis, $\beta_{\arch_i} \models
    \btr{\phi_i}{\B_i}$, for $i=1,2$. Since $D_1 \cap D_2 =
    \emptyset$, we have $\beta_{\arch_1} \cup \beta_{\arch_2} \models
    \#(\B_1, \B_2)$ and, by Lemma \ref{lemma:composition}
    (\ref{it1:composition}) there exists a boolean valuation $\beta :
    \B \rightarrow \set{\bot,\top}$ such that $\beta \cup \beta_1 \cup
    \beta_2 \models \biguplus(\B,\B_1,\B_2)$ and $\arch = \arch _1
    \uplus \arch_2 \in \boolarch{\beta}$. Moreover, $\beta \models
    \btr{\phi_1 * \phi_2}{\B}$. Since $\arch \in
    \boolarch{\beta_\arch}$, by Fact \ref{fact:eq-bval}, we conclude.
    ``$\Leftarrow$'' If $\beta_\arch \models \btr{\phi_1 *
      \phi_2}{\B}$, there exists valuations $\beta_i : \B_i
    \rightarrow \set{\bot,\top}$ such that $\beta_1 \cup \beta_2
    \models \arch(\B_1) \wedge \arch(\B_2) \wedge \#(\B_1,\B_2) \wedge
    \biguplus(\B,\B_1,\B_2)$ and $\beta_i \models \btr{\phi_1}{\B_i}$,
    for $i=1,2$. By Lemma \ref{lemma:composition}
    (\ref{it2:composition}) there exist architectures $\arch_i \in
    \boolarch{\beta_i}$, such that $\arch = \arch_1 \uplus \arch_2$.
    Since $\arch_i \in \boolarch{\beta_{\arch_i}}$, by Fact
    \ref{fact:eq-bval}, we obtain $\beta_i = \beta_{\arch_i}$, for
    $i=1,2$. By the inductive hypothesis, we obtain $\arch_i \models
    \phi_i$, for $i=1,2$. Since, by Lemma \ref{lemma:disjoint},
    $\arch_1$ and $\arch_2$ are disjoint, we obtain $\arch \models
    \phi_1 * \phi_2$.
    \vspace*{\baselineskip}
  \item $\phi_1 \wand \phi_2$: ``$\Rightarrow$'' Let $\beta_1 : \B_1
    \rightarrow \set{\bot,\top}$ be any valuation such that
    $\beta_\arch \cup \beta_1 \models \arch(\B_1) \wedge \#(\B,\B_1)
    \wedge \btr{\phi_1}{\B_1}$. By the induction hypothesis, there
    exists an architecture $\arch_1 \in \boolarch{\beta_1}$ and,
    moreover, since $\beta_\arch \cup \beta_1 \models \#(\B,\B_1)$, by
    Lemma \ref{lemma:disjoint}, $\arch$ and $\arch_1$ are
    disjoint. Since $\arch \models \phi_1 \wand \phi_2$, we have
    $\arch \uplus \arch_1 \models \phi_2$. By the inductive
    hypothesis, we have $\beta_{\arch \uplus \arch_1} \models
    \btr{\phi_2}{\B''}$. Moreover, by Lemma \ref{lemma:composition}
    (\ref{it1:composition}), there exists a valuation $\beta'' : \B''
    \rightarrow \set{\bot,\top}$ such that $\beta'' \cup \beta_\arch
    \cup \beta_1 \models \biguplus(\B'',\B,\B_1)$ and, by Fact
    \ref{fact:eq-bval}, $\beta''$ and $\beta_{\arch \uplus \arch_1}$
    are the same. Since the choice of $\beta_1$ was arbitrary, we
    obtain $\beta_\arch \models \btr{\phi_1 \wand
      \phi_2}{\B}$. ``$\Leftarrow$'' Let $\arch_1$ be any architecture
    disjoint from $\arch$, such that $\arch_1 \models \phi_1$. By the
    induction hypothesis, $\beta_{\arch_1} \models
    \btr{\phi_1}{\B_1}$. Moreover, by Lemma \ref{lemma:disjoint}, we
    have $\beta_\arch \cup \beta_{\arch_1} \models \#(\B,\B_1)$, thus, since $\beta_\arch
    \models \btr{\phi_1 \wand \phi_2}{\B}$, there exists a valuation
    $\beta'' : \B'' \rightarrow \set{\bot,\top}$ such that $\beta_\arch \cup
    \beta_{\arch_1} \cup \beta'' \models \arch(\B'') \wedge
    \biguplus(\B'',\B,\B_1) \wedge \btr{\phi_2}{\B''}$. Then, by Lemma
    \ref{lemma:composition} (\ref{it2:composition}), there exists an
    architecture $\arch'' \in \boolarch{\beta''}$ such that $\arch'' =
    \arch \uplus \arch_1$. By the induction hypothesis, because
    $\beta'' \models \btr{\phi_2}{\B''}$, we obtain $\arch \uplus
    \arch_1 \models \phi_2$ and, since the choice of $\arch_1$ was
    arbitrary, we obtain $\arch \models \phi_1 \wand \phi_2$. \qed
  \end{compactitem}}

We remind that, since there are at most $2^k$ sequences $1 \leq i_1 <
\ldots < i_\ell \leq k$, the size of each of the formulae $\arch(\B)$,
$\#(\B,\B')$, $\biguplus(\B,\B',\B'')$ and $\closure(\B,\B')$ is
$2^{\bigO(k)}$. It is easy to check that, given any \psil\ formula
$\phi$ such that $\portsof{\phi} \subseteq \set{p_1, \ldots, p_k}$,
its translation to \qbf\ takes $\len{\phi} \cdot 2^{\bigO(k)}$ time.

In the following, we provide a tight complexity result by bounding the
number of ports that occur in a boolean term $b$ from an atomic
proposition $p \inter b$, $p \exinter b$, $p \closeinter b$ or $p
\closexinter b$, by a constant $n\geq1$, independent of the input. We
shall denote by \psilk{n}\ the fragment of \psil\ formulae that meets
this condition.

\begin{corollary}\label{cor:psil-complexity}
  The satisfiability and entailment problems for \psil\ are
  in \expspace. If $n\geq1$ is a constant not part of the input, the
  satisfiability and entailment problems for \psilk{n}
  are \pspace-complete.
\end{corollary}
\proof{ The \expspace\ upper bound for satisfiability is immediate,
  since the \qbf\ translation of any \psil\ formula $\phi$, such that
  $\portsof{\phi} \subseteq \set{p_1, \ldots, p_k}$ takes time
  $\len{\phi} \cdot 2^{\bigO(k)}$. For the entailment problem, let
  $\phi$ and $\psi$ be two \psil\ formulae, such that $\portsof{\phi}
  \cup \portsof{\psi} \subseteq \set{p_1,\ldots,p_k}$ and assume that
  there exists an architecture $\arch$ such that $\arch \models \phi$
  and $\arch \not\models \psi$. By Theorem \ref{thm:psil-qbf}, there
  exists a boolean valuation $\beta : \B \rightarrow \set{\bot,\top}$,
  such that $\arch \in \boolarch{\beta}$ and $\beta \models
  \btr{\phi}{\B}$. Moreover, since $\arch \not\models \psi$, for every
  boolean valuation $\beta' : \B \rightarrow \set{\bot,\top}$, such
  that $\arch \in \boolarch{\beta'}$, we have $\beta' \models
  \neg\btr{\psi}{\B}$. By Fact \ref{fact:eq-bval}, since $\arch \in
  \boolarch{\beta} \cap \boolarch{\beta'}$, for any such valuation
  $\beta'$, we have that $\beta$ and $\beta'$ are the same, thus
  $\beta \models \btr{\phi}{\B} \wedge
  \neg\btr{\psi}{\B}$. Since \expspace\ is closed under complement, by
  Savitch's Theorem, we obtain the \expspace\ upper bound for
  entailment.

  \vspace*{\baselineskip}\noindent For the second point, the upper
  bound is established noticing that the number of sequences $1 \leq
  i_1 < \ldots < i_\ell \leq k$, for $\ell \leq n$ is bounded by
  $k\choose{n}$, thus the translation of a \psilk{n} formula in
  \qbf\ takes polynomial time. For the \pspace-hard lower bound, we
  reduce from the validity of \qbf\ sentences $\forall x_1 \exists y_1
  \ldots \forall x_k \exists y_k ~.~ F$, where $F$ is a propositional
  formula with free variables $x_1, y_1, \ldots, x_k, y_k$, written in
  positive normal form (note that this is w.l.o.g.). To this end, we
  consider, for each variable $x \in \set{x_1, y_1, \ldots, x_k, y_k}$
  two ports $x_t$ and $x_f$. Let $\mathsf{false}$ be a shorthand for
  $\emp \wedge p \inter p$, where $p$ is a port which is not a member
  of $\set{x_t,x_f \mid x \in \set{x_1,y_1,\ldots,x_k,y_k}}$.
  Intuitively, $\has{x_t} \isdef x_t \inter x_t \wand \mathsf{false}$
  (resp. $\has{x_f} \isdef x_f \inter x_f \wand \mathsf{false}$)
  encodes the fact that $x$ is true (resp. false). Given a set $S
  \subseteq \set{x_1, y_1, \ldots, x_k, y_k}$, we write $A_S \isdef
  \Asterisk_{x \in S} x_t \inter x_t \vee x_f \inter x_f$. Considering
  the total order $x_1 \prec y_1 \prec \ldots \prec x_k \prec y_k$, we
  write $A_{\set{\preceq x}} \isdef A_{\set{x' \mid x' \preceq x}}$. The
  reduction from \qbf\ to \psilk{n} is implemented by the following
  recursive function:
  \[\begin{array}{rcl}
  \tau(x) & \isdef & \has{x_t} \\
  \tau(\neg x) & \isdef & \has{x_f} \\
  \tau(F_1 \wedge F_2) & \isdef & \tau(F_1) \wedge \tau(F_2) \\
  \tau(F_1 \vee F_2) & \isdef & \tau(F_1) \vee \tau(F_2) \\
  \tau(\forall x_i ~.~ G) & \isdef & A_{\set{x_i}} \wand \tau(G) \\
  \tau(\exists y_i ~.~ G) & \isdef & [A_{\set{\preceq x_i}} \wedge
    ((A_{\set{\preceq y_i}} \wedge \tau(G)) \wand \mathsf{false})] \wand \mathsf{false}
  \end{array}\]
  We show that any \qbf\ sentence $\forall x_1 \exists y_1 \ldots
  \forall x_k \exists y_k ~.~ F$ is valid if and only if $\emp \wedge
  \tau(\forall x_1 \exists y_1 \ldots \forall x_k \exists y_k ~.~ F)$
  is satisfiable, or equivalently, the entailment between $\emp$ and
  $\tau(\forall x_1 \exists y_1 \ldots \forall x_k \exists y_k ~.~ F)$
  holds. The encoding of the universal quantifier is directly via
  $A_{\set{x}} \wand \tau(G)$ that asserts the validity of $\tau(G)$
  under any extension of the model (architecture) with domain either
  $\set{x_t}$ or $\set{x_f}$. The existential quantifier is encoded
  using a double negation. If $\arch \models (P \wedge Q) \wand
  \mathsf{false}$, then for any extension $\arch' \models P$ we have
  $\arch' \not\models Q$. Now assume that $\arch
  \models[A_{\set{\preceq x_i}} \wedge ((A_{\set{\preceq y_i}} \wedge
    \tau(G)) \wand \mathsf{false})] \wand \mathsf{false}$. Then, for
  any extension $\arch'$ of $\arch$, such that $\arch' \models
  A_{\set{\preceq{x_i}}}$, we have $\arch' \not\models
  (A_{\set{\preceq y_i}} \wedge \tau(G)) \wand \mathsf{false}$. This
  means that there exists an extension $\arch''$ such that $\arch
  \uplus \arch' \uplus \arch'' \models \tau(G)$, which captures the
  fact that some valuation of $y_i$ makes the sentence valid. A
  similar encoding is used in \cite[Proposition
    6]{CalcagnoYangOHearn01}. \qed}

\subsection{Decidability of \ssil}
\label{sec:ssil}

We recall that \ssil\ is the fragment of \sil\ in which negation is
allowed, but not the magic wand. The proof of decidability for
\ssil\ follows a very similar pattern to the decidability proof for
\psil (\S\ref{sec:psil}). Just as before, we first define an
equivalence relation on architectures, then we characterize the
equivalence classes of this relation by test formulae. As a
consequence, each formula of \ssil\ is equivalent to a boolean
combination of test formulae from a finite set and, moreover, based on
this fact, we obtain a small model property that implies the
decidability of \ssil. 

The main difficulty here is that \ssil\ has negation, which allows to
describe architectures with invisible ports in the domain. For
instance, the formula $\neg\emp * \neg\emp$ states the existence of at
least two ports, none of them corresponding to a port symbol. When
composing such architectures, these invisible ports can determine
which interactions are kept and which are lost, based on their
\emph{visible interaction type}, which is formally defined next.

\begin{example}\label{ex:visible-interaction}
  Consider the architectures $\arch_1 = \tuple{\set{p,\alpha},
    \set{\set{p,\alpha,\beta}}}$ and $\arch_2 = \tuple{\set{\beta},
    \set{\set{p,\beta}}}$, where the set of visible ports is $P =
  \set{p}$. Because $\set{p,\alpha,\beta} \in \intf{\arch_1}$ has a
  non-empty intersection with $\dom{\arch_2} = \set{\beta}$, we obtain
  $\arch_1 \uplus \arch_2 = \tuple{\set{p,\alpha,\beta},
    \emptyset}$. \hfill$\blacksquare$
\end{example}

Let $P \subseteq \ports$ be a set of visible ports and $\arch =
\tuple{D,\I}$ be an architecture. For an invisible port $x \in D
\setminus P$, we define its visible interaction type as the set of
interactions involving $x$, restricted to their visible ports:
\(\vtype{x}{P}{\arch} \isdef \interproj{\I}{\set{x}} \sqcap P\).  The
function $\vmap{\arch}{P} : 2^{2^P} \rightarrow 2^\ports$ gives the
set of invisible ports with a given visible interaction type from the
domain of $\arch$:
\[\vmap{\arch}{P}(\mathcal{S}) = \set{x \in D \setminus P
  \mid \vtype{x}{P}{\arch} = \mathcal{S}} \text{, for any $\mathcal{S}
  \in 2^{2^P}$}\] Consider further the function $b_P : \nat \times
2^{2^P} \rightarrow \nat$, defined by the recurrence relation:

\[b_P(1,\mathcal{S}) \isdef 1 \text{ and } \forall n > 1 ~.~ b_P(n,\mathcal{S}) \isdef
2 \cdot \sum_{\mathcal{S} \subseteq \mathcal{S}'} b_P(n-1,\mathcal{S}')\]

\begin{definition}\label{def:veq}
  Given architectures $\arch_i = \tuple{D_i, \I_i}$, for $i=1,2$, a
  finite set of ports $P \subseteq \ports$ and an integer $n \geq 0$,
  we have $\arch_1 \veq{P}{n} \arch_2$ if and only if the following
  hold:
  \begin{compactenum}
  \item\label{it1:veq} $D_1 \cap P = D_2 \cap P$,
  \item\label{it2:veq} $\interproj{\I_1}{(D_1 \cap P)} \sqcap P =
    \interproj{\I_2}{(D_2 \cap P)} \sqcap P$,
  \item\label{it3:veq} for all $\mathcal{S} \in 2^{2^P}$, we have:
    \begin{compactenum}
    \item\label{it31:veq} $\card{\vmap{\arch_1}{P}(\mathcal{S})} <
      b_P(n,\mathcal{S}) \Rightarrow \card{\vmap{\arch_2}{P}(\mathcal{S})} =
      \card{\vmap{\arch_1}{P}(\mathcal{S})}$,
    \item\label{it32:veq} $\card{\vmap{\arch_1}{P}(\mathcal{S})} \geq
      b_P(n,\mathcal{S}) \Rightarrow \card{\vmap{\arch_2}{P}(\mathcal{S})} \geq
      b_P(n,\mathcal{S})$.
    \end{compactenum}
  \end{compactenum}
\end{definition}
It is easy to prove that $\veq{P}{n}$ is an equivalence relation, for
any $P \subseteq \ports$ and any $n \geq 1$. Moreover, given any set
of ports $P' \subseteq P$ and any integer $n' \leq n$, we have
$\arch_1 \veq{P}{n} \arch_2 \Rightarrow \arch_1 \veq{P'}{n'} \arch_2$.
The following lemma proves that $\veq{P}{n}$ is compatible with the
composition of architectures: 

\begin{lemma}\label{lemma:ssil-star}
  Let $\arch = \tuple{D,\I}$ and $\arch' = \tuple{D',\I'}$ be
  architectures and $P \subseteq \ports$ be a set of ports, such that
  $\arch \veq{P}{n} \arch'$, for some $n \geq 2$. Then for any
  architectures $\arch_i = \tuple{D_i,\I_i}$, such that $\arch =
  \arch_1 \uplus \arch_2$ there exist architectures $\arch'_i =
  \tuple{D'_i, \I'_i}$, such that $\arch' = \arch'_1 \uplus \arch'_2$
  and $\arch_i \veq{P}{n-1} \arch'_i$, for $i=1,2$.
\end{lemma}
\proof{
    We define two mappings $\mu_i : 2^{2^P} \rightarrow 2^\ports$
  describing how the ports from $D' \setminus P$ occur in the
  interactions of $\arch'_i$, for $i=1,2$, respectively. The idea is
  to define the architectures $\arch'_i$ such that $\mu_i =
  \vmap{\arch'_i}{P}$, for $i=1,2$. Let $\mathcal{S} \in 2^{2^P}$ be
  an arbitrary set of interactions involving only visible ports. We
  distinguish the cases below: \begin{compactenum}
  \item If $\card{\vmap{\arch}{P}(\mathcal{S})} =
    \card{\vmap{\arch'}{P}(\mathcal{S})}$ then there exists a
    bijection $\pi_{\mathcal{S}} : \vmap{\arch'}{P}(\mathcal{S})
    \rightarrow \vmap{\arch}{P}(\mathcal{S})$. In this case, for each
    $x \in \vmap{\arch}{P}(\mathcal{S})$ and each $\mathcal{S}'
    \supseteq \mathcal{S}$, we require that:
    \begin{equation}\label{eq:same-card}
      x \in \mu_i(\mathcal{S}') \iff \pi_{\mathcal{S}}(x) \in
      \vmap{\arch_i}{P}(\mathcal{S}') \text{, for all $i=1,2$}
    \end{equation} 
  \item Else $\card{\vmap{\arch}{P}(\mathcal{S})} \neq
    \card{\vmap{\arch'}{P}(\mathcal{S})}$, thus necessarily
    $\card{\vmap{\arch}{P}(\mathcal{S})} \geq b_P(n,\mathcal{S})$ and
    $\card{\vmap{\arch'}{P}(\mathcal{S})} \geq b_P(n,\mathcal{S})$,
    because we assumed that $\arch \veq{P}{n} \arch'$. Because $\arch
    = \arch_1 \uplus \arch_2$, we have that $(D_1,D_2)$ is a partition
    of $D$ and define $E_i \isdef \vmap{\arch}{P}(\mathcal{S}) \cap
    D_i$, a partition of $\vmap{\arch}{P}(\mathcal{S})$, for $i=1,2$.
        We distinguish the cases below: \begin{compactenum}
    \item if $\card{E_1} < \frac{b_P(n,\mathcal{S})}{2}$ and $\card{E_2}
      \geq \frac{b_P(n,\mathcal{S})}{2}$ let $(E'_1,E'_2)$ be a
      partition of $\vmap{\arch'}{P}(\mathcal{S})$ such that
      $\card{E'_1} = \card{E_1}$ and $\card{E'_2} \geq
      \frac{b_P(n,\mathcal{S})}{2}$. Such a partition exists because
      $\card{\vmap{\arch'}{P}(\mathcal{S})} \geq
      b_P(n,\mathcal{S})$. Since $\card{E'_1} = \card{E_1}$, there
      exists a bijection $\rho_{\mathcal{S}} : E'_1 \rightarrow
      E_1$. Then for each $x \in E_1$ and each $\mathcal{S}'
      \supseteq \mathcal{S}$, we require:
      \begin{equation}\label{eq:same-card-one}
        x \in \mu_1(\mathcal{S}') \iff \rho_{\mathcal{S}}(x) \in \vmap{\arch_1}{P}(\mathcal{S}')
      \end{equation}
      Further, we split $E'_2$ between the sets
      $\set{\mu_2(\mathcal{S}') \mid \mathcal{S} \subseteq
        \mathcal{S}'}$ such that, for each $\mathcal{S}' \supseteq
      \mathcal{S}$, the following hold:
      \begin{equation}\label{eq:diff-card-one}
        \begin{array}{rcl}
          \card{\vmap{\arch_2}{P}(\mathcal{S}')} < b_P(n-1,\mathcal{S}') & \Rightarrow &
          \card{\mu_2(\mathcal{S}') \cap E'_2} = \card{\vmap{\arch_2}{P}(\mathcal{S}') \cap E_2} \\
          \card{\vmap{\arch_2}{P}(\mathcal{S}')} \geq b_P(n-1,\mathcal{S}') & \Rightarrow &
          \card{\mu_2(\mathcal{S}') \cap E'_2} \geq b_P(n-1,\mathcal{S}')
        \end{array}        
      \end{equation}
      \noindent
      Note that, since $\card{E'_2} \geq \sum_{\mathcal{S} \subseteq
        \mathcal{S}'} b_P(n-1,\mathcal{S}')$, such a partitioning of
      $E'_2$ is always possible.
    \item else, if $\card{E_1} \geq \frac{b_P(n,\mathcal{S})}{2}$ and
      $\card{E_2} < \frac{b_P(n,\mathcal{S})}{2}$, we partition
      $\vmap{\arch'}{P}(\mathcal{S})$ symmetrically.
    \item otherwise, if $\card{E_1} \geq \frac{b_P(n,\mathcal{S})}{2}$
      and $\card{E_2} \geq \frac{b_P(n,\mathcal{S})}{2}$, then let
      $(E'_1, E'_2)$ be a partition of $\vmap{\arch'}{P}(\mathcal{S})$
      such that $\card{E'_1} \geq \frac{b_P(n,\mathcal{S})}{2}$ and
      $\card{E'_2} \geq \frac{b_P(n,\mathcal{S})}{2}$. Such a
      partitioning exists because
      $\card{\vmap{\arch'}{P}(\mathcal{S})} \geq
      b_P(n,\mathcal{S})$. Then we split $E'_i$ between the sets
      $\set{\mu_i(\mathcal{S}') \mid \mathcal{S} \subseteq
        \mathcal{S}'}$ such that, for each $\mathcal{S}' \subseteq
      \mathcal{S}$, the following hold, for $i=1,2$:
      \begin{equation}\label{eq:diff-card-two}
        \begin{array}{rcl}
          \card{\vmap{\arch_i}{P}(\mathcal{S}')} < b_P(n-1,\mathcal{S}') & \Rightarrow &
          \card{\mu_i(\mathcal{S}') \cap E'_i} = \card{\vmap{\arch_i}{P}(\mathcal{S}') \cap E_i} \\
          \card{\vmap{\arch_i}{P}(\mathcal{S}')} \geq b_P(n-1,\mathcal{S}') & \Rightarrow &
          \card{\mu_i(\mathcal{S}') \cap E'_i} \geq b_P(n-1,\mathcal{S}') 
        \end{array}
      \end{equation}
      \noindent
      Note that, since $\card{E'_i} \geq \sum_{\mathcal{S} \subseteq
        \mathcal{S}'} b_P(n-1,\mathcal{S}')$, $i=1,2$, such a
      partitioning is always possible.
    \end{compactenum}
  \end{compactenum}
  Moreover, nothing else is in $\mu_i(\mathcal{S}')$, for any
  $\mathcal{S}' \in 2^{2^P}$, for $i=1,2$. We define now the domains
  of $\arch'_1$ and $\arch'_2$ as follows:
  \begin{equation}\label{eq:dom}
    D'_i \isdef (D_i \cap P) \cup \bigcup_{\mathcal{S} \in 2^{2^P}}
    \mu_i(\mathcal{S}) \text{, for $i=1,2$}
  \end{equation}
  Because the sets $\set{\vmap{\arch_i}{P}(\mathcal{S}) \mid
    \mathcal{S} \in 2^{2^P}}$ form a partition of $D_i \setminus P$,
  by the definition of $\mu_i$, the sets $\set{\mu_i(\mathcal{S}) \mid
    \mathcal{S} \in 2^{2^P}}$ form a partition of $D'_i \setminus P$,
  for $i=1,2$, respectively. Then we can define mappings $\lambda_i :
  D'_i \setminus P \rightarrow 2^{2^P}$ as $\lambda_i(x) \isdef
  \mathcal{S} \iff x \in \mu_i(\mathcal{S})$, for all $x \in D'_i
  \setminus P$, for $i=1,2$. Similarly, since
  $\set{\vmap{\arch'}{P}(\mathcal{S}) \mid \mathcal{S} \in 2^{2^P}}$
  is a partition of $D' \setminus P$, we can define the mapping
  $\lambda : D' \setminus P \rightarrow 2^{2^P}$ as $\lambda(x) \isdef
  \mathcal{S} \iff x \in \vmap{\arch'}{P}(\mathcal{S})$. Next, we
  define the interaction sets of $\arch'_i$ as:
  \[\begin{array}{rcl}
  \I'_i & \isdef & \interproj{\I'}{D'_i} \cup X_i \cup Y_i \\
  \text{where:} \\
  X_i & \isdef & \set{I \cup \set{x,\alpha_i} \mid x \in D'_i \setminus P,~ I \in \lambda_i(x) \setminus \lambda(x),~ I \cap D'_{3-i} \neq \emptyset} \cup \\
  && \set{I \cup \set{x,\alpha_i} \cup (D'_{3-i} \setminus P) \mid x \in D'_i \setminus P,~ I \in \lambda_i(x) \setminus \lambda(x),~ I \cap D'_{3-i} = \emptyset} \\
  \\
  Y_i & \isdef & \set{I \cup \set{\beta_i} \mid I \in (\interproj{\I_i}{D_i \cap P} \sqcap P) \setminus \interproj{\I'}{D'_i},~ I \cap D'_{3-i} \neq \emptyset} \cup \\
  && \set{I \cup (D'_{3-i} \setminus P) \cup \set{\beta_i} \mid I \in (\interproj{\I_i}{D_i \cap P} \sqcap P) \setminus \interproj{\I'}{D'_i},~ I \cap D'_{3-i} = \emptyset}
  \end{array}\]
  and $\alpha_i, \beta_i \in \ports \setminus (D' \cup \bigcup \I')$
  are pairwise distinct ports that do not occur in $\arch'$,
  respectively, for $i=1,2$. Next, we prove the following facts:

  \begin{fact}\label{fact:vmap}
    For any $\mathcal{S} \in 2^{2^P}$, we have
    $\vmap{\arch'_i}{P}(\mathcal{S}) = \mu_i(\mathcal{S})$, for
    each $i=1,2$.
  \end{fact}
  \proof{ We prove the case $i=1$, the case $i=2$ being identical. Let
    $x \in D'_1 \setminus P$ be an arbitrary port and $\mathcal{S} \in
    2^{2^P}$ be a set of visible interactions. We have:
    \[\begin{array}{rcl}
    x \in \vmap{\arch'_1}{P}(\mathcal{S}) & \iff & \gtype{x}{P}{\I'_1} = \mathcal{S} \\
    & \iff & \set{I \cap P \mid I \in \I'_1,~ x \in I} = \mathcal{S} \\
    & \iff & \set{I \cap P \mid I \in \interproj{\I'}{D'_1},~ x \in I} \cup \set{I \cap P \mid I \in X_1,~ x \in I} \cup \set{I \cap P \mid I \in Y_1,~ x \in I} = \mathcal{S} \\
    & \iff & \set{I \cap P \mid I \in \I',~ x \in I} \cup \set{I \cap P \mid I \in X_1,~ x \in I} = \mathcal{S} \text{, since $x \in D'_1 \setminus P$} \\
    & \iff & \lambda(x) \cup (\lambda_1(x)\setminus\lambda(x)) = \mathcal{S} \\
    & \iff & \lambda(x) \cup \lambda_1(x) = \mathcal{S}
    \end{array}\]
    It is sufficient to prove $\lambda(x) \subseteq \lambda_1(x)$ in
    order to obtain $x \in \vmap{\arch'_1}{P}(\mathcal{S}) \iff
    \lambda_1(x)=\mathcal{S} \iff x \in \mu_1(\mathcal{S})$, as
    required. Since $x \in D'_1$, by the definition of $\mu_1$, it
    must be the case that $x \in \vmap{\arch'}{P}(\mathcal{S}')$, for
    some $\mathcal{S}' \subseteq \mathcal{S}$. Then $\lambda(x) =
    \mathcal{S}' \subseteq \mathcal{S} = \lambda_1(x)$ follows. \qed}

  \noindent
  Next, we prove that $\arch_i \veq{P}{n-1} \arch'_i$, for
  $i=1,2$. Again, we consider only the case $i=1$, the other case
  being identical. We consider the three points of Definition
  \ref{def:veq} below:
  
  \vspace*{\baselineskip}\noindent 
  (\ref{it1:veq}) $D'_1 \cap P = (D_1 \cap P) \cup
  \bigcup_{\mathcal{S} \in 2^{2^P}} (\mu_1(\mathcal{S}) \cap P) = D_1
  \cap P$, because $\mu_1(\mathcal{S}) \subseteq D'_1 \setminus P$,
  and thus $\mu_1(\mathcal{S}) \cap P = \emptyset$, for any
  $\mathcal{S} \in 2^{2^P}$.

  \vspace*{\baselineskip}\noindent (\ref{it2:veq}) We need to show
  that $\interproj{\I_1}{(D_1 \cap P)} \sqcap P = \interproj{\I'_1}{(D'_1
    \cap P)} \sqcap P$. Note that $\I_1 = \interproj{\I}{D_1} \cup A_1
  \cup B_1$, where:
  \[\begin{array}{rcl}
  A_1 & = & \set{I \mid I \in \I_1 \setminus \I,~ I \cap (D_1 \setminus P) \neq \emptyset} \\
  B_1 & = & \set{I \mid I \in \I_1 \setminus \I,~ I \cap D_1 \subseteq P} 
  \end{array}\]
  This is because $\I = (\I_1 \cap \I_2) \cup (\I_1 \cap
  2^{\overline{D}_2}) \cup (\I_2 \cap 2^{\overline{D}_1})$, hence
  $\interproj{\I}{D_1} = \I \cap \I_1$ and $\I_1 = (\I \cap \I_1) \cup
  (\I_1 \setminus \I) = \interproj{\I}{D_1} \cup (A_1 \cup B_1)$.

  \begin{fact}\label{fact:decomp1}
    $\interproj{\I}{(D_1 \cap P)} \sqcap P = \interproj{\I'}{(D_1 \cap P)}
    \sqcap P$
  \end{fact}
  \proof{ ``$\subseteq$'' Let $I \in \intersect{D_1 \cap P}{\I}$ be an
    interaction. Then $I \cap D_1 \cap P \neq \emptyset$ and
    consequently $I \cap D \cap P \neq \emptyset$. But then $I \cap P
    \in (\intersect{D \cap P}{\I}) \sqcap P = (\intersect{D' \cap
      P}{\I'}) \sqcap P$, because $\arch \veq{P}{n} \arch'$, by
    Definition \ref{def:veq} (\ref{it2:veq}). Then there exists $I'
    \in \intersect{D' \cap P}{\I'}$ such that $I \cap P = I' \cap P$.
    Hence $I' \cap P \cap D_1 \neq \emptyset$ and $I' \in
    \intersect{D_1 \cap P}{\I'}$, which implies $I \cap P = I' \cap P
    \in \intersect{D_1 \cap P}{\I'} \sqcap P$. The other direction
    is symmetric. \qed}

  \begin{fact}\label{fact:decomp2}
    $\intersect{D_1 \cap P}{(\I \cup A_1)} \sqcap P =
    \intersect{D_1 \cap P}{(\I' \cup X_1)} \sqcap P$
  \end{fact}
  \proof{ ``$\subseteq$'' Let $I \in \intersect{D_1 \cap P}{(\I \cup
      A_1)}$ be an interaction. If $I \in \intersect{D_1 \cap P}{\I}$
    then $I \cap P \in \intersect{D_1 \cap P}{\I'} \sqcap P$, by
    Fact \ref{fact:decomp1}. Assume that $I \in \intersect{D_1 \cap
      P}{A_1}$, then $I \in \I_1 \setminus \I$, $I \cap (D_1 \setminus
    P) \neq \emptyset$ and $I \cap D_1 \cap P \neq \emptyset$. Since
    $I \cap (D_1 \setminus P) \neq \emptyset$, there exists $x \in I
    \cap (D_1 \setminus P)$ and let $\mathcal{S} \isdef
    \vtype{x}{P}{\arch_1}$. Then $x \in \vmap{\arch_1}{P}(\mathcal{S})$
    and, by the definition of $\mu_1$, there exists $x' \in D'_1
    \setminus P$ such that $x' \in \mu_1(\mathcal{S})$ and,
    consequently $\lambda_1(x') = \mathcal{S}$. Since $x \in I \cap
    (D_1 \setminus P)$ and $\mathcal{S} = \vtype{x}{P}{\arch_1}$ we have
    $I \cap P \in \mathcal{S}$ and thus $I \cap P \in
    \lambda_1(x')$. We distinguish the following
    cases: \begin{compactitem}
    \item if $I \cap P \not\in \lambda(x')$ then $I \cap P \in
      \intersect{D_1 \cap P}{X_1} \sqcap P$.
    \item else, $I \cap P \in \lambda(x')$ and, because $x' \in D'_1
      \setminus P \subseteq D' \setminus P$, there exists $I' \in \I'$
      such that $I \cap P = I' \cap P$. Moreover, since $(I \cap P)
      \cap (D_1 \cap P) \neq \emptyset$, we obtain $I \cap P = I' \cap
      P \in \intersect{D_1 \cap P}{\I'} \sqcap P$.
    \end{compactitem}
            
    ``$\supseteq$'' Let $I \in \intersect{D_1 \cap P}{(\I' \cup X_1)}$
    be an interaction. If $I \in \intersect{D_1 \cap P}{\I'}$ then $I
    \cap P \in \intersect{D_1 \cap P}{\I}$, by Fact
    \ref{fact:decomp1}. Assume that $I \in \intersect{D_1 \cap
      P}{X_1}$, then there exists $x' \in D'_1 \setminus P$, such that
    $I \cap P \in \lambda_1(x') \setminus \lambda(x')$. By the
    definition of $\lambda_1$, there exists $x \in D_1 \setminus P$,
    such that $I \cap P \in \gtype{x}{P}{\I_1}$. Then there exists an
    interaction $J \in \I_1$ such that $x \in J$ and $J \cap P = I
    \cap P$. Since, moreover, $(I \cap P) \cap (D_1 \cap P) \neq
    \emptyset$, we have $J \cap D_1 \cap P \neq \emptyset$, hence $J
    \in \intersect{D_1 \cap P}{\I_1}$. Since $x \in J$ and $x \in D_1
    \setminus P$, we have $J \cap (D_1 \setminus P) \neq \emptyset$,
    hence $J \not\in B_1$. Then $J \in \intersect{D_1 \cap P}{(\I \cup
      A_1)}$ and $I \cap P = J \cap P \in \intersect{D_1 \cap P}{(\I
      \cup A_1)} \sqcap P$. \qed}

  \noindent
  Back to point (\ref{it2:veq}) of Definition \ref{def:veq}, it
  suffices to show the following points: \begin{compactitem}
  \item $\intersect{D_1 \cap P}{B_1} \subseteq \intersect{D_1 \cap
    P}{\I'_1} \sqcap P$: Let $I \in \intersect{D_1 \cap P}{B_1}$ be an
    interaction. Then $I \cap P = I$ and $I \cap D_1 \cap P \neq
    \emptyset$. If $I \in \intersect{D_1 \cap P}{(\I \cup A_1)}$ then
    $I = I \cap P \in \intersect{D_1 \cap P}{(\I \cup A_1)} \sqcap P =
    \intersect{D_1 \cap P}{(\I' \cup X_1)} \sqcap P$, by Fact
    \ref{fact:decomp2}. Then $I \in \intersect{D_1 \cap P}{\I'_1}
    \sqcap P$. Otherwise, $I \not\in \intersect{D_1 \cap P}{(\I \cup
      A_1)}$ then $I \not\in \intersect{D_1 \cap P}{(\I \cup A_1)}
    \sqcap P = \intersect{D_1 \cap P}{(\I' \cup X_1)} \sqcap P$, by
    Fact \ref{fact:decomp1}, thus $I \not\in \intersect{D_1 \cap
      P}{(\I' \cup X_1)}$. But because $I \in \intersect{D_1 \cap
      P}{B_1}$, we have $I \in \intersect{D_1 \cap P}{\I_1}$, thus $I
    \in Y_1 \sqcap P \subseteq \intersect{D_1 \cap P}{\I'_1} \sqcap
    P$.
  \item $\intersect{D_1 \cap P}{Y_1} \sqcap P \subseteq
    \intersect{D_1 \cap P}{\I_1} \sqcap P$: If $I \in \intersect{D_1
    \cap P}{Y_1}$, then $I \cap P \in (\intersect{D_1 \cap P}{\I_1})
    \sqcap P$, by the definition of $Y_1$.
  \end{compactitem}

  \vspace*{\baselineskip}\noindent
  (\ref{it3:veq}) by Fact \ref{fact:vmap}, it is sufficient to prove
  that, for all $\mathcal{S} \in 2^{2^P}$: \begin{compactenum}
    \vspace*{\baselineskip}
  \item $\card{\vmap{\arch_1}{P}(\mathcal{S})} < b_P(n-1, \mathcal{S})
    \Rightarrow \card{\mu_1(\mathcal{S})} =
    \card{\vmap{\arch_1}{P}(\mathcal{S})}$:

        \vspace*{\baselineskip}\noindent
    Let $\mathcal{S} \in 2^{2^P}$ be an arbitrary set of interactions
    such that $\card{\vmap{\arch_1}{P}(\mathcal{S})} < b_P(n-1,
    \mathcal{S})$ and let $x \in \mu_1(\mathcal{S})$ be a port. We
    shall exhibit a unique port $y \in \vmap{\arch_1}{P}(\mathcal{S})$
    in order to prove that $\card{\mu_1(\mathcal{S})} \leq
    \card{\vmap{\arch_1}{P}(\mathcal{S})}$. By the definition of
    $\mu_1$, there exists a set $\mathcal{S}' \subseteq \mathcal{S}$
    such that $x \in \vmap{\arch'}{P}(\mathcal{S}')$. We distinguish
    the following cases: \begin{compactitem}
    \item if $\card{\vmap{\arch}{P}(\mathcal{S}')} =
      \card{\vmap{\arch'}{P}(\mathcal{S}')}$ then let $y =
      \pi_{\mathcal{S}'}(x) \in \vmap{\arch_1}{P}(\mathcal{S})$, where
      $\pi_{\mathcal{S}'} : \vmap{\arch'}{P}(\mathcal{S}') \rightarrow
      \vmap{\arch}{P}(\mathcal{S}')$ is the bijection from
      (\ref{eq:same-card}).
    \item else, if $\card{\vmap{\arch}{P}(\mathcal{S}')} \neq
      \card{\vmap{\arch'}{P}(\mathcal{S}')}$ and
      $\card{\vmap{\arch}{P}(\mathcal{S}') \cap D_1} <
      \frac{b_P(n,\mathcal{S}')}{2}$ then let $y =
      \rho_{\mathcal{S}'}(x) \in \vmap{\arch_1}{P}(\mathcal{S})$,
      where $\rho_{\mathcal{S}'}$ is the bijection from
      (\ref{eq:same-card-one}).
    \item otherwise, if $\card{\vmap{\arch}{P}(\mathcal{S}')} \neq
      \card{\vmap{\arch'}{P}(\mathcal{S}')}$ and
      $\card{\vmap{\arch}{P}(\mathcal{S}') \cap D_1} \geq
      \frac{b_P(n,\mathcal{S}')}{2}$ then, because
      $\card{\vmap{\arch_1}{P}(\mathcal{S})} < b_P(n-1,\mathcal{S})$, we
      obtain, by (\ref{eq:diff-card-two}) that:
      \[\card{\mu_1(\mathcal{S}) \cap E'_1} =
      \card{\vmap{\arch_1}{P}(\mathcal{S}) \cap
        \vmap{\arch}{P}(\mathcal{S}') \cap D_1}\] where $E'_1
      \subseteq \vmap{\arch'}{P}(\mathcal{S}')$ is such that $x \in
        E'_1$. Then there exists a bijection $\xi : \mu_1(\mathcal{S})
        \cap E'_1 \rightarrow \vmap{\arch_1}{P}(\mathcal{S}) \cap
        \vmap{\arch}{P}(\mathcal{S}') \cap D_1$ and let $y = \xi(x)
        \in \vmap{\arch_1}{P}(\mathcal{S})$.
    \end{compactitem}
    The unique $y \in \vmap{\arch_1}{P}(\mathcal{S})$ is defined as
    the image of $x$ via a bijection that choses among the above
    disjoint cases. Moreover, since these are the only cases that
    explain why $x \in \mu_1(\mathcal{S})$, i.e.\ nothing else is in
    $\mu_1(\mathcal{S})$, we obtain that $\card{\mu_1(\mathcal{S})} =
    \card{\vmap{\arch_1}{P}(\mathcal{S})}$, as required.
    \vspace*{\baselineskip}
  \item $\card{\vmap{\arch_1}{P}(\mathcal{S})} \geq b_P(n-1, \mathcal{S})
    \Rightarrow \card{\mu_1(\mathcal{S})} \geq b_P(n-1, \mathcal{S})$:

   \vspace*{\baselineskip}\noindent Let $\mathcal{S} \in 2^{2^P}$ be
    a set of interactions such that
    $\card{\vmap{\arch_1}{P}(\mathcal{S})} \geq b_P(n, \mathcal{S})$.
    By the definition of $\mu_1$, for each $x \in \mu_1(\mathcal{S})$
    there exists $\mathcal{S}' \subseteq \mathcal{S}$ such that $x \in
    \vmap{\arch'}{P}(\mathcal{S}')$ and let $\mathcal{S}_1, \ldots,
    \mathcal{S}_k \subseteq \mathcal{S}$ be all the sets of
    interactions such that $\vmap{\arch'}{P}(\mathcal{S}_i) \cap
    \mu_1(\mathcal{S}) \neq \emptyset$. Moreover, the sets
    $\vmap{\arch'}{P}(\mathcal{S}_i)$ are pairwise disjoint and
    $\mu_1(\mathcal{S}) = \bigcup_{i=1}^k \mu_1(\mathcal{S}) \cap
    \vmap{\arch'}{P}(\mathcal{S}_i)$, leading to
    $\card{\mu_1(\mathcal{S})} = \sum_{i=1}^k \card{\mu_1(\mathcal{S})
      \cap \vmap{\arch'}{P}(\mathcal{S}_i)}$. For an arbitrary $1 \leq
    i \leq k$, we distinguish the following cases: \begin{compactitem}
    \item if $\card{\vmap{\arch}{P}(\mathcal{S}_i)} =
      \card{\vmap{\arch'}{P}(\mathcal{S}_i)}$, then by
      (\ref{eq:same-card}), we have:
      \[\card{\mu_1(\mathcal{S}) \cap \vmap{\arch'}{P}(\mathcal{S}_i)}
      = \card{\vmap{\arch_1}{P}(\mathcal{S}) \cap
        \vmap{\arch}{P}(\mathcal{S}_i)}\]
    \item else, if $\card{\vmap{\arch}{P}(\mathcal{S}_i)} \neq
      \card{\vmap{\arch'}{P}(\mathcal{S}_i)}$ and
      $\card{\vmap{\arch}{P}(\mathcal{S}_i) \cap D_1} <
      \frac{b_P(n,\mathcal{S}_i)}{2}$, then by (\ref{eq:same-card-one}), we have: \[\card{\mu_1(\mathcal{S})
        \cap \vmap{\arch'}{P}(\mathcal{S}_i)} =
      \card{\vmap{\arch_1}{P}(\mathcal{S}) \cap
        \vmap{\arch}{P}(\mathcal{S}_i)}\]
    \item othwerwise, if $\card{\vmap{\arch}{P}(\mathcal{S}_i)} \neq
      \card{\vmap{\arch'}{P}(\mathcal{S}_i)}$ and
      $\card{\vmap{\arch}{P}(\mathcal{S}_i) \cap D_1} \geq
      \frac{b_P(n,\mathcal{S}_i)}{2}$, since
      $\card{\vmap{\arch_1}{P}(\mathcal{S})} \geq b_P(n-1,
      \mathcal{S})$, by (\ref{eq:diff-card-two}) we obtain:
     \[\card{\mu_1(\mathcal{S}) \cap
        \vmap{\arch'}{P}(\mathcal{S}_i)} \geq b_P(n-1,\mathcal{S})\]     
    \end{compactitem}
     If $\card{\mu_1(\mathcal{S}) \cap \vmap{\arch'}{P}(\mathcal{S}_i)}
    \geq b_P(n-1,\mathcal{S})$ for some $1 \leq i \leq k$, then
    $\card{\mu_1(\mathcal{S})} \geq b_P(n-1,\mathcal{S})$ and we are
    done. Otherwise, we compute:
    \[\card{\mu_1(\mathcal{S})} = \sum_{i=1}^k \card{\mu_1(\mathcal{S})
      \cap \vmap{\arch'}{P}(\mathcal{S}_i)} \\ = \sum_{i=1}^k
    \card{\vmap{\arch_1}{P}(\mathcal{S}) \cap
      \vmap{\arch}{P}(\mathcal{S}_i)} =
    \card{\vmap{\arch_1}{P}(\mathcal{S})} \geq b_P(n-1,\mathcal{S})\]
    The latter equality is by definition of $\mu_1$ and the assumption
    $\card{\mu_1(\mathcal{S}) \cap \vmap{\arch'}{P}(\mathcal{S}_i)} <
    b_P(n-1,\mathcal{S})$, for all $1 \leq i \leq k$.
  \end{compactenum}

    \vspace*{\baselineskip}\noindent Finally, we must prove that
  $\arch'_1 \uplus \arch'_2 = \arch'$. We compute:
  \[\begin{array}{rcl}
  D'_1 \cup D'_2 & = & ((D_1 \cup D_2) \cap P) \cup \bigcup_{\mathcal{S} \in 2^{2^P}} \mu_1(\mathcal{S}) \cup \bigcup_{\mathcal{S} \in 2^{2^P}} \mu_2(\mathcal{S}) \\
  & = & (D \cap P) \cup (D' \setminus P) \text{, because $\set{\mu_1(\mathcal{S})}_{\mathcal{S} \in 2^{2^P}} \cup \set{\mu_2(\mathcal{S})}_{\mathcal{S} \in 2^{2^P}}$ partitions $ D' \setminus P$} \\
  & = & (D' \cap P) \cup (D' \setminus P) \text{, since $\arch \veq{n}{P} \arch'$ thus $D' \cap P = D \cap P$} \\
  & = & D' 
  \end{array}\]
  By the definition of $\I'_i$, $i=1,2$, we have: \[\I'_1 \cap \I'_2 =
  (\intersect{D'_1}{\I'} \cup X_1 \cup Y_1) \cap
  (\intersect{D'_2}{\I'} \cup X_2 \cup Y_2)\] and prove that $\I'_1
  \cap \I'_2 = \intersect{D'_1}{\I'} \cap \intersect{D'_2}{\I'}$, by
  showing the following: \begin{compactitem}
  \item $\intersect{D'_1}{\I'} \cap X_2 = \emptyset$: if there exists
    an interaction $I \in \intersect{D'_1}{\I'} \cap X_2$, then
    $\alpha_2 \in I$, by the definition of $X_2$, and $I \not\in \I'$,
    because $\alpha_2 \not\in \bigcup\I'$, contradiction.
  \item $\intersect{D'_1}{\I'} \cap Y_2 = \emptyset$: if there exists
    an interaction $I \in \intersect{D'_1}{\I'} \cap Y_2$, then
    $\beta_2 \in I$, by the definition of $Y_2$ and $I \not\in \I'$,
    because $\beta_2 \not\in \bigcup\I'$, contradiction.
  \item $X_1 \cap X_2 = \emptyset$: if there exists an interaction $I
    \in X_1 \cap X_2$, then $\alpha_1 \in I$, by the definition of
    $X_1$ and thus $I \not\in X_2$, by the definition of $X_2$,
    contradiction.
  \item $Y_1 \cap X_2 = \emptyset$: if there exists an interaction $I
    \in Y_1 \cap X_2$, then $\beta_1 \in I$, by the definition of
    $Y_1$ and thus $I \not\in X_2$, by the definition of $X_2$,
    contradiction.
  \item $Y_1 \cap Y_2 = \emptyset$: if there exists an interaction $I
    \in Y_1 \cap Y_2$, then $\beta_1 \in Y_1$, by the definition of
    $Y_1$ and thus $I \not\in Y_2$, by the definition of $Y_2$,
    contradiction.
  \end{compactitem}
   Proving the emptiness of the remaining sets $X_1 \cap
  \intersect{D'_2}{\I'}$, $X_1 \cap Y_2$ and $Y_1 \cap
  \intersect{D'_2}{\I'}$ is done symmetrically. Also, by the
  definition of $X_i$ and $Y_i$, for $i=1,2$, we have that $I \cap
  D'_1 \neq \emptyset$ and $I \cap D'_2 \neq \emptyset$, for all $I
  \in \bigcup_{i=1,2} X_i \cup Y_i$. Consequently, we obtain:
  \[\I'_i \cap 2^{\overline{D'}_{3-i}} = \intersect{D'_i}{\I'} \text{, for $i=1,2$}\]
  and conclude the proof as follows:
  \[(\I'_1 \cap \I'_2) \cup (\I'_1 \cap 2^{\overline{D'}_2}) \cup (\I'_2
  \cap 2^{\overline{D'}_1}) = (\intersect{D'_1}{\I'} \cap
  \intersect{D'_2}{\I'}) \cup (\intersect{D'_1}{\I'} \cap
  2^{\overline{D'}_2}) \cup (\intersect{D'_2}{\I'} \cap
  2^{\overline{D'}_1}) = \I'\] \qed}

The next theorem proves that the architectures which are equivalent in
the sense of Definition \ref{def:veq} cannot be distinguished by
\ssil\ formulae up to a given bound, defined recursively on the
structure of formulae: 

\[\begin{array}{rclcrcl}
\bound{\emp} & \isdef & 1 & \hspace*{1cm} & \bound{p \inter b} & \isdef & 1 \\
\bound{p \closeinter b} & \isdef & 1 && \bound{p \closexinter b} & \isdef & 1 \\
\bound{\psi_1 \wedge \psi_2} & \isdef & \max(\bound{\psi_1}, \bound{\psi_2}) && \bound{\neg\psi_1} & \isdef & \bound{\psi_1} \\
\bound{\psi_1 * \psi_2} & \isdef & \max(\bound{\psi_1},\bound{\psi_2})+1
\end{array}\]

\begin{theorem}\label{thm:ssil-equiv}
  Let $\arch = \tuple{D, \I}$ and $\arch' = \tuple{D', \I'}$ be
  architectures, $P \in 2^\ports$ be a set of ports and $n \geq 1$ be
  an integer, such that $\arch \veq{P}{n} \arch'$. Then, for any
  formula $\psi$ of \ssil, such that $\portsof{\psi} \subseteq P$ and
  $\bound{\psi} \leq n$, we have $\arch \models \psi$ if and only if
  $\arch' \models \psi$.
\end{theorem}
\proof{ By induction on the structure of $\psi$. We consider the
  cases: \begin{compactitem}
    \vspace*{\baselineskip}
  \item $\emp$: if $\arch \models \emp$ then $D = \emptyset$ and $\I =
    \emptyset$. Since $\arch \veq{P}{n} \arch'$, we have $D \cap P =
    D' \cap P = \emptyset$. Suppose, for a contradiction, that there
    exists a port $x \in D' \setminus P$. Then there exists a set
    $\mathcal{S} \in 2^{2^P}$ such that $x \in
    \vmap{\arch'}{P}(\mathcal{S})$, hence
    $\card{\vmap{\arch'}{P}(\mathcal{S})} \geq 1$. Since $\arch
    \veq{P}{n} \arch'$, it must be that
    $\card{\vmap{\arch}{P}(\mathcal{S})} \geq 1$, which contradicts
    with $\card{\vmap{\arch}{P}(\mathcal{S})} = 0$, a consequence of
    $D = \emptyset$. Hence $D' = \emptyset$ and $\I' = \emptyset$
    follows, since $\arch'$ is an architecture, thus $\arch' \models
    \emp$. The other direction is symmetrical. 
    \vspace*{\baselineskip}
  \item $p \inter b$: if $\arch \models p \inter b$, we have $D =
    \set{p}$ and $I \vdash b$, for all $I \in \I$. Since $\arch
    \veq{P}{n} \arch'$ and $p \in \portsof{p \inter b} \subseteq P$,
    we obtain $D \cap P = D' \cap P = \set{p}$. Moreover, $D'
    \setminus P = \emptyset$ follows in the same way as above and thus
    $D' = \set{p}$. Since $\arch \veq{P}{n} \arch'$, we have
    $\interproj{\I}{P} \sqcap P = \interproj{\I'}{P} \sqcap P$. Let
    $I' \in \I'$ be an interaction. Then $I' = J \cup U'$, where $J
    \subseteq P$ and $U' \cap P = \emptyset$. Consequently, there
    exists an interaction $I = J \cap U$, for some $U \cap P =
    \emptyset$. Moreover, since $I \vdash b$ and $\portsof{b}
    \subseteq P$, we have $J \vdash b$, thus $I' \vdash b$ and $\arch'
    \models p \inter b$ follows. 
    \vspace*{\baselineskip}
  \item $p \closeinter b$: by an argument similar to the point above. 
    \vspace*{\baselineskip}
  \item $p \closexinter b$: by an argument similar to the point above. 
    \vspace*{\baselineskip}
  \item $\psi_1 \wedge \psi_2$: if $\arch \models \psi_1 \wedge
    \psi_2$ then $\arch \models \psi_i$, for $i=1,2$. By the induction
    hypothesis, since $\bound{\psi_i} \leq \max(\bound{\psi_1},
    \bound{\psi_2}) = \bound{\psi_1 \wedge \psi_2}$, we obtain $\arch'
    \models \psi_i$, for $i=1,2$, hence $\arch' \models \psi_1 \wedge
    \psi_2$. 
    \vspace*{\baselineskip}
  \item $\neg\psi_1$: by a direct application of the induction
    hypothesis.
    \vspace*{\baselineskip}
  \item $\psi_1 * \psi_2$: if $\arch \models \phi_1 * \phi_2$ then
    there exist architectures $\arch_i \models \phi_i$, for $i=1,2$,
    such that $\arch = \arch_1 \uplus \arch_2$. By Lemma
    \ref{lemma:ssil-star}, because $\arch \veq{P}{n} \arch'$, there
    exist architectures $\arch'_i$, such that $\arch_i \veq{P}{n-1}
    \arch'_i$, for $i=1,2$ and $\arch' = \arch'_1 \uplus \arch'_2$. By
    the induction hypothesis, since $\bound{\psi_i} \leq
    \max(\bound{\psi_1}, \bound{\psi_2}) \leq n - 1$, we obtain that
    $\arch'_i \models \psi_i$, for $i=1,2$, and thus $\arch' \models
    \psi_1 * \psi_2$. \qed
  \end{compactitem}}

Next, we move on to the definition of test formulae for \ssil: 

\begin{definition}\label{def:ssil-test-formulae}
  Given a set of port symbols $P \subseteq \psym$ and an integer $n
  \geq 1$, we denote by $\stest{P}{n}$ the following set of formulae,
  for each $p,q_1,\ldots,q_k \in P$ and each $1 \leq m \leq
  b_P(n,\emptyset)$:
  \[\begin{array}{lcl}
  \has{p} \isdef p \inter p \wand \bot && p \ieinter q_1 \ldots q_k \isdef p \exinter q_1 \ldots q_k * ~\top \\[2mm]
  \type{q_1, \ldots, q_k}{m} & \isdef & \underbrace{(\exists x ~.~
    x \exinter q_1 \ldots q_k) * \ldots * (\exists x ~.~ x \exinter
    q_1 \ldots q_k)}_{m \text{ times}}
  \end{array}\]
  Given architectures $\arch_1$ and $\arch_2$, we write $\arch_1
  \stfeq{P}{n} \arch_2$ for $\arch_1 \models \psi \iff \arch_2 \models
  \psi$, for all $\psi \in \stest{P}{n}$.
\end{definition}

The following lemma proves that the equivalence of architectures via
test formulae is a refinement of the equivalence relation introduced
by Definition \ref{def:veq}. 

\begin{lemma}\label{lemma:ssil-tf-veq}
  Given a set of ports $P \in 2^\ports$ and an integer $n \geq 1$, for
  any two architectures $\arch_i = \tuple{D_i, \I_i}$, for $i=1,2$, we
  have $\arch_1 \veq{P}{n} \arch_2$ if $\arch_1 \stfeq{P}{n} \arch_2$.
\end{lemma}
\proof{
  We prove the three points of Definition \ref{def:veq}: 
  
  \vspace*{\baselineskip}\noindent (\ref{it1:veq}) Suppose, for a
  contradiction, that $D_1 \cap P \not\subseteq D_2 \cap P$, thus
  there exists a port $p \in D_1 \cap P$ such that $p \not\in
  D_2$. Then $\arch_1 \models \has{p}$ and $\arch_2
  \not\models\has{p}$. Since $p \in P$ thus $\has{p} \in \stest{P}{n}$,
  we reached a contradiction with $\arch_1 \stfeq{P}{n}
  \arch_2$. Hence $D_1 \cap P \subseteq D_2 \cap P$ and the other
  direction is symmetrical.

  \vspace*{\baselineskip}\noindent (\ref{it2:veq}) Suppose, for a
  contradiction, that $\interproj{\I_1}{(D_1 \cap P)} \sqcap P
  \not\subseteq \interproj{\I_2}{(D_2 \cap P)} \sqcap P$. Then, there
  exists an interaction $J \in \interproj{\I_1}{(D_1 \cap P)} \sqcap
  P$ such that $J \not\in \interproj{\I_2}{(D_2 \cap P)} \sqcap
  P$. Let $p \in J \cap D_1$ be a port (we know that one exists
  because $J \in \interproj{\I_1}{(D_1 \cap P)} \sqcap P$) and let
  $\set{q_1, \ldots, q_k} \isdef J \setminus \set{p}$.  Since $p \in J
  \cap D_1$ and $J \subseteq P$, by the previous point, we have $p \in
  D_2 \cap P$. We have $\arch_1 \models p \ieinter q_1 \ldots q_k$ and
  since $\arch_1 \stfeq{P}{n} \arch_2$, we obtain $\arch_2 \models p
  \ieinter q_1 \ldots q_k$. But then we obtain $J \in
  \interproj{\I_2}{(D_2 \cap P)} \sqcap P$, contradiction. Hence
  $\interproj{\I_1}{(D_1 \cap P)} \sqcap P \subseteq
  \interproj{\I_2}{(D_2 \cap P)} \sqcap P$ and the other direction is
  symmetrical. 

  \vspace*{\baselineskip}\noindent (\ref{it3:veq}) Let $\mathcal{S}
  \in 2^{2^\ports}$ be a set of visible
  ports. We distinguish the following cases: \begin{compactitem}
  \item if $\card{\vmap{\arch_1}{P}(\mathcal{S})} <
    b_P(n,\mathcal{S})$ and $\card{\vmap{\arch_2}{P}(\mathcal{S})}
    \neq \card{\vmap{\arch_1}{P}(\mathcal{S})}$ then let $m =
    \card{\vmap{\arch_1}{P}(\mathcal{S})}$. We have $\arch_1 \models
    \type{\mathcal{S}}{m} \wedge \neg \type{\mathcal{S}}{m+1}$
    and $\arch_2 \not\models \type{\mathcal{S}}{m} \wedge \neg
    \type{\mathcal{S}}{m+1}$. Since $m + 1 \leq b_P(n,\mathcal{S})
    \leq b_P(n,\emptyset)$, we obtain that $\type{\mathcal{S}}{m},
    \type{\mathcal{S}}{m+1} \in \stest{P}{n}$, thus $\arch_1
    \not{\stfeq{P}{n}} \arch_2$, contradiction. Then
    $\card{\vmap{\arch_1}{P}(\mathcal{S})} < b_P(n,\mathcal{S})
    \Rightarrow \card{\vmap{\arch_2}{P}(\mathcal{S})} =
    \card{\vmap{\arch_1}{P}(\mathcal{S})}$.
    \item if $\card{\vmap{\arch_1}{P}(\mathcal{S})} \geq
      b_P(n,\mathcal{S})$ and $\card{\vmap{\arch_2}{P}(\mathcal{S})} <
      b_P(n,\mathcal{S})$ then let
      $\card{\vmap{\arch_2}{P}(\mathcal{S})} = m$. We have $\arch_2
      \models \type{\mathcal{S}}{m} \wedge \neg
      \type{\mathcal{S}}{m+1}$ and $\arch_1 \not\models
      \type{\mathcal{S}}{m} \wedge \neg \type{\mathcal{S}}{m+1}$.
      Since $m+1 \leq b_P(n,\mathcal{S}) \leq b_P(n,\emptyset)$, we
      obtain that $\arch_1 \not{\stfeq{P}{n}} \arch_2$,
      contradiction. Then $\card{\vmap{\arch_1}{P}(\mathcal{S})} \geq
      b_P(n,\mathcal{S}) \Rightarrow
      \card{\vmap{\arch_2}{P}(\mathcal{S})} \geq b_P(n,\mathcal{S})$. \qed
  \end{compactitem}}

A first consequence of this result is that every formula of \ssil\ is
equivalent to a finite boolean combination of test formulae.

\begin{corollary}\label{cor:ssil-boolean-test-formulae}
   Each formula $\psi$ of \ssil\ is equivalent to a finite boolean
   combination of test formulae from $\stest{\portsof{\psi}}{\bound{\psi}}$.
\end{corollary}
\proof{ The proof is the same as for Corollary
  \ref{cor:boolean-test-formulae}.  \qed}

The other consequence is a small model property for the
\ssil\ fragment, which entails the decidability of its satisfiability
problem. 

\begin{corollary}
  If $\psi$ is a satisfiable \ssil\ formula has a model $\arch =
  \tuple{D,\I}$ such that $\card{D} \leq B$ and $\card{I} \leq B$, for
  each $I \in \I$, where $B = \bigO(2^{\card{\portsof{\psi}}}) \cdot
  b_{\portsof{\psi}}(\bound{\psi}, \emptyset)$.
\end{corollary}
\proof{ Let $\arch' = \tuple{D',\I'}$ be the architecture obtained
  from $\arch$ as follows : \begin{compactitem}
    \item remove from $D$ and from each $I \in \I$ enough many ports
      $p \in D \setminus \portsof{\psi}$, such that
      $\card{\vmap{\arch'}{\portsof{\psi}}}(\mathcal{S}) \leq
      b_{\portsof{\psi}}(\bound{\psi}, \emptyset)$, for each
      $\mathcal{S} \in 2^{2^\ports}$, and
    \item remove from each from each $I \in \I$ all ports $p \in I
      \setminus (D \cup \portsof{\psi})$.
  \end{compactitem}
  It is easy to check that $\arch
  \veq{\portsof{\psi}}{b_{\portsof{\psi}}(\bound{\psi}, \emptyset)}
  \arch'$ thus, by Theorem \ref{thm:ssil-equiv}, we obtain that
  $\arch' \models \psi$. Further, we compute:
  \[\begin{array}{rcl}
  \card{D'} & = & \card{D' \cap \portsof{\psi}} + \card{D' \setminus \portsof{\psi}} \\
  & = & \card{D' \cap \portsof{\psi}} + \sum_{\mathcal{S} \in 2^{2^\ports}} \card{\vmap{\arch}{\portsof{\psi}}}(\mathcal{S}) \\
  & \leq & \card{\portsof{\psi}} +  2^{\card{\portsof{\psi}}} \cdot b_{\portsof{\psi}}(\bound{\psi}, \emptyset) = B 
  \end{array}\]
  Let $I \in \I'$ be an interaction. We compute:
  \[\begin{array}{rcl}
  \card{I} & = & \card{D \cap I} + \card{I \setminus D} \\
  & \leq & \card{D} + \card{\portsof{\psi}} = B \text{ \qed}
  \end{array}\]}

\subsection{Decidability of Component-based Extensions of \sil}
\label{sec:comp-sil}

\todo{This section is under construction}

In this section we extend the decidability results from
\S\ref{sec:ssil} and \S\ref{sec:psil} to fragments of the logic
\slarch\ obtained by considering variables $i,j \in \ivars$ ranging
over component identifiers and function symbols $\pport,\qport \in
\pfunc$, interpreted as functions mapping component identifiers to
ports.  Moreover, we allow equality atoms $i = j$ and port terms
$\pport(i)$ to occur anywhere a port symbol $p \in \psym$ is allowed
to occur in \ssil\ and \psil, respectively.

\section{Behaviors of Component-based Systems}
\label{sec:behaviors}

\todo{This section is under construction}

In this section we define the \emph{behaviors} of an architecture,
which are the sequences of interaction events, ordered by the moment
in time when the events occur. We consider systems consisting of
finitely many components, with no \`a~priori bound on their number,
that are replicas of a small number of finite-state
machines.

Formally, a \emph{finite-state machine} (FSM) is a pair $M = (\states,
\rules)$, where $\states$ is a finite set of states and $\rules
\subseteq \states \times 2^{\ports} \times \states$ is a transition
relation, where $q \arrow{I}{} q'$ stands for $(q,I,q') \in \rules$.
We denote by $\Sigma(M) \isdef \set{I \mid \exists q, q' \in Q ~.~ q
  \arrow{I}{} q'}$ the set of transition labels of $M$. We write
$(\states_1, \rules_1) \subseteq (\states_2, \rules_2)$ for $\states_1
= \states_2$ and $\rules_1 \subseteq \rules_2$. 

The \emph{asynchronous product} of two FSMs $M_i =
(\states_i,\rules_i)$, for $i=1,2$, is the FSM $M_1 \otimes M_2 \isdef
(\states_1 \times \states_2, \rules)$ where, for all $q_i,q'_i\in
Q_i$, $i=1,2$ and all $I \in 2^\ports$, $(q_1,q_2) \arrow{I}{}
(q'_1,q'_2)$ if and only if one of the following
holds: \begin{compactitem}
\item $q_1 \arrow{I}{1} q'_1$ and $q_2 \arrow{I}{2} q'_2$,
\item $q_1 \arrow{I}{1} q'_1$, $q_2 = q'_2$ and $I \not\in \Sigma(M_2)$, 
\item $q_2 \arrow{I}{1} q'_2$, $q_1 = q'_1$ and $I \not\in \Sigma(M_1)$.
\end{compactitem}

\begin{definition}\label{def:arch-fsm}
Given an architecture $\arch = \tuple{D,\I}$ and a FSM $M =
(\states,\rules)$, such that $I \cap D \neq \emptyset$, for all $I \in
\Sigma(M)$. We define $\behave{\arch}{M} \isdef
(\states,\rules_\arch)$ where, for all $q,q' \in \states$ and all $I
\in 2^\ports$ we have, $q \arrow{I}{\arch} q'$ if and only if $q
\arrow{I}{} q'$ and $I \in \I$.
\end{definition}
Note that $\behave{\arch}{M}$ is undefined if $\Sigma(M)$ contains
interactions that do not intersect with $\dom{\arch}$. The following
theorem relates the composition of architectures with the asynchronous
product of their behaviors. 

\begin{theorem}\label{thm:arch-behavior}
  Let $\arch_i = \tuple{D_i, \I_i}$ be architectures and $M =
  (\states_i, \rules_i)$ be FSMs, such that $\behave{\arch_i}{M_i}$ is
  defined, for all $i=1,2$. Then the following hold: \begin{compactenum}
    \item\label{it1:arch-behavior} $\behave{\arch_1 \uplus
      \arch_2}{M_1 \otimes M_2} \subseteq \behave{\arch_1}{M_1}
      \otimes \behave{\arch_2}{M_2}$,
    \item\label{it2:arch-behavior} $\behave{\arch_1 \uplus
      \arch_2}{M_1 \otimes M_2} = \behave{\arch_1}{M_1} \otimes
      \behave{\arch_2}{M_2}$ if, moreover, $I \cap D_{3-i} \neq
      \emptyset \Rightarrow I \in \I_{3-i}$, for all $I \in \I_i$,
      $i=1,2$.
  \end{compactenum}
\end{theorem}
\proof{ In the following, we denote:
  \[\begin{array}{rcl}
  \arch_1 \uplus \arch_2 & \isdef & \tuple{D_1 \cup D_2, \I_\uplus} \\
  M_1 \otimes M_2 & \isdef & (Q_1 \times Q_2, \rules_{12}) \\
  \behave{\arch_1 \uplus \arch_2}{M_1 \otimes M_2} & \isdef & (Q_1 \times Q_2, \rules) \\
  \behave{\arch_i}{M_i} & \isdef & (Q_i, \rules_{\arch_i}) \text{, for $i=1,2$} \\
  \behave{\arch_1}{M_1} \otimes \behave{\arch_2}{M_2} & \isdef & (Q_1 \times Q_2, \rules_\otimes)
  \end{array}\]
  Note that, because $\behave{\arch_i}{M_i}$ is defined, for $i=1,2$,
  for each $I \in \Sigma(M_1 \otimes M_2) = \Sigma(M_1) \cup
  \Sigma(M_2)$ we have $I \cap (D_1 \cup D_2) \neq \emptyset$, thus
  $\behave{\arch_1 \uplus \arch_2}{M_1 \otimes M_2}$ is defined.
  
  \vspace*{\baselineskip}\noindent\framebox{\ref{it1:arch-behavior}} Let
  $(q_1,q_2) \arrow{I}{} (q'_1,q'_2)$ be a transition of
  $\behave{\arch_1 \uplus \arch_2}{M_1 \otimes M_2}$, for some $I \in
  2^\ports$. Then $I \in \I_\uplus$ and $(q_1,q_2) \arrow{I}{12}
  (q'_1,q'_2)$. Since $\I_\uplus = (\I_1 \cap \I_2) \cup (\I_1 \cap
  2^{\overline{D}_2}) \cup (\I_2 \cap 2^{\overline{D}_1})$, we
  distinguish the following cases: \begin{inparaenum}[(1)]
  \item\label{it1:arch} $I \in \I_1 \cap \I_2$, 
  \item\label{it2:arch} $I \in \I_1$ and $I \cap D_2 = \emptyset$ and
  \item\label{it3:arch} $I \in \I_2$ and $I \cap D_1 = \emptyset$.
  \end{inparaenum}
  Moreover, based on the definition of $M_1 \otimes M_2$, we
  distinguish the following cases: \begin{inparaenum}[(a)]
  \item\label{it1:fsm} $q_i \arrow{I}{i} q'_i$, for $i=1,2$,
  \item\label{it2:fsm} $q_1 \arrow{I}{1} q'_1$, $q_2 = q'_2$ and $I
    \not\in \Sigma(M_2)$,
  \item\label{it3:fsm} $I \not\in \Sigma(M_1)$, $q_1 = q'_1$ and $q_2
    \arrow{I}{2} q'_2$.
  \end{inparaenum}
  We give the proof in the following composed cases:

  \vspace*{\baselineskip}\noindent(\ref{it1:arch}\ref{it1:fsm}) Since
  $I \in \I_i$ and $q_i \arrow{I}{i} q'_i$, we obtain $q_i
  \arrow{I}{\arch_i} q'_i$, for both $i=1,2$, thus $(q_1,q_2)
  \arrow{I}{\otimes} (q'_1,q'_2)$.

  \vspace*{\baselineskip}\noindent(\ref{it1:arch}\ref{it2:fsm}) Since
  $I \in \I_1$ and $q_1 \arrow{I}{1} q'_1$, we obtain $q_1
  \arrow{I}{\arch_1} q'_1$. Moreover, $q_2 = q'_2$ and $I \not\in
  \Sigma(M_2) \supseteq \Sigma(\behave{\arch_2}{M_2})$, thus
  $(q_1,q_2) \arrow{I}{\otimes} (q'_1,q_2)$.

  \vspace*{\baselineskip}\noindent(\ref{it1:arch}\ref{it3:fsm}) This
  case is symmetrical to (\ref{it1:arch}\ref{it2:fsm}).

  \vspace*{\baselineskip}\noindent(\ref{it2:arch}\ref{it1:fsm}) Since
  $I \in \I_1$ and $q_1 \arrow{I}{1} q'_1$, we obtain $q_1
  \arrow{I}{\arch_1} q'_1$. Moreover, because $I \cap D_2 = \emptyset$
  and since $\behave{\arch_2}{M_2}$ is defined, we obtain $I \not\in
  \Sigma(M_2)$. If $q_2 = q'_2$, we obtain that $(q_1,q_2)
  \arrow{I}{\otimes} (q'_1,q_2)$. Else, $q_2 \neq q'_2$ and $q_2
  \arrow{I}{2} q'_2$ contradicts $I \not\in \Sigma(M_2)$.

  \vspace*{\baselineskip}\noindent(\ref{it2:arch}\ref{it2:fsm})
  Similar to (\ref{it2:arch}\ref{it1:fsm}), using directly that
  $q_2=q'_2$ and $I \not\in \Sigma(M_2)$.

  \vspace*{\baselineskip}\noindent(\ref{it2:arch}\ref{it3:fsm})
  Because $I \cap D_2 = \emptyset$ and since $\behave{\arch_2}{M_2}$
  is defined, we obtain $I \not\in \Sigma(M_2)$, which contradicts
  $q_2 \arrow{I}{2} q'_2$. 

  \vspace*{\baselineskip}\noindent The cases
  (\ref{it3:arch}\ref{it1:fsm}), (\ref{it3:arch}\ref{it2:fsm}) and
  (\ref{it3:arch}\ref{it3:fsm}) are symmetrical to
  (\ref{it2:arch}\ref{it1:fsm}), (\ref{it2:arch}\ref{it2:fsm}) and
  (\ref{it2:arch}\ref{it3:fsm}), respectively.
  
  \vspace*{\baselineskip}\noindent\framebox{\ref{it2:arch-behavior}}
  To show that $\behave{\arch_1 \uplus \arch_2}{M_1 \otimes M_2}
  \supseteq \behave{\arch_1}{M_1} \otimes \behave{\arch_2}{M_2}$, let
  $(q_1,q_2) \arrow{I}{\otimes} (q'_1,q'_2)$ be a transition of
  $\behave{\arch_1 \uplus \arch_2}{M_1 \otimes M_2} =
  \behave{\arch_1}{M_1} \otimes \behave{\arch_2}{M_2}$. We distinguish
  the following cases: \begin{compactenum}[(1)]
  \item if $q_i \arrow{I}{\behave{\arch_i}{M_i}} q'_i$, then $q_i
    \arrow{I}{i} q'_i$ and $I \in \I_i$, for both $i=1,2$. We obtain
    $(q_1, q_2) \arrow{I}{12} (q'_1, q'_2)$ and $I \in \I_1 \cap \I_2
    \subseteq \I_\uplus$, thus $(q_1, q_2) \arrow{I}{} (q'_1, q'_2)$. 
  \item if $q_1 \arrow{I}{\behave{\arch_1}{M_1}} q'_1$, $q_2 = q'_2$
    and $I \not\in \Sigma(\behave{\arch_2}{M_2})$, we consider two
    cases: \begin{compactenum}[(a)]
    \item if $I \in \I_2 \setminus \Sigma(M_2)$ then $I \in \I_1 \cap
      \I_2 \subseteq \I_\uplus$ and $(q_1,q_2) \arrow{I}{12}
      (q'_1,q_2)$, thus $(q_1,q_2) \arrow{I}{} (q'_1,q_2)$. 
    \item else $I \not\in \I_2$ and since $I \in \I_1$, by the
      hypothesis $I \cap D_2 \neq \emptyset \Rightarrow I \in \I_2$,
      we deduce that $I \cap D_2 = \emptyset$. Then $I \in \I_1 \cap
      2^{\overline{D}_2} \subseteq \I_\uplus$. Moreover, since
      $\behave{\arch_2}{M_2}$ is defined, we obtain $I \not\in
      \Sigma(M_2)$, thus $(q_1,q_2) \arrow{I}{12} (q'_1,q_2)$ and
      $(q_1,q_2) \arrow{I}{} (q'_1,q_2)$ follows.
    \end{compactenum} 
  \item This case is symmetrical to the above one.
  \end{compactenum}  
  \qed}

\section{Dynamic Reconfigurability}
\label{sec:reconf}

\todo{This section is under construction}

We extend architectures to capture reconfigurability, by
distinguishing between the \emph{architecture layer}, describing the
components that are active in the system and their interactions, and
the \emph{map layer}, which is the graph onto which the components are
\emph{deployed}. Formally, the component layer consists
of: \begin{compactitem}
  \item a countably infinite set $\ids$ of \emph{component
    identifiers}, ranged over by the set of identifier variables
    $\ivar = \set{i,j,\ldots}$,
  \item a finite set of total \emph{port functions} of type $\ids
    \rightarrow \ports$, denoted by the set of function symbols
    $\pfunc = \set{\pport,\qport,\ldots}$.
\end{compactitem}
The map layer consists of: \begin{compactitem}
\item a countably infinite set $\nodes$ of \emph{map nodes}, with a
  designated element $\nil \in \nodes$ and ranged over by the node
  variables $\nvar = \set{n,m,\ldots}$,
\item a partial \emph{map} $M : \nodes \finmap \nodes^k$, with finite
  domain $\dom{M}$, where $\nil \not\in \dom{M}$. We assume that the
  image of each node $n \in \dom{M}$ consists of exactly $k\geq1$
  nodes $M(n) = (n_1, \ldots, n_k)$.
\end{compactitem}
The link between the layers is established by a finite partial
\emph{deployment} function $\Delta : \ids \finmap \dom{M} \cup
\set{\nil}$. By $\archset$ we denote the set of architectures $\arch =
\tuple{D,\I}$, with $D \subseteq \ports$ and $\I \subseteq 2^\ports$,
such that $D \cap I \neq \emptyset$, for all $I \in \I$. Moreover, by
$\mapset^k$ we denote the set of maps $M : \nodes \finmap \nodes^k$.

We describe such systems using a combination of two resource logics,
defined in the following. Given a constant $k\geq1$, the formulae of
the \emph{Separation Logic of Maps} (\slmap{k}) are defined by the
following syntax:
\[\begin{array}{rcl}
t & := & \mathit{nil} \mid n \in \nvar \\
\phi & := & t_1 = t_2 \mid \emp_m \mid i \deploy t, i \in \ivar \mid t_0 \pto (t_1,\ldots,t_k) \mid Q(t_1,\ldots,t_{\#Q}) \mid \\
&& \phi_1 \wedge \phi_2 \mid \neg \phi_1 \mid \exists i ~.~ \phi_1,~ i \in \ivar \mid \exists n ~.~ \phi_1,~ n \in \nvar \mid 
\phi_1 *_m \phi_2 \mid \phi_1 \wand_m \phi_2 
\end{array}\]
where $Q(t_1, \ldots, t_{\#(Q)}) \in \npred$ is a predicate symbol of
type $\nodes^{\#(Q)} \rightarrow \set{\bot,\top}$. By $\top_m$ we
denote the equality $n=n$, the choice of $n \in \nvar$ being
unimportant. A \slmap\ \emph{sentence} is a formula in which all
variables occur within the scope of a quantifier.

Since \slmap{k}\ formulae contain two types of variables, we consider
extended valuations $\nu : \ivar \cup \nvar \rightarrow \ids \cup
\nodes$, such that $\nu(x) \in \ids$, when $x \in \ivar$ and $\nu(n)
\in \nodes$, when $n \in \nvar$. For a term $t$, we write $\nu(t)$ to
denote the node $\nil$ if $t=\mathit{nil}$ and the node $\nu(n)$ if $t
= n \in \nvar$. The semantics of \slmap{k}\ is defined by a satisfaction
relation $\tuple{\Delta,M} \models_\nu^\X \phi$ between pairs of
deployments and maps and formulae, parameterized by a valuation $\nu :
\ivar \cup \nvar \rightarrow \ids \cup \nodes$ and an interpretation
of predicate symbols $\X : \npred \rightarrow \bigcup_{k\geq1}
2^{\nodes^k \times \mapset}$, such that $\X(Q) \subseteq
\nodes^{\#(Q)} \times \mapset$.
\[\begin{array}{lcl}
\tuple{\Delta,M} \models^\X_\nu t_1 = t_2 & \iff & \nu(t_1) = \nu(t_2) \\
\tuple{\Delta,M} \models^\X_\nu \emp_m & \iff & M = \emptyset \\
\tuple{\Delta,M} \models^\X_\nu x \deploy t & \iff & \Delta(\nu(x)) = \nu(t) \\
\tuple{\Delta,M} \models^\X_\nu t_0 \pto (t_1,\ldots,t_k) & \iff & \dom{M} = \set{\nu(t_0)} \text{ and } 
M(\nu(t_0))_\ell = (\nu(t_1), \ldots, \nu(t_k)) \\
\tuple{\Delta,M} \models^\X_\nu Q(t_1, \ldots, t_{\#(Q)}) & \iff & (\tuple{\nu(t_1), \ldots, \nu(t_{\#(Q)})},M) \in \X(Q) \\
\tuple{\Delta,M} \models^\X_\nu \exists x ~.~ \phi_1 & \iff & \tuple{D,M} \models_{\nu[x \leftarrow i]}^\X \phi_1
\text{, for some node $i \in \ids$} \\
\tuple{\Delta,M} \models^\X_\nu \exists n ~.~ \phi_1 & \iff & \tuple{D,M} \models_{\nu[n \leftarrow v]}^\X \phi_1
\text{, for some node $v \in \nodes$} \\
\tuple{\Delta,M} \models^\X_\nu \phi_1 *_m \phi_2 & \iff & \text{there exists maps $M_1, M_2$ such that
  $\dom{M_1} \cap \dom{M_2} = \emptyset$,} \\
&& \text{$M = M_1 \cup M_2$ and $\tuple{\Delta,M_i} \models^\X_\nu \phi_i$, for $i=1,2$} \\
\tuple{\Delta,M} \models^\X_\nu \phi_1 \wand_m \phi_2 & \iff & \text{for all maps $M_1$ such that
  $\dom{M_1} \cap \dom{M} = \emptyset$} \\
&& \text{and $\tuple{\Delta,M_1} \models^\X_\nu \phi_1$, we have $\tuple{\Delta,M_1 \cup M} \models^\X_\nu \phi_2$}
\end{array}\]
The semantics of the boolean connectives is standard, thus we omit it
for brevity.

The combined \emph{Separation Logic of Architectures and Maps} (\slam)
is the extension of \slarch\ which allows \slmap{k}\ sentences to occur
as atomic propositions. A \slam\ formula is interpreted over
structures $(\arch,\Delta,M)$, where $\arch = \tuple{D,\I}$ is an
architecture, $\Delta$ is a deployment and $M$ is a map, by a
satisfaction relation $(\arch,\Delta,M) \models_\nu^\X \phi$
parameterized by a valuation $\nu : \ivar \cup \nvar \rightarrow \ids
\cup \nodes$, as before, and an interpretation $\X : \ipred \cup
\npred \rightarrow \bigcup_{k\geq1} 2^{\ids^k} \cup 2^{\ids^k}$, such
that $\X(P) \subseteq \ids^{\#(P)}$, when $P \in \ipred$, and $\X(P)
\subseteq \nodes^{\#(P)}$, when $P \in \npred$.

Both \slarch\ and \slmap\ use predicate symbols, whose interpretation
is the least solution of a system of inductive definitions of the
form:
\[\begin{array}{rcl}
P(x_1, \ldots, x_{\#P}) & \leftarrow & \rho_a,~ \text{ where } P \in \ipred \text{ and } x_1, \ldots, x_{\#(P)} \in \ivar \\
Q(n_1, \ldots, n_{\#Q}) & \leftarrow & \rho_m,~ \text{ where } Q \in \npred \text{ and } n_1, \ldots, n_{\#(Q)} \in \nvar
\end{array}\]
where the logical fragments used for the rules of the system are given by the syntax:
\[\begin{array}{rcl}
\rho_a & := & x=y \mid x\neq y \mid \emp_a \mid p(x) \inter b \mid  p(x) \exinter b \mid p(x) \closeinter b \mid p(x) \closexinter b \mid \\
&& P(x_1, \ldots, x_{\#P}) \mid \rho'_a * \rho''_a \mid \exists x ~.~ \rho'_a \\
\rho_m & := & t_1=t_2 \mid t_1 \neq t_2 \mid \emp_m \mid t_1 \mapsto t_2 \mid Q(t_1,\ldots,t_{\#Q}) \mid \rho'_m *_m \rho''_m \mid \exists n ~.~ \rho'_m
\end{array}\]
The main restriction here is that all predicate symbols occur at
positive polarity within the rules, which ensures the monotonicity of
the rules and the existence of least solutions. The following example
defines two common structures, used in many applications.

\begin{example}\label{ex:pipe-lseg}
  A pipeline architecture, starting with $x$ and ending with $y$,
  where $x_p$ refers to the component previous to $x$ and $y_n$ to the
  component next to $y$:
  \[\begin{array}{rcl}
  \mathsf{pipe}(x,x_p,y_n,y) & \leftarrow & \emp * x=y_n * x_p=y \\
  \mathsf{pipe}(x,x_p,y_n,y) & \leftarrow & \exists z ~.~ \mathit{in}(x) 
  \closeinter \mathit{out}(x_p) * \mathit{out}(x) \exinter \mathit{in}(z) * \mathsf{pipe}(z,x,y_n,y)
  \end{array}\]  
  An acyclic list map, stretching between nodes $n$ and $m$: 
  \[\begin{array}{rcl}
  \mathsf{alist}(n,m) & \leftarrow & \emp *_m n=m \\
  \mathsf{alist}(n,m) & \leftarrow & \exists n' ~.~ n \mapsto n' *_m n\neq m *_m \mathsf{alist}(n',m)
  \end{array}\]
  \hfill$\blacksquare$
\end{example}

\subsection{Synchronization and Deployment Rules} 

We consider universally quantified sentences that describe the
interactions of the architecture, of the form \(\forall \vec{x} ~.~
\phi_m(\vec{x}) \rightarrow \phi_a(\vec{x})\), where $\vec{x} = x_1,
\ldots, x_m \in \ivar$, $\phi_m$ is a formula of \slmap\ and $\phi_a$
is a formula of \slarch. We call these sentences \emph{synchronization
  rules} in the following.

\begin{example}\label{ex:sync-rule}
  The synchronization rule below requires that each two components
  deployed on neighbouring map nodes interact via their $\mathit{in}$
  and $\mathit{out}$ ports, respectively: 
  \[\begin{array}{rcr}
  \forall x \forall y & . & (\exists n \exists m ~.~ n \neq m \wedge n \mapsto m *_m \top_m \wedge x \deploy n \wedge y \deploy m) \rightarrow \\ 
  && \mathit{in}(x) \closexinter \mathit{out}(y) * \mathit{out}(y) \closexinter \mathit{in}(x) * \top_a
  \end{array}\]
  \hfill$\blacksquare$
\end{example}

The following formula states that an identifier belongs to an existing
component, i.e.\ is allocated. We recall that the set $\pfunc$ of port
symbols is finite is finite.
\[\mathsf{alloc}_a(x) \isdef \left(\bigvee_{p \in \pfunc} p(x) \inter p(x)\right) \wand \bot_a\]
We write $\exists_a x ~.~ \phi$ as a shorthand for $\exists x ~.~
\mathsf{alloc}_a(x) \wedge \phi$ and $\forall_a x ~.~ \phi$ for
$\forall x ~.~ \mathsf{alloc}_a(x) \rightarrow \phi$. 

In a similar way, we define the set of nodes that are part of the
domain of the map:
\[\mathsf{alloc}_m(n) \isdef n \mapsto n \wand \bot_m\]
and write $\exists_m n ~.~ \phi$ (resp. $\forall_m n ~.~ \phi$) for
$\exists n ~.~ \mathsf{alloc}_m(n) \wedge \phi$ (resp. $\forall n ~.~
\mathsf{alloc}_m(n) \rightarrow \phi$).

We can now specify \emph{deployment rules}, which are sentences such
as:
\[\begin{array}{rl}
\forall_a x \exists_m n ~.~ x \deploy n & \text{ (every component is deployed)} \\
\forall_m n \exists_a x ~.~ x \deploy n & \text{ (every node has a deployed component)} \\
\forall_m n \forall_a x \forall_a y ~.~ x \deploy n \wedge y \deploy n \rightarrow x = y & 
\text{ (at most one component is deployed on each node)}
\end{array}\]
Note that we need \slam\ to write syncrhonization rules, whereas
deployment rules can be written using only \slmap. We state the
following synthesis problem:

\begin{definition}[Architecture Synthesis]
Given a \slmap\ sentence $\phi$, synchronization rules $\Phi_1,
\ldots, \Phi_k$ and deployment rules $\Psi_1, \ldots, \Psi_\ell$, find
a \slarch\ sentence $\psi$ such that the following \slam\ formula is
valid:
\[\left(\phi \wedge \bigwedge_{i=1}^k\Phi_i \wedge \bigwedge_{i=1}^\ell\Psi_i\right) 
\rightarrow \psi\]
\end{definition}

\begin{example}\label{ex:inv-synth}
  Considering the formula $\exists n \exists m ~.~ \mathsf{alist}(n,m)$ (Example \ref{ex:pipe-lseg}), the syncrhonization rule: 
  \[\begin{array}{rcr}
  (\Phi)~ \forall x \forall y & . & (\exists n \exists m ~.~ n \neq m \wedge n \mapsto m * \top_m \wedge x \deploy n \wedge y \deploy m) \rightarrow \\ 
  && \mathit{in}(x) \closexinter \mathit{out}(y) * \mathit{out}(y) \closexinter \mathit{in}(x) * \top_a
  \end{array}\]
  and the deployment rules: 
  \[\begin{array}{rl}
  (\Psi_1) & \forall_a x \exists_m n ~.~ x \deploy n \\
  (\Psi_2) & \forall_m n \exists_a x ~.~ x \deploy n \\
  (\Psi_3) & \forall_m n \forall_a x \forall_a y ~.~ x \deploy n \wedge y \deploy n \rightarrow x = y
  \end{array}\]
  a solution to the architecture synthesis problem is $\exists x
  \exists x_p \exists y_n \exists y ~.~ \mathsf{pipe}(x,x_p,y_n,y)$
  (Example \ref{ex:pipe-lseg}). \hfill$\blacksquare$
\end{example}

\subsection{Reconfiguration Axioms}

In this section we tackle the problem of defining the reconfiguration
actions. First, we describe their operational semantics, in terms of
updates of the map and the deployment and then we give their axiomatic
semantics in terms of Hoare triples. The latter is derived from a set
of \emph{local axioms}, describing the changes to the (small set of)
cells affected by the reconfiguration and a \emph{frame rule} enabling
a general weakest pre- (strongest post-) condition calculus. 

A \emph{reconfiguration sequence} is a set of actions written in the
following syntax: 

\[\begin{array}{rcl}
i & \in & \ivar,~ n \in \nvar, \ell \in \set{1,\ldots,k} \\
\mathit{term} & := & n \mid \nil \\
\mathit{action} & := & n.\ell = \mathit{term} \mid \mathsf{deploy}(i,n) 
\mid \mathsf{delete}(n) \mid n = \mathsf{new} \mid n = \emph{term} \mid n = m.\ell
\end{array}\]

The operational semantics is given in terms of steps
$(\sigma,\Delta,M) \leadsto (\sigma',\Delta',M')$ where $\sigma :
\nvar \rightarrow \nodes$ is a \emph{store}, $\Delta$ and $M$ are the
deployment and map functions, and $\sigma'$, $\Delta'$ and $M'$ denote
the next values of $\sigma$, $\Delta$ and $M$, respectively. Given a
tuple $\tau \in \nodes^k$ and $1 \leq \ell \leq k$, we denote by
$\tau_\ell$ its $\ell$-th element and by $\tau_{\set{\ell \leftarrow
    \alpha}}$ the tuple with the same elements as $\tau$ except for
its $\ell$-th element, who is set to $\alpha$. The following rules
define the reconfiguration steps:
\[\begin{array}{ccc}
\infer[n.\ell = t]{(\sigma,\Delta,M) \leadsto (\sigma,\Delta,M[\sigma(n) \leftarrow \tau])}{
  \sigma(n) \in \dom{M} \hspace*{2mm} \tau = M(\sigma(n))_{\set{\ell \leftarrow \sigma(t)}}} 
\\\\
\infer[\mathsf{deploy}(i,n)]{(\sigma,\Delta,M) \leadsto (\sigma,\Delta[i \leftarrow \sigma(n)],M)}{} 
\\\\
\infer[\mathsf{free}(n)]{(\sigma,\Delta,M) \leadsto (\sigma,\Delta,M')}{
  \sigma(n) \in \dom{M} \hspace*{2mm} M' = M \setminus \set{\tuple{\sigma(n),M(n)}}}
\\
\infer[n = \mathsf{new}]{(\sigma,\Delta,M) \leadsto (\sigma[n \leftarrow v],\Delta,M')}{
  v \not\in \dom{M} \hspace*{2mm} M' = M \cup \set{\tuple{v,(\overbrace{\nil,\ldots,\nil}^k)}}}
\\\\
\infer[n = t]{(\sigma,\Delta,M) \leadsto (\sigma[n \leftarrow \sigma(t)],\Delta,M)}{}
\\\\
\infer[n = m.\ell]{(\sigma,\Delta,M) \leadsto (\sigma[n \leftarrow M(\sigma(m))_\ell],\Delta,M)}{\sigma(m) \in \dom{M}}
\end{array}\]

In order to carry out deductive verification of reconfiguration
sequences, we define the semantics of the reconfiguration actions by
the following set of local axioms, the encompass the principle of
local reasoning:
\[\begin{array}{rcl}
\set{\exists m_1 \ldots \exists m_k ~.~ n \mapsto (m_1,\ldots,m_k) } & n.\ell = t & \set{n \mapsto (m_1, \ldots, m_{\ell-1}, t, m_{\ell+1}, \ldots, m_k)} 
\\[2mm]
\set{\emp_m} & \mathsf{deploy}(i,n) & \set{i \deploy n \wedge \emp_m} 
\\[2mm]
\set{\exists m ~.~ n \mapsto m} & \mathsf{free}(n) & \set{\emp_m} 
\\
\set{\emp_m} & n = \mathsf{new} & \set{n \mapsto (\overbrace{\nil, \ldots, \nil}^k)} 
\\[2mm]
\set{n = m \wedge \emp_m} & n = t & \set{n=t[m/n] \wedge \emp_m} 
\\[2mm]
\set{m \mapsto (m_1, \ldots, m_k)} & n := m.\ell & \set{n = m_\ell \wedge m \mapsto (m_1, \ldots, m_k)} 
\end{array}\]
These small axioms define a full predicate transformer calculus by
means of the following frame rule, that captures the idea of local
reasoning: 
\[
\infer[\modif{C} \cap \fv{F} = \emptyset]{\set{\phi *_m F} C \set{\psi *_m C}}{\set{\phi} C \set{\psi}}
\]
where $\modif{n := \mathsf{new}} = \modif{n := t} = \modif{n := [m]} =
\set{n}$ and $\modif{[n] := t} = \modif{\mathsf{free}(n)} = \emptyset$
denotes the set of variables whose values are altered by the action.

%% file: roadmap.pdf_t
\begin{picture}(0,0)%
\includegraphics{roadmap.pdf}%
\end{picture}%
\setlength{\unitlength}{2960sp}%
\begingroup\makeatletter\ifx\SetFigFont\undefined%
\gdef\SetFigFont#1#2#3#4#5{%
  \reset@font\fontsize{#1}{#2pt}%
  \fontfamily{#3}\fontseries{#4}\fontshape{#5}%
  \selectfont}%
\fi\endgroup%
\begin{picture}(7659,2166)(3889,-1915)
\put(10576,-1246){\makebox(0,0)[b]{\smash{{\SetFigFont{9}{10.8}{\rmdefault}{\mddefault}{\updefault}{\color[rgb]{0,0,0}\S\ref{sec:reconf}. Reconfigurability}%
}}}}
\put(4801,-136){\makebox(0,0)[b]{\smash{{\SetFigFont{9}{10.8}{\rmdefault}{\mddefault}{\updefault}{\color[rgb]{0,0,0}\S\ref{sec:arch}. Architectures}%
}}}}
\put(6826,-136){\makebox(0,0)[b]{\smash{{\SetFigFont{9}{10.8}{\rmdefault}{\mddefault}{\updefault}{\color[rgb]{0,0,0}\S\ref{sec:sil}. \sil}%
}}}}
\put(6811,-1246){\makebox(0,0)[b]{\smash{{\SetFigFont{9}{10.8}{\rmdefault}{\mddefault}{\updefault}{\color[rgb]{0,0,0}\S\ref{sec:decidability}. Decidability}%
}}}}
\put(8626,-136){\makebox(0,0)[b]{\smash{{\SetFigFont{9}{10.8}{\rmdefault}{\mddefault}{\updefault}{\color[rgb]{0,0,0}\S\ref{sec:slarch}. \slarch}%
}}}}
\put(8611,-1246){\makebox(0,0)[b]{\smash{{\SetFigFont{9}{10.8}{\rmdefault}{\mddefault}{\updefault}{\color[rgb]{0,0,0}\S\ref{sec:behaviors}. Behaviors}%
}}}}
\end{picture}%

%% file: ex-intro-parametric.pdf_t
\begin{picture}(0,0)%
\includegraphics{ex-intro-parametric.pdf}%
\end{picture}%
\setlength{\unitlength}{3108sp}%
\begingroup\makeatletter\ifx\SetFigFont\undefined%
\gdef\SetFigFont#1#2#3#4#5{%
  \reset@font\fontsize{#1}{#2pt}%
  \fontfamily{#3}\fontseries{#4}\fontshape{#5}%
  \selectfont}%
\fi\endgroup%
\begin{picture}(3946,1512)(2239,-751)
\put(4996, 29){\makebox(0,0)[lb]{\smash{{\SetFigFont{8}{9.6}{\rmdefault}{\mddefault}{\updefault}{\color[rgb]{0,0,0}$\mathsf{t}(k)$}%
}}}}
\put(4546,-736){\makebox(0,0)[lb]{\smash{{\SetFigFont{9}{10.8}{\rmdefault}{\mddefault}{\updefault}{\color[rgb]{0,0,0}...}%
}}}}
\put(5986,-736){\makebox(0,0)[lb]{\smash{{\SetFigFont{9}{10.8}{\rmdefault}{\mddefault}{\updefault}{\color[rgb]{0,0,0}...}%
}}}}
\put(5446, 29){\makebox(0,0)[lb]{\smash{{\SetFigFont{8}{9.6}{\rmdefault}{\mddefault}{\updefault}{\color[rgb]{0,0,0}$\mathsf{l}(k)$}%
}}}}
\put(3916, 29){\makebox(0,0)[lb]{\smash{{\SetFigFont{8}{9.6}{\rmdefault}{\mddefault}{\updefault}{\color[rgb]{0,0,0}$\mathsf{l}(j)$}%
}}}}
\put(2296,-646){\makebox(0,0)[lb]{\smash{{\SetFigFont{7}{8.4}{\rmdefault}{\mddefault}{\itdefault}{\color[rgb]{0,0,0}$\mathsf{Semaphore}_i$}%
}}}}
\put(3421,-646){\makebox(0,0)[lb]{\smash{{\SetFigFont{7}{8.4}{\rmdefault}{\mddefault}{\itdefault}{\color[rgb]{0,0,0}$\mathsf{Task}_i$}%
}}}}
\put(4996,-646){\makebox(0,0)[lb]{\smash{{\SetFigFont{7}{8.4}{\rmdefault}{\mddefault}{\itdefault}{\color[rgb]{0,0,0}$\mathsf{Task}_k$}%
}}}}
\put(2296, 29){\makebox(0,0)[lb]{\smash{{\SetFigFont{8}{9.6}{\rmdefault}{\mddefault}{\updefault}{\color[rgb]{0,0,0}$\mathsf{p}(i)$}%
}}}}
\put(2746, 29){\makebox(0,0)[lb]{\smash{{\SetFigFont{8}{9.6}{\rmdefault}{\mddefault}{\updefault}{\color[rgb]{0,0,0}$\mathsf{v}(i)$}%
}}}}
\put(3466, 29){\makebox(0,0)[lb]{\smash{{\SetFigFont{8}{9.6}{\rmdefault}{\mddefault}{\updefault}{\color[rgb]{0,0,0}$\mathsf{t}(j)$}%
}}}}
\end{picture}%

%% file: controller.pdf_t
\begin{picture}(0,0)%
\includegraphics{controller.pdf}%
\end{picture}%
\setlength{\unitlength}{1973sp}%
\begingroup\makeatletter\ifx\SetFigFont\undefined%
\gdef\SetFigFont#1#2#3#4#5{%
  \reset@font\fontsize{#1}{#2pt}%
  \fontfamily{#3}\fontseries{#4}\fontshape{#5}%
  \selectfont}%
\fi\endgroup%
\begin{picture}(6024,2724)(2089,-2773)
\put(7651,-2161){\makebox(0,0)[b]{\smash{{\SetFigFont{6}{7.2}{\rmdefault}{\mddefault}{\updefault}{\color[rgb]{0,0,0}$\mathsf{q}(j_n)$}%
}}}}
\put(5251,-361){\makebox(0,0)[b]{\smash{{\SetFigFont{6}{7.2}{\rmdefault}{\mddefault}{\updefault}{\color[rgb]{0,0,0}$\mathsf{Controller}$}%
}}}}
\put(5251,-811){\makebox(0,0)[b]{\smash{{\SetFigFont{6}{7.2}{\rmdefault}{\mddefault}{\updefault}{\color[rgb]{0,0,0}$\mathsf{p}(i)$}%
}}}}
\put(2551,-2611){\makebox(0,0)[b]{\smash{{\SetFigFont{6}{7.2}{\rmdefault}{\mddefault}{\updefault}{\color[rgb]{0,0,0}$\mathsf{Slave}_1$}%
}}}}
\put(4051,-2611){\makebox(0,0)[b]{\smash{{\SetFigFont{6}{7.2}{\rmdefault}{\mddefault}{\updefault}{\color[rgb]{0,0,0}$\mathsf{Slave}_2$}%
}}}}
\put(7651,-2611){\makebox(0,0)[b]{\smash{{\SetFigFont{6}{7.2}{\rmdefault}{\mddefault}{\updefault}{\color[rgb]{0,0,0}$\mathsf{Slave}_n$}%
}}}}
\put(2551,-2161){\makebox(0,0)[b]{\smash{{\SetFigFont{6}{7.2}{\rmdefault}{\mddefault}{\updefault}{\color[rgb]{0,0,0}$\mathsf{q}(j_1)$}%
}}}}
\put(4051,-2161){\makebox(0,0)[b]{\smash{{\SetFigFont{6}{7.2}{\rmdefault}{\mddefault}{\updefault}{\color[rgb]{0,0,0}$\mathsf{q}(j_2)$}%
}}}}
\end{picture}%

%% file: draft.bbl
\begin{thebibliography}{1}
\providecommand{\url}[1]{\texttt{#1}}
\providecommand{\urlprefix}{URL }

\bibitem{CalcagnoYangOHearn01}
Calcagno, C., Yang, H., O’hearn, P.W.: Computability and complexity results
  for a spatial assertion language for data structures. In: FST TCS 2001,
  Proceedings, pp. 108--119. Springer (2001)

\end{thebibliography}
